\documentclass{aa} 

\usepackage{graphicx}
\usepackage{tabularx}
\usepackage{placeins}
\usepackage{abbriv}
\usepackage{acro}
\usepackage[english]{babel}
\usepackage{array}
\usepackage{capt-of}
\usepackage{natbib}
\newcolumntype{L}[1]{>{\raggedright\let\newline\\\arraybackslash\hspace{0pt}}m{#1}}
\newcolumntype{C}[1]{>{\centering\let\newline\\\arraybackslash\hspace{0pt}}m{#1}}
\newcolumntype{R}[1]{>{\raggedleft\let\newline\\\arraybackslash\hspace{0pt}}m{#1}}

\usepackage{txfonts}
\usepackage[colorlinks=true, allcolors=blue]{hyperref}
\def\kms{km~s$^{-1}$}
\graphicspath{{folder/}{Figures/}}

\begin{document}

   \title{VLT/X-shooter spectroscopy of massive young stellar objects \\ in the 30 Doradus region of the Large Magellanic Cloud\thanks{Based on observations at the European Southern Observatory under ESO program 090.C-0346(A).}}

   \author{
          M. L. van Gelder\inst{1,2}\fnmsep\thanks{\email{vgelder@strw.leidenuniv.nl}}
          \and
          L. Kaper\inst{1}
          \and
          J. Japelj\inst{1}
          \and
          M. C. Ram\'irez-Tannus\inst{1,3}
          \and
          L. E. Ellerbroek\inst{1}
          \and
          R. H. Barbá\inst{4}
          \and
          \\J. M. Bestenlehner\inst{5}
          \and
          A. Bik\inst{6}
          \and 
          G. Gr\"afener\inst{7}
          \and
          A. de Koter\inst{1,8}
          \and
          S. E. de Mink\inst{1,9}
          \and
          E. Sabbi\inst{10}
           \and 
          H. Sana\inst{8}
          \and
          \\M. Sewiło\inst{11,12}
          \and
          J. S. Vink\inst{13}
          \and
          N. R. Walborn\inst{10}\fnmsep\thanks{We regret to say that Dr. Nolan Walborn passed away early 2018. He was one of the initiators of this program.}
          }

\institute{
         Anton Pannekoek Institute, University of Amsterdam, Science Park 904, 1098XH Amsterdam, The Netherlands
         \and
         Leiden Observatory, Leiden University, PO Box 9513, 2300RA Leiden, The Netherlands
         \and
         Max-Plank-Institute for Astronomy, Königstuhl 17, 69117 Heidelberg, Germany
         \and
         Departamento de Física y Astronomía, Universidad de La Serena, Av. cisternas 1200 norte, La Serena, Chile
         \and
         Department of Physics and Astronomy, University of Sheffield, Sheffield S3 7RH, UK
         \and
         Department of Astronomy, Stockholm University, AlbaNova University Centre, 106 91 Stockholm, Sweden
         \and
         Argelander-Institut für Astronomie der Universität Bonn, Auf dem Hügel 71, 53121 Bonn, Germany
         \and
         Institute of Astrophysics, KU Leuven, Celestijnenlaan 200D, 3001 Leuven, Belgium
         \and
         Center for Astrophysics, Harvard-Smithsonian, 60 Garden Street, Cambridge, MA 02138, USA
         \and
         Space Telescope Science Institute, 2700 San Martin Drive, MD 21218, Baltimore, USA 
         \and
         Department of Astronomy, University of Maryland, College Park, MD 20742, USA
         \and
         CRESST II and Exoplanets and Stellar Astrophysics Laboratory, NASA Goddard Space Flight Center, Greenbelt, MD 20771, USA
         \and
         Armagh Observatory, College Hill, Armagh BT61 9DG, UK
             }

   \date{Received XXX; accepted XXX}
   
   \titlerunning{Massive young stellar objects in 30~Doradus}

\abstract{
The process of massive star ($M\geq8~M_\odot$) formation is still poorly understood. Observations of \aclp{MYSO} (MYSOs) are challenging due to their rarity, short formation timescale, large distances, and high circumstellar extinction. {Here, we present the results of a spectroscopic analysis of a population of \ac{MYSO}s in the \ac{LMC}.} We took advantage of the spectral resolution and wavelength coverage of X-shooter (300-2500~nm), mounted on the \acl*{ESO} \acl*{VLT}, to detect characteristic spectral features in a dozen \ac{MYSO} candidates near \acl{30Dor}, the largest starburst region in the Local Group hosting the most massive stars known. The X-shooter spectra are strongly contaminated by nebular emission. We used a scaling method to subtract the nebular contamination from our objects. We detect { H$\alpha,\beta$}, [O\,{\sc i}]~630.0~nm, \acl{CaII}, {[Fe\,{\sc ii}]~1643.5~nm}, fluorescent Fe\,{\sc ii}~1687.8~nm, H$_2$~2121.8~nm, Br$\gamma$, and CO bandhead emission in the spectra of multiple candidates. This leads to the spectroscopic confirmation of 10 candidates as {bona fide} \ac{MYSO}s. We compare our observations with photometric observations from the literature and find all \ac{MYSO}s to have a strong near-infrared excess. We compute lower limits to the brightness and luminosity of the \ac{MYSO} candidates, confirming the near-infrared excess and the massive nature of the objects. 
{No clear correlation is seen between the Br$\gamma$ luminosity and metallicity. }
Combining our sample with other \ac{LMC} samples results in a combined detection rate of {disk features such as} fluorescent Fe\,{\sc ii} and CO bandheads which is consistent with the Galactic rate (40\%).
Most of our \ac{MYSO}s show outflow features.
}


   \maketitle
%

\section{Introduction}
The formation process of massive stars ($M\geq8~M_\odot$) is still poorly understood \citep[e.g.][]{Zinnecker2007, Beuther2007, Tan2014}. Due to their short formation timescale ($\sim10^{4-5}$~yr) and the severe extinction ($A_\mathrm{V}\sim10-100$~mag) by the surrounding gas and dust, observations of \aclp{MYSO} (MYSOs) are challenging. Additionally, massive stars are rare and therefore typically located at larger distances. 

Already before reaching the \ac{ZAMS}, a \ac{MYSO} is expected to produce significant amounts of \ac{UV} radiation creating an expanding hyper- or \acl{UCHII} region \citep{Churchwell2002}. Despite the strong \ac{UV} radiation counteracting the accretion process through e.g. radiation pressure or photo-ionization \citep[e.g.][]{Wolfire1987,Krumholz2009,Kuiper2011,Kuiper2018}, the current belief is that mass accretes onto the central (proto-)star via an accretion disk similarly to low-mass stars.

If most of the mass is accreted through an accretion disk, \ac{MYSO}s are expected to be surrounded by massive, extended disks \citep{Beltran2016}. These disks have been observed (spectroscopically) around low ($M \lesssim 2~M_\odot$) and intermediate ($2 \lesssim M \lesssim 8~M_\odot$) mass stars \citep[e.g.][]{Ellerbroek2011,Alcala2014}. At \ac{NIR} wavelengths, disks around \ac{MYSO}s are commonly observed \citep[e.g.][]{Bik2006,Wheelwright2010,Ilee2013}, and recently disks around \ac{MYSO}s have been detected at sub-mm and cm wavelengths \citep[e.g.][]{Ilee2016,Ilee2018_2}. However, observations in the optical are scarce due to the high extinction. In the Galactic open cluster M17, \citet{Macla2017} identified a population of \ac{MYSO}s with disks by observing strong infrared excess and detecting double-peaked spectral lines in the optical. {Additionally, they see CO bandhead emission which can be produced in a Keplerian disk \citep[e.g.][]{Blum2004,Bik2004,Bik2006,Wheelwright2010,Ilee2013}, and seems to be highly dependent on the accretion rate \citep{Ilee2018_1}.} The Galactic \ac{RMS} survey has shown that the luminosity of accretion tracers such as Br$\gamma$ is correlated to the mass of the \acs{YSO}, and that disk tracing features such as CO bandheads and fluorescent Fe\,{\sc ii} emission are present in $\sim40\%$ of the \ac{MYSO}s \citep{Cooper2013,Pomohaci2017}.

Outflows are common in \ac{MYSO}s \citep[e.g.][]{Zhang2001,Zhang2005}. In the earliest stage of star formation, {they} are mostly molecular in origin \citep{Bachiller1996}. {At later stages, the outflow contains low-density atomic material and hence shows forbidden lines of e.g. O, N, S or Fe, either ionized or not \citep{Ellerbroek2013_1}}. Additionally, the [O\,{\sc i}]~630.0~nm line has been observed to originate from disk winds or the regions where the stellar \ac{UV} radiation impinges on the disk surface \citep[e.g.][]{Finkenzeller1985,Vanderplas2008}.

The Magellanic Clouds are interesting systems for studying massive star formation. 
The lower metallicity in the \ac{LMC} and \ac{SMC} \citep[about 1/2 and 1/5 of solar, respectively;][]{Peimbert2000,Rolleston2002} may influence the process of massive star formation. 
Most spectroscopic observations of \ac{MYSO}s in the \ac{LMC} and \ac{SMC} have been in the \ac{MIR} with the \textit{Spitzer}/\acl{IRS} \citep[\acs{IRS}; e.g.][]{vanloon2005,Oliveira2009,Oliveira2013,Seale2009,Seale2011,Woods2011,Ruffle2015,Jones2017}. 
In the \ac{NIR}, \ac{MYSO}s in the \ac{LMC} have also been observed by {\it AKARI} Infrared Camera \citep{Shimonishi2008,Shimonishi2010}.
The \ac{VLT} allow us now to spectroscopically observe (apparent) single \ac{MYSO}s in the Magellanic Clouds \citep[e.g.][]{Ward2016,Ward2017,Ward2017PhD,Rubio2018,Reiter2019}.

30~Doradus (\acs{30Dor}; also known as the Tarantula Nebula) is the most prominent massive star forming region in the Local Group. It is situated in the \ac{LMC} at a distance of about 50~kpc \citep{Pietrzynski2013}. Its dense massive core, \ac{R136}, has a total mass of up to $10^5~M_\odot$ and a cluster age of $\sim1.5$~Myr \citep{Selman2013, Crowther2016}. \ac{R136} hosts the most massive stars known with masses up to $300~M_\odot$ \citep{deKoter1997,Crowther2010}. The strong \ac{UV} radiation originating from the hot stars in \ac{R136} ionizes the surrounding cluster medium creating the largest H\,{\sc ii} region of the \ac{LMC} and the Local Group in general. Recent \ac{IMF} measurements show \ac{30Dor} to host an excess of about 30~\% in massive stars compared to the Salpeter \ac{IMF} \citep{Salpeter1955,Schneider2018_1}. \ac{30Dor} has been observed extensively in the \ac{VFTS} by obtaining high resolution spectra of about 800 O and B stars \citep{Evans2011}, and in the \ac{HTTP}, a panchromatic imaging survey with \ac{HST} of \ac{30Dor}'s stellar population down to masses of 0.5~$M_\odot$ \citep{Sabbi2013,Sabbi2016}. 

The massive star formation rate in \ac{30Dor} apparently rapidly increased about $7-8$~Myr ago \citep{Cigoni2015,Schneider2018_2}, but seems to have diminished about 1~Myr ago (though this may be an extinction effect; heavily extincted stars are not in the \ac{VFTS} and \ac{HTTP} samples).
Nevertheless, in the nebular region surrounding \ac{R136}, continuing massive star formation was first suggested by \citet{Hyland1992}, who indicated four candidate protostars with masses of $15-20~M_\odot$, and \citet{Rubio1992}, who discovered 17 \ac{NIR} sources to the north and west of \ac{R136}. Later investigations showed \ac{30Dor} to be a two-stage starburst region \citep{Walborn1992}, with substantial star formation going on in the surrounding region \citep{Walborn1999,Brandner2001}. More recently, \cite{Walborn2013} reported the top 10 \ac{MYSO} candidates using the \textit{Spitzer}/\ac{IRAC} 3--8~$\mu$m wavelength range from the \acl{SAGE} \citep[\acs{SAGE};][]{Meixner2006} program combined with the \acl{VMC} \citep[\acs{VMC};][]{Cioni2011} photometric observations. They derive masses and luminosities of about $10-30~M_\odot$ and $10^{4-5}~L_\odot$, respectively, by fitting the \ac{SED} to the \acs{YSO} models of \citet{Robitaille2006}. 
{Additionally, they conclude that all apparently single \ac{MYSO} candidates are Class I sources \citep[using the classification scheme based on the \ac{MIR} spectral index of][]{Andre2000}. Throughout this paper, we will refer to the empirically defined Classes and Types \citep[based on the appearance of the \ac{SED};][]{Chen2009}, and the theoretically defined Stages of \citet{Robitaille2006}. Since Class 0 objects may only be distinguished from Class I objects at sub-mm wavelengths we will combine these as Class 0/I.}

In this paper, we report the results of optical (300~nm) to \ac{NIR} (2500~nm) follow-up observations with \ac{VLT}/X-shooter \citep{Vernet2011} of the top 10 {\it Spitzer} \ac{MYSO} candidates of \citet{Walborn2013}. The aim is to confirm their \ac{MYSO} nature using optical and \ac{NIR} emission features. In Section~\ref{sec:targsamp} we introduce the target sample and photometry from the literature. Our \ac{VLT}/X-shooter observations, data reduction, and methods of subtracting nebular contamination are described in Section~\ref{sec:VLTxshoo}. Our analysis of the spectra and classification of the targets, leading to the confirmation of {10} candidates as \ac{MYSO}s, are presented in Section~\ref{sec:spectarg}. We discuss the results in Section~\ref{sec:discussion}. Section~\ref{sec:conclusion} provides a summary. 


\section{Target sample}
\label{sec:targsamp}
Our targets were selected based on the top 10 \ac{MYSO} candidates of \citet{Walborn2013}. They selected the 10 brightest targets in the {\it Spitzer}/\ac{IRAC} bands (labeling them as S1--S10), and combined these with \ac{VMC} photometry. In this work we adopt the same names. In Fig.~\ref{fig:30Dor} we show the positions of our targets in a \ac{VMC} $Y$ (1.02~$\mu$m), $J$ (1.25~$\mu$m), and $K_\mathrm{s}$ (2.15~$\mu$m) three-color image. In the \ac{VLT}/X-shooter observation of S5, a total of 6 objects could be identified within the slit range which we labeled S5-A,B,C,D,E,F (see Fig.~\ref{fig:S5_targets}). S8 is an unresolved double-system, which we discuss as one single target. S10-B and S10-C were also unresolved and are labeled as S10-BC in this work. Supplementary to S1 to S10 (and additional targets on the slit), our target sample includes R135; a \ac{WR} star of \ac{SpT}~WN7h+OB \citep{Evans2011} located in the vicinity of S3 and S3-K. A log of our VLT/X-shooter observations of a total of 23 sources is presented in Table~\ref{tab:obslog}. 

\subsection{Photometry}
\label{sec:phot}
With the {\ac{VISTA} IR camera \citep[][]{Dalton2006}} $J$-band and $K_\mathrm{s}$-band and \acl{IRSF} \citep[IRSF;][]{Kato2007} $H$-band (1.63~$\mu$m) photometric observations {presented by} \citet{Walborn2013}, we constructed a \ac{NIR} color-magnitude and color-color diagram of our targets; see Fig.~\ref{fig:photfigure}. For S5-A and R135, all magnitudes are from the \acl{2MASS} \citep[2MASS;][]{Cutri2003}. We assumed a $J>19$ lower limit for S3, S9, and S10-BC. We lack (part of) the relevant photometric observations of S1-SE, S5-B, S5-C, S5-D, and S5-F, hence these objects do not appear in Fig.~\ref{fig:photfigure}. The \ac{ZAMS} is computed using \ac{MIST} models \citep{Paxton2011, Paxton2013, Paxton2015, Dotter2016, Choi2016} with half the solar metallicity \citep{Rolleston2002} and a distance to the \ac{LMC} of 50~kpc \citep{Pietrzynski2013}. We also plot the positions of O~V stars of \citet{Martins2006}. Using the \citet{MaizApellaniz2014} extinction law for \ac{30Dor}, we draw the reddening lines of an O3~V star for $R_\mathrm{V}=3.1$ (average Galactic value) and $R_\mathrm{V}=5.0$ \citep[values observed in \ac{30Dor}, e.g.][]{Bestenlehner2011,Bestenlehner2014}. 

\begin{figure}[t]
\centering
\includegraphics[width=1\linewidth]{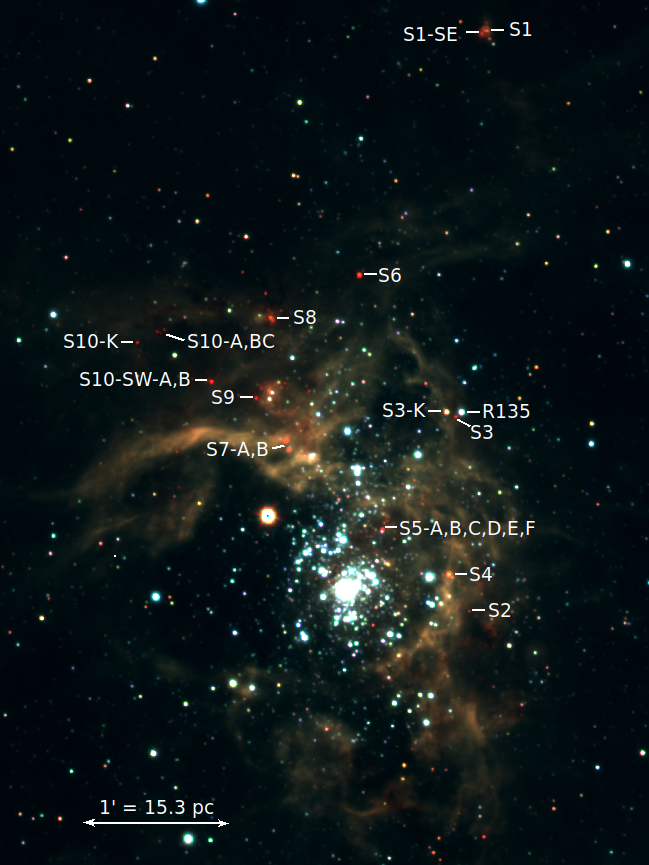}
\caption{The \ac{30Dor} nebula seen in the $Y$ (blue), $J$ (green) and $K_\mathrm{s}$ (red) bands with the VMC \protect\citep{Cioni2011}. North is up and east is to the left. The observed X-shooter targets are labeled in white (see also Table~\protect\ref{tab:obslog}). The central massive cluster \ac{R136} is the bright cluster of stars in the {middle}. }
\label{fig:30Dor}
\end{figure}

\begin{table*}[t]
\begin{center}
\caption[Log of VLT/X-shooter observations]{A list of the VLT/X-shooter observations used in this paper.}
\footnotesize{
\begin{tabular}{llllccccl}
\hline
\hline
Object & RA (J2000) & Dec (J2000) & Date & \multicolumn{3}{c}{Exp.time (s)} & Seeing & Remarks \\ \cline{5-7}
 & (hh:mm:ss.ss) & (dd:mm:ss.s) &  (dd-mm-yyyy) & UVB & VIS & NIR & (") & \\ 
\hline
S1 & 05:38:31.62 & -69:02:14.6 & 30-11-2012 & 4$\times$670 & 4$\times$700 & 4$\times$50 &  0.7 &  \\
S1-SE & 05:38:32.25 & -69:02:14.0 & 30-11-2012 & 4$\times$670 & 4$\times$700 & 4$\times$50 &  1.0 & \\
S2 & 05:38:33.09 & -69:06:11.7 & 28-01-2013 & 4$\times$670 & 4$\times$700 & 4$\times$50 & 1.2 & \\
S3 & 05:38:34.05 & -69:04:52.2 & 26-01-2013 & 4$\times$670 & 4$\times$700 & 4$\times$50 & 1.4 & \\
S3-K & 05:38:34.69 & -69:04:50.0 & 27-01-2013 & 4$\times$700 & 8$\times$330 & 4$\times$50 & 2.1 & \\
S4 & 05:38:34.60 & -69:05:56.8 & 28-01-2013 & 4$\times$670 & 4$\times$700 & 4$\times$50 & 1.2 & \\
S5-A,B,C,D,E,F & 05:38:39.68\tablefootmark{1} & -69:05:37.9\tablefootmark{1} & 26-01-2013 & 4$\times$700 & 8$\times$330 & 4$\times$50 & 0.8 & 6 objects on slit \\
S6 & 05:38:41.36 & -69:03:54.0 & 24-01-2013 & 4$\times$670 & 4$\times$700 & 4$\times$50 & 1.5 & \\
S7-A,B & 05:38:46.84\tablefootmark{2} & -69:05:05.4\tablefootmark{2} & 24-01-2013 & 4$\times$670 & 4$\times$700 & 4$\times$50 & 2.2 & 2 objects on slit \\
S8 & 05:38:48.17 & -69:04:11.7 & 28-01-2013 & 4$\times$670 & 4$\times$700 & 4$\times$50 & 0.8 & 2 unresolved objects\\
S9 & 05:38:49.27 & -69:04:44.4 & 24-01-2013 & 4$\times$670 & 4$\times$700 & 4$\times$50 & 1.5 & \\
S10-A,BC & 05:38:56.31\tablefootmark{3} & -69:04:16.1\tablefootmark{3} & 29-01-2013 & 4$\times$670 & 4$\times$700 & 4$\times$50 & 1.5 & S10-B$\&$C unresolved\\
S10-K & 05:38:58.38 & -69:04:21.6 & 28-01-2013 & 4$\times$670 & 4$\times$700 & 4$\times$50 & 1.2 & \\
S10-SW-A,B & 05:38:52.72\tablefootmark{4} & -69:04:37.5\tablefootmark{4} & 28-01-2013 & 4$\times$670 & 4$\times$700 & 4$\times$50 & 1.3 & 2 objects on slit \\
R135 & 05:38:33.62 & -69:04:50.4 & 14-01-2013 & 2$\times$210 & 2$\times$240 & 2$\times$50 & 1.2 & Wolf-Rayet star \\
\hline
\end{tabular}
}
\label{tab:obslog}
\tablefoot{All observations were carried out under ESO program 090.C-0346(A). {The object names are the same as defined by \protect\citet{Walborn2013}, where we introduced additional letters if multiple or additional objects were identified on the X-shooter slit}. \\ 
\tablefoottext{1}{Position of S5-E.} \tablefoottext{2}{Position of S7-A.} \tablefoottext{3}{Position of S10-A.} \tablefoottext{4}{Position of S10-SW-A.}}
\end{center}
\end{table*}

Almost all our \ac{MYSO} candidates are located far above the reddening line for an O3~V star indicating the presence of a strong \ac{NIR} excess. For our brightest $K_\mathrm{s}$-band target, S4, this excess may be $\gtrsim$5~mag suggesting the object to be $\gtrsim$100~times brighter in the $K_\mathrm{s}$-band than the central (proto)star would be.
{The \ac{NIR} excess is considerably stronger compared to Galactic \ac{MYSO} observations of \citet{Bik2006} and \citet{Macla2017}, and have on average a bluer $J-K$ color and brighter $K_\mathrm{s}$-band magnitude than most objects in the sample of \citet{Cooper2013}. This is an observational bias; our targets were selected as the brightest \ac{NIR} and \ac{MIR} targets in \ac{30Dor}. 

In star forming regions the extinction is highly dependent on the line of sight \citep{Ellerbroek2013_2,Macla2018}. \citet{DeMarchi2016} determined an average total-to-selective extinction $R_\mathrm{V}$ of 4.5 towards the \ac{30Dor} region, which is about midway in between the two reddening lines in Fig.~\ref{fig:photfigure}. \citet{Walborn2013} measure an extinction of $A_\mathrm{V} \lesssim10$~mag for all apparently single \ac{MYSO} candidates (i.e. S2, S3, S3-K, S4, S6, S7-A, and S10-K). They find S3 as the most extincted object with $A_\mathrm{V}=10$ and S4 as one of the least extincted objects with $A_\mathrm{V}=1.8$. {Since we are confronted with a \ac{NIR} excess, we can not get an accurate estimate of the extinction from the color-color diagram in Fig.~\ref{fig:photfigure}. However, the positions of our targets in the color-color diagram suggest a $>$5~mag higher extinction than the values computed by \citet{Walborn2013}.}

\begin{figure*}[t]
\centering
\includegraphics[width=1.0\linewidth]{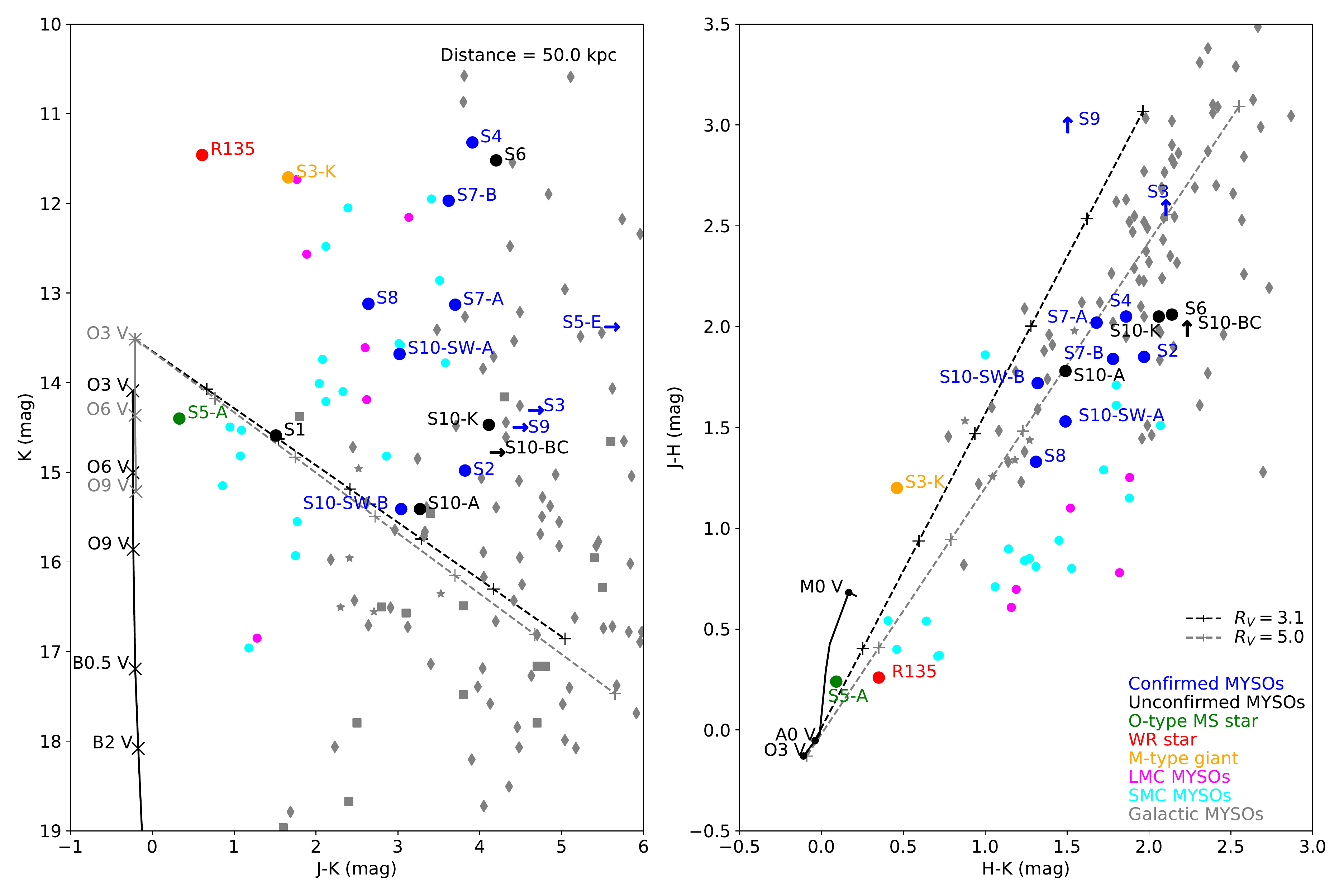}
\caption[Near-infrared color-magnitude and color-color diagrams of our targets]{{\it Left:} \ac{NIR} color-magnitude diagram of our targets. With dots we show the objects with accurate photometry \protect\citep{Cutri2003,Walborn2013}; with arrows we show the positions of objects based on a $J>19$ lower limit. Errors on the datapoints were omitted for clarity but are typically $\sim$0.05~mag. In blue we show the \ac{MYSO}s confirmed in this work, and in black the unconfirmed \ac{MYSO} candidates. The green, red, and yellow dots indicate a MS, {M-type giant}, and \ac{WR} star, respectively. {The magenta dots are other LMC MYSOs \protect\citep{Ward2016,Reiter2019}, and the cyan dots are SMC MYSOs \protect\citep{Ward2017,Rubio2018}.} The gray stars, squares, and diamonds are respectively the \ac{MYSO}s of \protect\citet{Macla2017}, \protect\citet{Bik2006}, and \protect\citet{Cooper2013} projected at a distance of 50~kpc. The ZAMS is shown as a black line, and the gray line indicates the position of O~V stars of \protect\citet{Martins2006}. {The dashed black and gray lines are the reddening lines of an O3~V star for a $R_\mathrm{V}$ of $3.1$ and $5.0$ respectively, where the crosses from left to right represent a visual extinction $A_\mathrm{V}$ of $5$, $10$, $15$, $20$, $25$, and $30$~mag, respectively.} Note that almost all our \ac{MYSO} candidates are located above the reddening lines, implying a strong \ac{NIR} excess. {\it Right:} \ac{NIR} color-color diagram of our targets. The colors and symbols are the same as in the left plot. Most of our targets show evidence of a \ac{NIR} excess due to their position at the right of the reddening lines. }
\label{fig:photfigure}
\end{figure*}

\begin{figure}[t]
\centering
\includegraphics[width=1.0\linewidth]{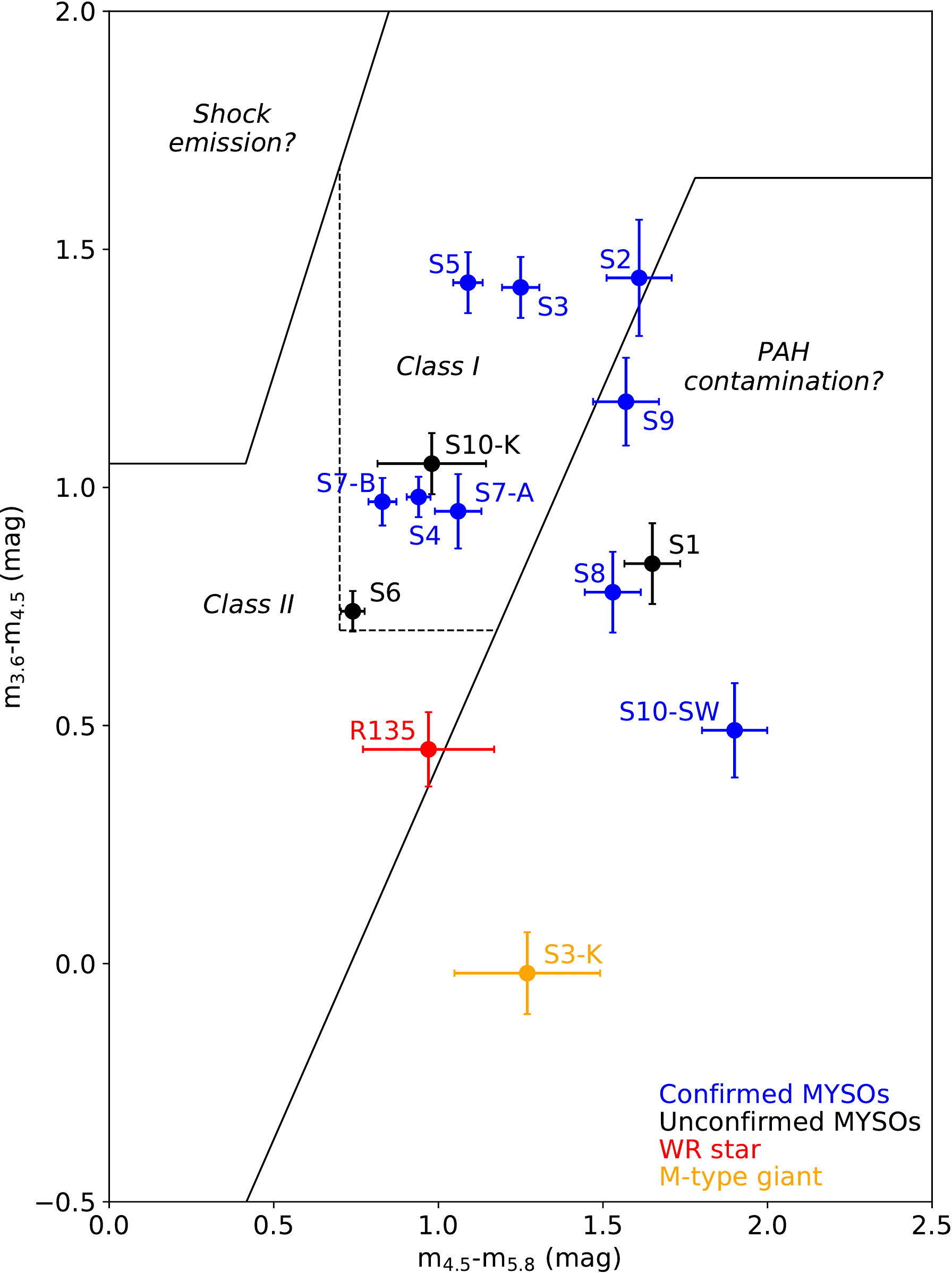}
\caption{A MIR color-color diagram showing the {\it Spitzer}/IRAC colors of our confirmed \ac{MYSO}s (blue dots), and the unconfirmed \ac{MYSO} candidates (black dots). The {\it Spitzer}/IRAC photometric points of the MYSO candidates and S3-K are from \protect\citet{Walborn2013}, the photometric points of R135 are from the {\it Spitzer}/SAGE catalog \mbox{\citep{Meixner2006}}. {Following the classification scheme of \protect\citet{Gutermuth2009}, we indicate the Class I and Class II regions, and the regions where the {\it Spitzer}/IRAC colors might include unresolved knots of shock emission or resolved structured PAH emission.}}
\label{fig:photIRAC}
\end{figure}

With the (available) {\it Spitzer} photometric points of \citet{Walborn2013} we created a \ac{MIR} color-color diagram in Fig.~\ref{fig:photIRAC}. {Following the classification scheme of \citet{Gutermuth2009}, we indicate the regions of Class I and Class II sources, and where the {\it Spitzer}/IRAC bands might be dominated by unresolved knots of shocked emission, or emission by resolved structures of \aclp{PAH} (PAHs).} Figure~\ref{fig:photIRAC} suggests that none of our targets should be Class II objects, and that some targets may in fact be \ac{PAH} dominated structures rather than \ac{MYSO}s. However, all objects in the \ac{PAH} contaminated region (except for S3-K and S9) are resolved into multiple components in higher angular resolution data, which could explain their position. We note that, according to the classification scheme of \citet{Megeath2004}, S6 should be a Class II \ac{MYSO}, whereas all other \ac{MYSO} candidates are Class I objects.

\section{Reduction of VLT/X-shooter observations}
\label{sec:VLTxshoo}
We took spectra of our targets using the X-shooter spectrograph mounted on the \ac{VLT} \citep{Vernet2011}. X-shooter is an intermediate resolution (R$\sim$4000-17\,000) slit spectrograph covering a wavelength range from 300~nm to 2500~nm, divided over three arms: \acf{UVB}, \acf{VIS}, and \acf{NIR}. 

In Table~\ref{tab:obslog} we present the log of our \ac{VLT}/X-shooter observations. {For the targets which have been previously resolved into multiple systems} \citep[S7, S8, S10 and S10-SW;][]{Hyland1992, Rubio1992, Walborn2013}, the X-shooter slit was positioned such that all targets would be observed within a single observation. 
Our spectra were taken in nodding mode, splitting the integration time of each observation (except R135 and S3-K) into 4 nodding observations of each 670~s, 700~s, and 50~s for the \ac{UVB}, \ac{VIS}, and \ac{NIR} arms, respectively. Given their brightness, the observation in the \ac{VIS} arm was split into two for R135, S3-K, and S5. The slit length was 11" for each arm, and the width was 1.0", 0.9" and 0.6" for the \ac{UVB}, \ac{VIS} and \ac{NIR} arms, respectively. This results in a resolving power of 5100, 8800 and 8100, respectively. For the objects S1, S3-K, S4, S5, S6, S8, and  R135, a slit width of 0.4" was used in the \ac{NIR} arm, corresponding to a resolving power of 11\,300. Unfortunately the \ac{ADC} was not working during our observations, which complicated the data reduction. The latter was especially an issue for the observations where we could not arrange the slit according to the parallactic angle (e.g. in the case multiple targets are included in one observation). 

\begin{figure*}[t]
\centering
\includegraphics[width=1\linewidth]{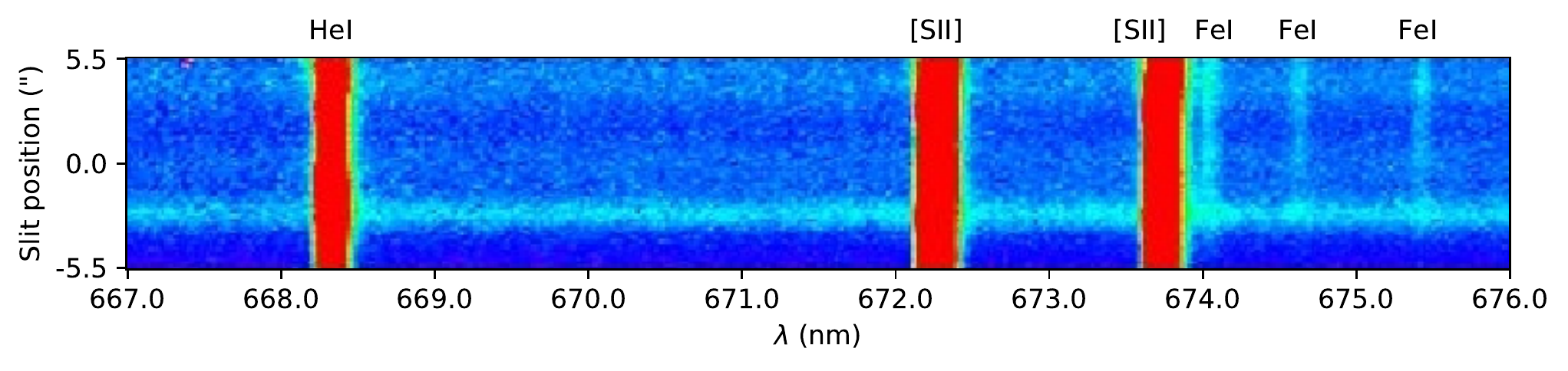}
\caption{Reduced 2D science frame of S4 as assessed from the pipeline. We show a single nodding position. The color indicates flux, {where red and blue are high and low flux, respectively}. The spatial position is with respect to the center of the slit. The figure is centered around the [S\,{\sc ii}]~671.6,673.1~{nm}  lines (the two bright lines on the right) and the He\,{\sc i}~667.8~{nm} line (the bright line on the left), all originating from the nebula. The three weak lines on the right are nebular Fe\,{\sc i} lines. The continuum of S4 is visible in the lower part of the 2D frame. Additionally, some nebular continuum is visible, and at the top some weak continuum from an object not analyzed in this work. }
\label{fig:slit_stare}
\end{figure*}
\begin{figure*}[t]
\centering
\includegraphics[width=1.0\linewidth]{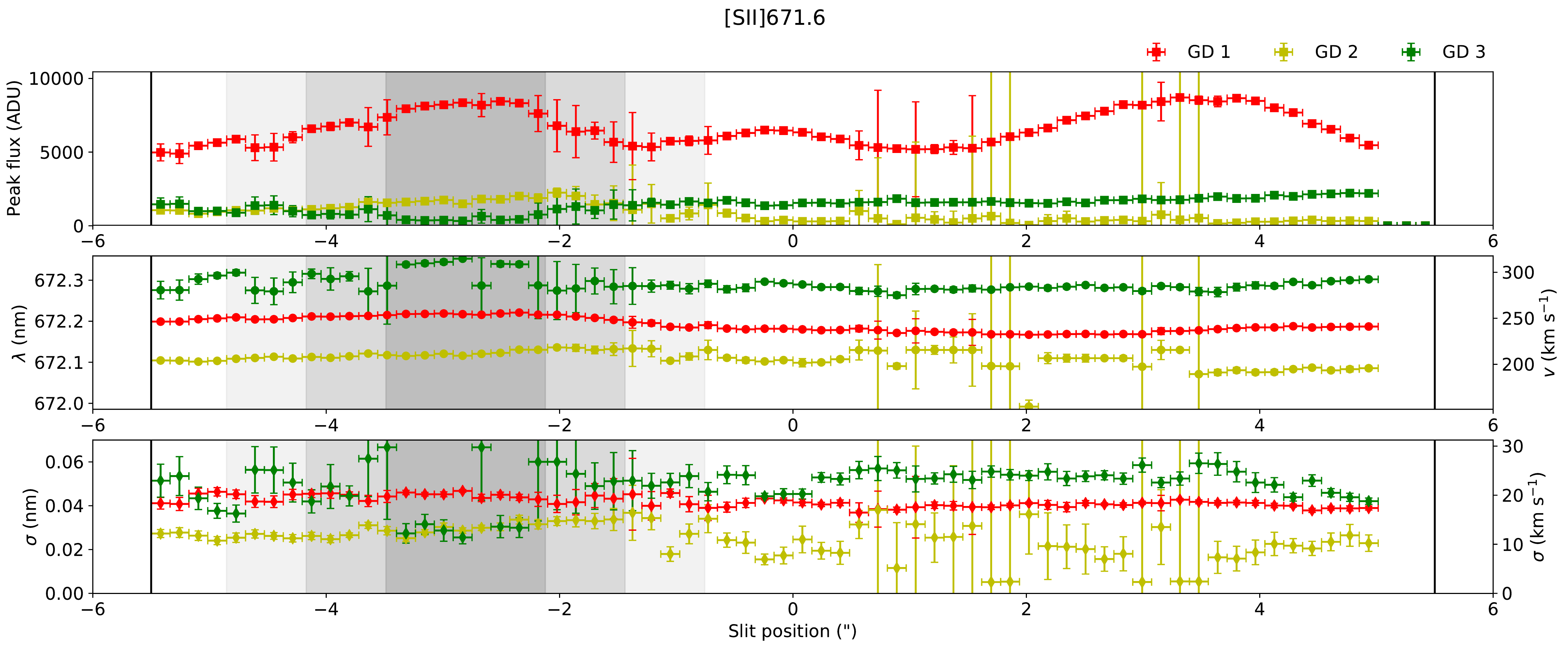}
\caption{Variations in the {[S\,{\sc ii}]}~671.6~{nm} nebular line along the slit, modeled with a triple Gaussian model. The variations are shown for one of the nodding mode positions of S4 (i.e. the middle red line in Fig.~\protect\ref{fig:slit_stare}). {\it Top:} Variations in the peak flux of each GD along the slit. {\it Middle:} Variations in the central wavelength (or RV shift) of each GD along the slit. {\it Bottom:} Variations in $\sigma$ of each GD along the slit. In the dark, lighter and lightest vertical gray zones we show the 1$\sigma$, 2$\sigma$ and 3$\sigma$ seeing ranges of the object, respectively. For more information, see the text.}
\label{fig:S671var}
\end{figure*}
\begin{figure*}[t]
\centering
\includegraphics[width=1.0\linewidth]{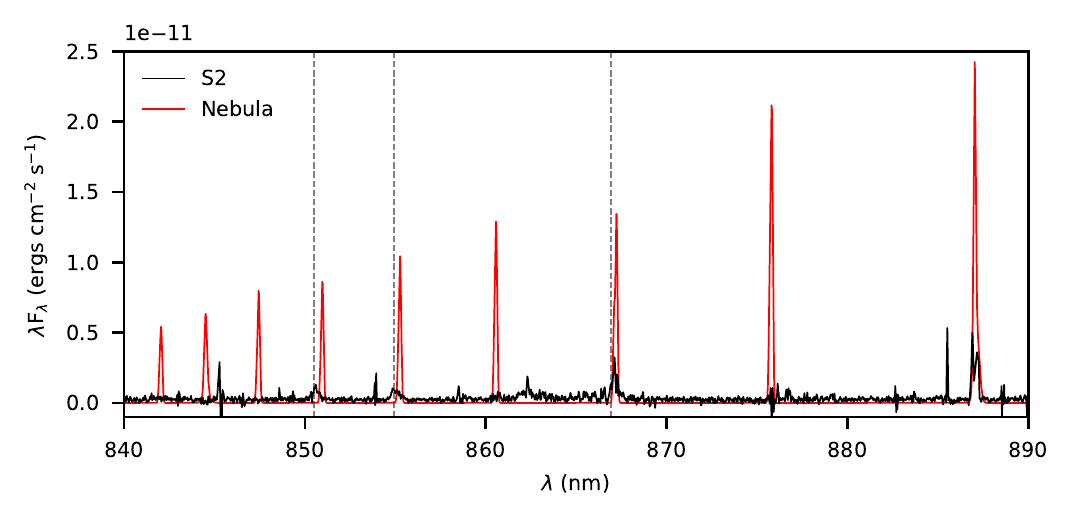}
\caption{The nebular and sky subtracted spectrum of S2 (black) and subtracted nebular contribution (red), centered around the Ca\,{\sc ii} IRT lines (positions indicated with vertical gray dashed lines). The nebular lines are Pa-11--19. Note that some residuals persist after the nebular subtraction. These features could (also) be produced by the object itself. }
\label{fig:S2_nebsub}
\end{figure*}

We reduced the data using the X-shooter Workflow for Physical Mode Date Reduction version 2.9.3. \citep{Modigliani2010}. The pipeline was implemented in the ESO-Reflex version 2.8 \citep{Freudling2013}. We performed a flux calibration using the spectrophotometric standard stars from the \ac{ESO} database. The \ac{UVB} and \ac{VIS} fluxes were scaled to match the absolute fluxes in the \ac{NIR} arm. We corrected our spectra for telluric features using the software tool \textsc{molecfit} version 1.2.0 \citep{Smette2015, Kausch2015}.

\subsection{Correcting for nebular emission}
All our spectra are contaminated by strong nebular emission lines, see Fig.~\ref{fig:slit_stare}. Early type stars show mostly H and He lines in the X-shooter wavelength range that also have a nebular counterpart. Since these nebular counterparts are very strong, they first need to be removed before the spectral features originating from the \ac{MYSO} can be analyzed. Fortunately, the spectral resolution of X-shooter allows us to discriminate between the nebular emission and spectral features originating from the \ac{MYSO}s \citep{Kaper2011}. 
 
Our data were acquired in nodding mode; however, the nodding mode sky reduction results in an erroneous subtraction of the nebular emission as the nebular emission lines vary in strength and position (i.e. \ac{RV}) along the slit. We investigated these variations by reducing the nodding mode data in staring mode. The atmospheric contribution has not yet been subtracted at this stage. 

\subsubsection{Modeling the nebular variations}
\label{subsubsec:nebmodeling}
To subtract the nebular lines, their variations along the X-shooter slit need to be modeled. We do this by extracting the spectrum from each spatial pixel along the slit, and fitting the nebular lines. {As line profile models we used mainly a \acf{GD}, \acl{FGD} \citep[\acs{FGD};][]{Blazquez2008}, and \acl{MD} \citep[\ac{MD};][]{Moffat1969}. The definitions of these functions can be found in Appendix~\ref{app:nebcor}.}

The \ac{30Dor} region consists of multiple velocity components along each line of sight with a velocity dispersion of up to several tens of \kms\text{} \citep{Torres-Flores2013,Mendes2017}. If we identified multiple components in a nebular line, we used a combination of the models introduced above (e.g. if the nebular line had two velocity components we used two \ac{GD}s to model these nebular lines). We assumed that the local continuum around a nebular line is roughly linear and therefore modeled it with a linear function. We fitted the lines using a minimizing $\chi^2$ fitting routine.

As nebular lines are typically narrow, the range around the nebular line through which the continuum was fitted was typically about $\sim$0.2~nm, $\sim$0.4~nm, $\sim$0.7~nm for the \ac{UVB}, \ac{VIS}, and \ac{NIR} arms, respectively. This corresponds to about $\sim$2, $\sim$5, and $\sim$4 velocity resolution elements, respectively (or to about $\sim$10, $\sim$20, $\sim$12 data points per range). The number of wavelength bins per nebular line is thus relatively low, making it difficult to fit the lines. Additional to nebular lines we fit known sky emission lines ([O\,{\sc i}]~557.7~{nm} in the VIS arm, O$_{2}$~1280.3~{nm} in the NIR arm, in the UVB arm no sky emission features are apparent) to monitor possible sky variations along the slit. {Sky variations were absent in all observations (but for the usual variation at $\gtrsim$2.25~$\mu$m).} In Fig.~\ref{fig:S671var} we show the modeled variation of the [S\,{\sc ii}]~671.6~{nm} nebular line (middle of the three red lines in Fig.~\ref{fig:slit_stare}) along the spatial direction of the X-shooter slit (y-axis in Fig.~\ref{fig:slit_stare}). {The line models for a few nebular lines can be found in Appendix~\ref{app:nebplot}}. 

In modeling the nebular lines we note that the measured variations in peak flux and position along the X-shooter slit differed between different ionization stages (per element). We determined empirically that the {different species} may be subdivided into two main categories. We find that low ionization species (e.g. [O\,{\sc i}], [O\,{\sc ii}], [N\,{\sc i}], [N\,{\sc ii}], [S\,{\sc ii}] and [Ca\,{\sc ii}]) are in one category, and high ionization species (e.g. [O\,{\sc iii}], [S\,{\sc iii}], [Ar\,{\sc iii}], [Ne\,{\sc iii}], [Fe\,{\sc ii}], [Ni\,{\sc ii}]) and non-forbidden transitions (e.g. He\,{\sc i}, O\,{\sc i}, Ba, Pa and Br) in the other. This difference in variations along the X-shooter slit was taken into account when subtracting the nebular contamination.

\subsubsection{Subtraction of nebular lines}
The angular resolution of our observations is limited by the seeing. To accurately model the nebular emission in the spectra of our targets, the spatial extent of the target due to seeing has to be taken into account. We corrected for this by summing all spatial pixels within the angular resolution range of the target (typically twice the angular resolution, but single angular resolution was used in crowded fields), and fitting the nebular lines in the spectrum extracted from the range of spatial pixels. Similarly we fit the nebular lines for all ranges of spatial pixels outside of the object range (i.e. all positions where we expect no flux of an \ac{MYSO} to contribute to the nebular line flux).  

We used a scaling method to subtract the nebular contamination from the \ac{MYSO} candidate spectrum. In this method we compute the nebular contribution in the object spectrum by scaling the nebular spectrum offset with respect to the object. The method goes as follows: we selected a region off-source to set as our "reference" nebular region, which was often a region relatively close to the object or a location where the nebular peak flux was approximately equally strong as the (at this point) approximated contribution on-source. At this off-source location, we set a reference line for each nebular line category, which was assumed to solely have a nebular (and thus no stellar) contribution. The forbidden lines are usually excellent candidates for this, however if these {could not be used} we used He\,{\sc i}, Ba, or Pa lines for scaling. The latter was often necessary for the subtraction in the \ac{NIR} arm as no strong forbidden lines are present in this wavelength range. The default scaling forbidden lines were [O\,{\sc ii}]~372.9~{nm} and [N\,{\sc ii}]~658.3~{nm} for the first scaling category in the \ac{UVB} and \ac{VIS} arm, respectively. For the second scaling category we used the [Ne\,{\sc iii}]~386.9~{nm} and [S\,{\sc iii}]~631.0~{nm} in the \ac{UVB} and \ac{VIS} arm, respectively. In the \ac{NIR} arm we generally only saw one category for which we used the Pa~1281.8~{nm} line. We note that forbidden lines may also originate from e.g. jets of the \ac{MYSO} candidates. {However, these jet lines are typically shifted (in \ac{RV}) with respect to the nebular lines} \citep{Ellerbroek2011, Ellerbroek2013_1, Mcleod2018}.

We determined the nebular contamination on-source by computing the scaling of the model parameters of the reference nebular line between the on-source spectrum and off-source spectrum (e.g. $N_\text{sca} = N_\text{on}/N_\text{off}$ for the peak flux $N$, and similar for the other model parameters). Next, we applied these scaling parameters to the off-source nebular line models, resulting in a model of the nebular lines in the on-source spectrum. We subtract the nebular emission simply by subtracting the scaled nebular model from the on-source data, giving us the object spectrum with solely a sky contribution left.

We estimate the sky contribution by also subtracting the nebular contamination at an off-source location. Note that this location is not necessarily the same position as the reference nebula. The local nebular contamination was subtracted without any scaling since we assume {that} no other emission sources {are} present at these distances from the object. This results in the sky continuum and emission lines at that spatial position. Some additional nebular continuum is present in the atmospheric spectra; however, if we assume this to be approximately constant along the spatial direction this nebular continuum contribution will also be present in the object spectrum. Assuming the atmospheric continuum and emission to be constant along the slit as well, we subtracted the sky of the object using this atmospheric spectrum. This gives us the object spectra free of nebular and atmospheric contamination but for some subtraction residuals. 

\begin{table*}[t]
\begin{center}
\caption{Detected spectral features and {target classification.} }
\footnotesize{
\begin{tabular}{@{\extracolsep{2pt}}llllllllllll@{}}
\hline
\hline
         &                      & \multicolumn{4}{c}{Accretion features}               & \multicolumn{2}{c}{Disk features} & \multicolumn{3}{c}{Outflow features} &                 \\ \cline{3-6} \cline{7-8} \cline{9-11}
Target   & Other names\tablefootmark{1} & H$\alpha,\beta$\tablefootmark{2} & Pa        & Ca\,{\sc ii}     & Br$\gamma$ & Fe\,{\sc ii}          & CO\tablefootmark{3}  & {[O\,{\sc i}]}   & {[Fe\,{\sc ii}]}  & H$_2$    & Classification  \\
         &                      &                 & series    & IRT       &            & 1687.8         &                  & 630.0       & 1643.5      & 2121.8   &                 \\ \hline
S1       & --                   & --              & A$^{w}$   & --        & --         & --             & --               & E$^{*}$     &  --         & --       & --              \\
S1-SE    & --                   & --              & --        & --        & --         & --             & --               & --          &  --         & E$^{w}$  & --              \\
\textbf{S2}       & IRSW-11, NIC07a      & E$^{rs}$        & --        & E         & E$^{*}$    & --             & --               & --          &  --         & --       & MYSO            \\
\textbf{S3}       & --                   & E               & E$^{w,*}$ & --        & E$^{w,*}$  & E$^{w}$        & --               & E$^{*}$     &  --         & E        & MYSO            \\
S3-K     & --                   & --              & A$^{w,*}$ & A         & --         & --             & A                & --          &  --         & --       & M-type giant    \\
\textbf{S4}      & IRSW-30, NIC03a, P3  & E$^{rs}$        & E$^{*}$   & E         & E          & E$^{w}$        & E$^{w}$          & E    &  E$^{w}$    & E        & MYSO            \\
S5-A     & IRSW-133, VFTS 476   & A               & A         & --        & --         & --             & --               & --          &  --         & --       & MS star         \\
S5-B     & --                   & --              & A$^{w}$   & --        & E$^{w}$    & --             & --               & --          &  --         & --       & MS star         \\
S5-C     & --                   & --              & A$^{w}$   & --        & --         & --             & --               & --          &  --         & --       & MS star         \\
S5-D     & --                   & --              & --        & A         & --         & --             & --               & E$^{*}$     &  --         & --       & Foreground star \\
\textbf{S5-E}     & IRSW-127             & --              & --        & E$^{w}$   & --         & --             & E$^{w}$          & --          &  --         & E        & MYSO            \\
S5-F     & --                   & --              & --        & --        & --         & --             & --               & --          &  --         & --       & MS star         \\
S6       & NIC16a, P2           & --              & E$^{*}$   & --        & --         & --             & --               & --          &  --         & --       & --              \\
\textbf{S7-A}     & IRSN-122, NIC12b, P1 & E               & E$^{w,*}$ & --        & E          & E              & E$^{w}$          & E$^{*}$     &  --         & E        & MYSO            \\
\textbf{S7-B}     & IRSN-126, NIC12d, P1 & E               & E$^{w,}$\tablefootmark{4} & --        & E$^{rs}$   & E              & --               & --          & E$^{w,}$\tablefootmark{4}     & E        & MYSO            \\
\textbf{S8}       & IRSN-137, NIC15b, P4 & E$^{rs}$        & E$^{*}$   & E$^{w}$   & E$^{w,*}$  & --             & --               & --          & --          & --       & MYSO\tablefootmark{5}     \\
\textbf{S9}       & IRSN-152             & E               & --        & --        & E$^{*}$    & E              & --               & --          & --          & E        & MYSO            \\
S10-A    & --                   & --              & E$^{*}$   & --        & --         & --             & --               & --          & --          & --       & --              \\
S10-BC   & --                   & --              & --        & --        & --         & --             & --               & --          & --          & E$^{w}$  & --              \\
S10-K    & --                   & E$^{w}$         & --        & E$^{w,*}$ & --         & --             & --               & --          & --          & --       & --              \\
\textbf{S10-SW-A} & IRSN-169, S11        & E$^{rs}$        & E         & --        & E          & E              & --               & --          & --          & E        & MYSO            \\
\textbf{S10-SW-B} & IRSN-170, S11        & E$^{w,rs}$      & --        & E$^{w}$   & --         & --             & --               & E$^{w}$     & --          & E$^{w}$  & MYSO            \\
R135     & VFTS 402             & E               & E         & --        & E          & --             & --               & --          & --          & --       & WR star         \\       
\hline
\end{tabular}
}
\tablefoot{A indicates an absorption feature, E an emission feature, $w$ a weak feature, $rs$ a red shoulder in the emission, $*$ a bad nebular residual, and -- the absence of the feature. The classification of a candidate as \ac{MYSO} was based on the presence of the listed spectral features. A question mark (?) indicates that the proposed classification is uncertain. {Some sources could note be classified.}\\ 
{
\tablefoottext{1}{IRS$x$-$xxx$ \protect\citep{Rubio1998}, NIC$xxx$ \protect\citep{Brandner2001}, P$x$ \protect\citep{Hyland1992}, and VFTS $xxx$ \protect\citep{Evans2011}.}} \tablefoottext{2}{Excluding the center of the line due to nebular saturation.} {\tablefoottext{3}{Bandheads.}} \tablefoottext{4}{Shows blueshifted emission component at about $-355$~\kms\text{} and $-265$~\kms\text{}.} \tablefoottext{5}{At least one (but possibly both) of the two components.}
}
\label{tab:specfeatures}
\end{center}
\end{table*}

The procedure described above was carried out at all nodding mode positions separately. The final, nebular corrected, spectra were then combined into a final object spectrum. We show the nebular corrected spectrum for S2 centered around the \ac{CaII} in Fig.~\ref{fig:S2_nebsub}. The \ac{CaII} does not show a nebular counterpart and is therefore assumed to originate from the object. Note that in Fig.~\ref{fig:S2_nebsub} we also performed the sky subtraction and telluric correction.

Residuals persist through the nebular subtraction process (see e.g. the right most Pa~line in Fig.~\ref{fig:S2_nebsub}). Typically these residuals are stronger for stronger nebular lines, and rather {easily} distinguishable from other spectral features. The nebular features are narrow, and since nebular lines are significantly stronger than the continuum, a poor subtraction results in large residuals.

\section{Spectral analysis and target classification} 
\label{sec:spectarg}
A source is classified as a \ac{MYSO} if it shows spectral emission features falling within 2 of the 3 following categories: (1) Accretion features (H$\alpha,\beta$, Pa series, \ac{CaII}, Br$\gamma$), (2) disk tracers (fluorescent Fe\,{\sc ii}~1697.8~nm, CO bandhead emission), and (3) outflowing material (H$_2$~2121.8~nm, [Fe\,{\sc ii}]~1643.5~nm, [O\,{\sc i}]~630.0~nm). Additionally, sources having three accretion features of which at least one has a red shoulder (indicative of inflowing material) are classified as \ac{MYSO}s. H$\alpha$ and H$\beta$ were often saturated in the center due to the nebular contamination. Therefore, the centers of these lines had to be omitted in this analysis. The broad wings in these lines, however, are of stellar origin and used for the detection of {potential} inflow. Most of our targets do not show photospheric absorption lines, hence these could not be used for classification purposes. However, when photospheric features were present we determined the \acf{SpT} using the classification scheme of \citet{Gray2009}. In the following subsections we present the spectroscopic results for all objects. A summary of all detected spectral features for each object and the final classification is shown in Table~\ref{tab:specfeatures}. In Appendix~\ref{app:lineplots}  we show the investigated spectral regions for all targets.

\subsection{S1 and S1-SE}
These objects are located relatively far away from \ac{30Dor}'s central cluster \ac{R136} (see Fig.~\ref{fig:30Dor}). The region (also called the Skull Nebula) is associated with the larger CO cloud \ac{30Dor}-06 of \citet{Johansson1998} and H\,{\sc ii} region No. 889 of \citet{Kastner2008}. It is located between an X-ray cavity possibly associated with 3 nearby \ac{WR} stars \citep[R144, R146 and R147;][]{Townsley2006}, and the older Hodge 301 cluster known to have hosted multiple supernovae \citep{Grebel2000, Cignoni2016}.

S1 is resolved into multiple objects with $I$-band (900~nm) magnitudes of $\sim$19--21 \citep{Walborn2013}. In our observation we did not resolve multiple components. We will therefore not probe this multiplicity and consider S1 as a single source. We observe weak continuum in the \ac{UVB} arm which gets stronger towards the \ac{VIS} and \ac{NIR} arms. We detect very weak Ba and Pa absorption lines, the  Pa jump, and [O\,{\sc i}]~630.0~{nm} emission. We cannot confirm a \ac{MYSO} nature. 

\citet{Walborn2013} suggest S1-SE to have two components; we however can not confirm this. S1-SE is fainter than S1 and shows no detectable continuum up to about 800~nm. The \ac{SN} is insufficient to detect any spectral features {except for some weak H$_2$~2121.8~{nm} emission}.

\subsection{S2}
S2 is a relatively faint source located southwest to \ac{R136} in the head of a dust pillar, and appears multiple in NICMOS data \citep{Walborn1999}. We do not resolve the object in our observation. Assuming a single (or compact multiple) origin, \mbox{\citet{Walborn2013}} estimated a luminosity $L = 3.4 \times 10^{4}~L_{\odot}$, effective temperature $T_{\text{eff}} = 12000$~K, stellar mass $M = 20.1~M_{\odot}$, and extinction $A_\mathrm{V} = 6.5$~mag. 

It is a very red source with relatively modest \ac{NIR} excess, we only detect continuum from about $\sim$1500~nm onwards increasing with wavelength. The Ba and Pa series are not detected except for broad H$\alpha$ emission with a red shoulder. Though this is an indication of a possible inflow, it could also be produced by the companion. The [O\,{\sc i}]~630.0~{nm} emission is contaminated by nebular subtraction residuals. S2 does show strong single-peaked \ac{CaII} emission (see e.g. Fig.~\ref{fig:S2_nebsub}), together with weak Br$\gamma$ emission. We classify S2 as a \ac{MYSO}. 

\subsection{S3, S3-K, and R135}
\label{subsec:S3}
The complex of S3, S3-K, and the isolated \ac{WR} star R135 lies to the northwest of \ac{R136} within a dust filament. \citet{Walborn2013} observe that at wavelengths shorter than the $K_\mathrm{s}$-band S3-K dominates over S3, whereas at longer wavelengths of 4.5~$\mu$m and 8.0~$\mu$m S3 dominates the entire region. Additionally, they determined $L = 8.2 \times 10^{4}~L_{\odot}$, $T_{\text{eff}} = 38000$~K, $M = 25.2~M_{\odot}$, and $A_\mathrm{V}=10.0$~mag for S3, and suggest S3-K to be a star with $T = 4750$~K and $A_\mathrm{V}=5.85$~mag rather than a \ac{YSO} by finding a better fit with photospheric models than with \ac{YSO} models.

Our spectrum of S3 has a relatively low \ac{SN} and is dominated by nebular subtraction residuals hampering the identification of intrinsic spectral features. The continuum becomes visible around $\sim$1600~nm and shows only a moderate increase in strength towards longer wavelengths. A broad emission feature is present around H$\alpha${, and we detect Fe\,{\sc ii}~1687.8~{nm} and H$_2$~2121.8~{nm} emission.} The Pa series is completely dominated by nebular subtraction residuals, only Pa$\beta$ shows some weak emission. Similarly, Br$\gamma$, and [O\,{\sc i}]~630.0~{nm} show emission features contaminated by residuals of the nebular subtraction. According to \citet{Walborn2013}, S3 is the second most massive \ac{MYSO} candidate of our sample. {Here, we confirm the \ac{MYSO} nature of S3}.

In our observation S3-K becomes visible from about 400~nm onwards and is very bright in the $K_\mathrm{s}$-band. The Ba series are not visible but for some subtraction residuals; the Pa series is weakly in absorption and [O\,{\sc i}]~630.0~{nm} is absent. The spectrum is dominated by CO, TiO, and other molecular absorption bands as well as various narrow absorption lines. Additionally, we see that the \ac{CaII} exhibits absorption. The \ac{SpT} should be early M (or possibly late K) due to the presence of many molecular bands and Ca\,{\sc ii}, Fe\,{\sc i} and Ti\,{\sc i} absorption features. The effective temperature of \citet{Walborn2013} indicates \ac{SpT} $\sim$K3. S3-K is, however, too bright to be a typical M/K-type  MS star at the distance of \ac{30Dor}. Fitting the \ac{CaII} yields a \ac{RV} of $261.4\pm0.8$~\kms\text{} which is consistent with the surrounding region \citep{Torres-Flores2013}. This excludes the suggestion of S3-K being a foreground star. {S3-K may be explained as a $\sim10~M_{\odot}$ M-type (super)giant, which is further supported by the fact that the flux does not strongly increase in the {\it Spitzer}/\ac{IRAC} bands.}

R135 is a \ac{WR} star of \ac{SpT}~WN7h+OB \citep[VFTS 402;][]{Evans2011}. In our X-shooter spectra we are not able to detect spectral features of a possible OB-type companion due to dilution by the \ac{WR} star. We identify broad emission in all hydrogen series (i.e. Ba, Pa and Br), and strong N\,{\sc iii} emission features. Using the classification scheme of \citet{Smith1968} we classify the \ac{WR} star as a WN7h star. This is in agreement with the earlier classification of \citet{Evans2011}. 

\subsection{S4}
S4 is located in the head of a bright-rimmed pillar oriented towards \ac{R136} \citep{Walborn1999, Walborn2002}. It is one of the most luminous sources in \ac{30Dor} at almost all \ac{NIR} wavelengths, and has the strongest \ac{NIR} excess of all the targets in our sample. \citet{Walborn2013} determined $L = 10.7 \times 10^{4}~L_{\odot}$, $T_{\text{eff}} = 39000$~K, $M = 27.4~M_{\odot}$, and $A_\mathrm{V} = 1.8$~mag. S4 has a companion \citep[IRSW-26;][]{Rubio1998} to the southwest which is $\gtrsim$3~mag fainter in the $K_\mathrm{s}$-band. We therefore analyze S4 as a single object.

\begin{figure*}[t]
\centering
\includegraphics[width=1.0\linewidth]{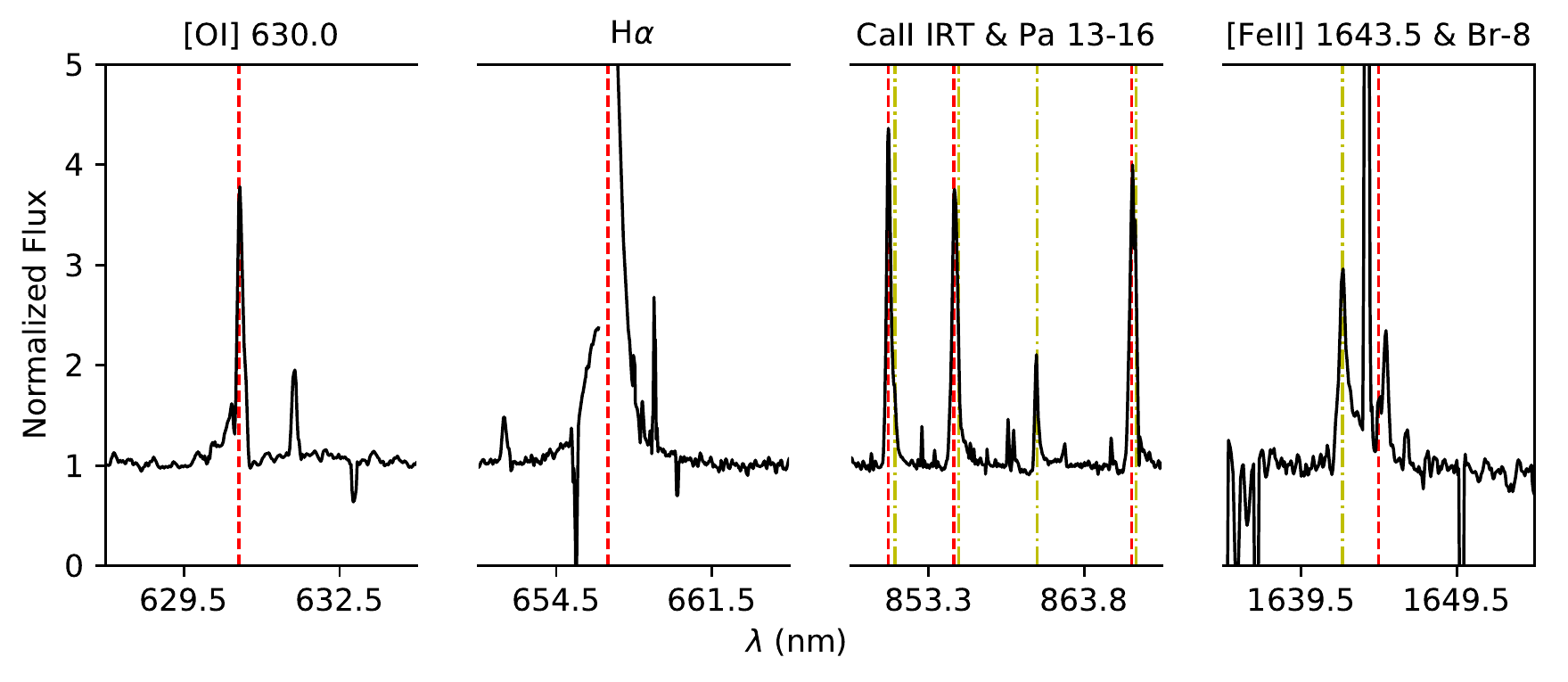}
\includegraphics[width=1.0\linewidth]{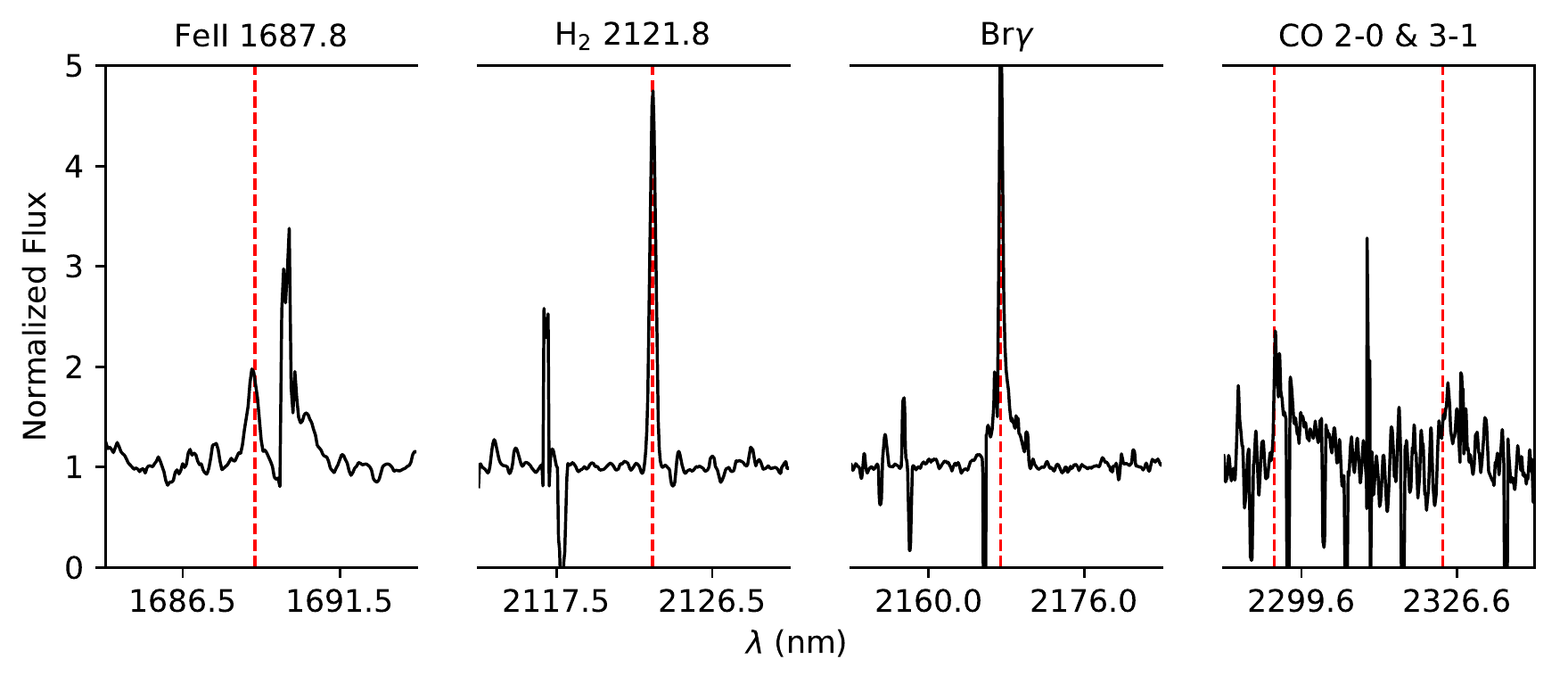}
\caption{The [O\,{\sc i}]~630.0~{nm}, H$\alpha$, \ac{CaII} and Pa-13--16, {[Fe\,{\sc ii}]~1643.5~{nm}}, {Fe\,{\sc ii}~1687.7~{nm}}, {H$_2$~2121.8~{nm}}, Br$\gamma$, and the 2--0 and 3--1 CO bandhead regions shown for S4. {For clarity, the {[Fe\,{\sc ii}]~1643.5~{nm}, {Fe\,{\sc ii}~1687.7~{nm}}, H$_2$~2121.8~{nm}}, Br$\gamma$ and CO bandhead regions have been enhanced by a factor of 5 and 15, respectively.}  We indicate the positions of the transitions by the red dashed lines. {The Pa~series and Br-8 line are marked by yellow dash-dotted lines for clarification.} The center of H$\alpha$ was saturated due to nebular emission and has been clipped. All narrow features are either telluric lines or residuals from the nebular or sky subtraction.}
\label{fig:S4lines}
\end{figure*}

The continuum of S4 is visible across the entire X-shooter wavelength range, but becomes substantially stronger from the $J$-band onwards. The nebular contamination was very strong resulting in the saturation of multiple nebular lines including H$\alpha,\beta$. Nevertheless we see clear signatures of in-falling material in the wings of these lines manifested by the red shoulders. In Fig.~\ref{fig:S4lines} we show {the spectral features used for the classification}. Note in particular the very strong \ac{CaII} lines and the strong red shoulder in H$\alpha$. Furthermore, we detect several Fe\,{\sc ii} emission features {and weak CO bandheads} indicative of a disk, and [Fe\,{\sc ii}] {and H$_2$} lines which are indications of a bipolar outflow \citep{Ellerbroek2013_1}. S4 is the most massive \ac{MYSO} in \citet{Walborn2013}, which agrees with the spectral features being the most prominent of all our targets. S4 is a \ac{MYSO}. 

\subsection{S5}
\label{subsec:S5}
S5 is situated in a boomerang-shaped molecular cloud located north of \ac{R136} \citep{Walborn1999,Kalari2018}. The X-shooter slit includes 6 targets, which we labeled A--F, see Fig.~\ref{fig:S5_targets}. 

\begin{figure}[t]
\centering
\includegraphics[width=1.0\linewidth]{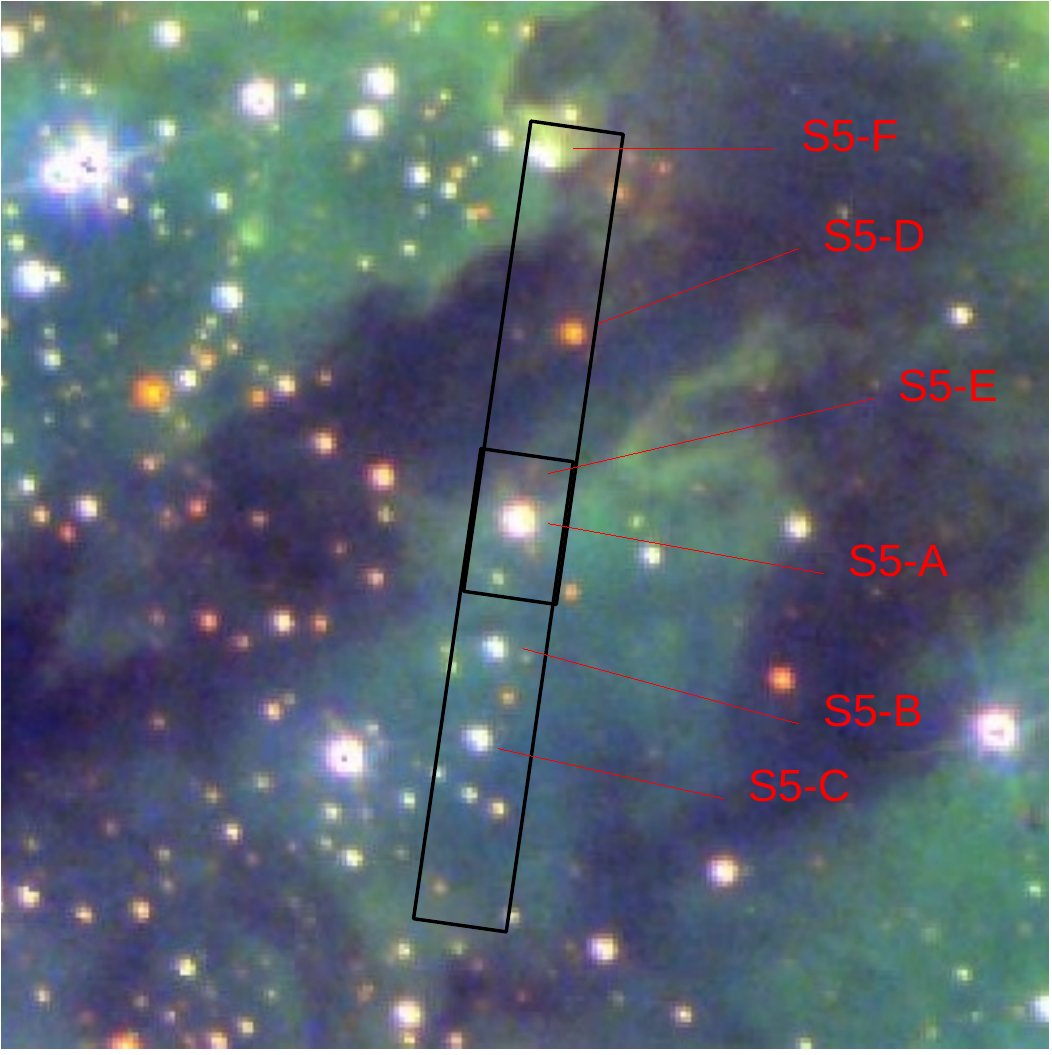}
\caption[S5 X-shooter slit position]{HST/WFPC2/F814W(red)+F555W(green)+F336W(blue) composite image (20"$\times$20") of the cluster field surrounding S5 \protect\citep{Walborn2002}. North is up and east is to the left. The two nodding positions of the X-shooter slit are shown with the two black rectangles. We identify a total of six continuum sources in our observation, which we labeled A--F (indicated in red). Note that S5-E is not visible in this image, nevertheless we still indicate its (presumed) location.}
\label{fig:S5_targets}
\end{figure}

S5-A is the brightest of the six objects at optical wavelengths. \citet{Walborn2014} identified S5-A as a O((n)) star, whereas we classify S5-A as an O6~V((f)) star. The difference with \citet{Walborn2014} is due to the nebular contamination (or larger residuals) still present in their observations whereas we subtracted it, allowing for a more precise spectral classification. We determined a \ac{RV} of $246.2\pm11.0$~\kms\text{}. This is in agreement with the \ac{RV} observations of \citet{Sana2013}.

S5-B  and S5-C are less bright. They do show strong Ba and He\,{\sc i} absorption, from which we determine a \ac{SpT} of B0~V for both objects. The \ac{SN} was not optimal hence this classification is rather uncertain. We find a \ac{RV} of $254.6\pm4.9$ and $247.1\pm10.9$~\kms\text{} for S5-B and S5-C, respectively.

S5-D is an object which only shows significant brightness in the \ac{VIS} arm of X-shooter. It shows \ac{CaII} absorption lines, which indicates S5-D to be a late-type star. We determine a \ac{RV} of $19.5\pm3.0$~\kms, implying S5-D is a foreground star. 

S5-F is located at the edge of the X-shooter slit and is only visible in the \ac{UVB} and blue part of the \ac{VIS} arms due to the \ac{ADC}s malfunctioning. In the \ac{UVB} arm we detect rising intensity towards longer wavelengths accompanied with weak Ba and He\,{\sc i} absorption features. Due to the low \ac{SN} we cannot further constrain the \ac{SpT} than early-B or late-O. We determined a \ac{RV} of $221.0\pm4.0$~\kms.

\begin{figure*}[t]
\centering
\includegraphics[width=1.0\linewidth]{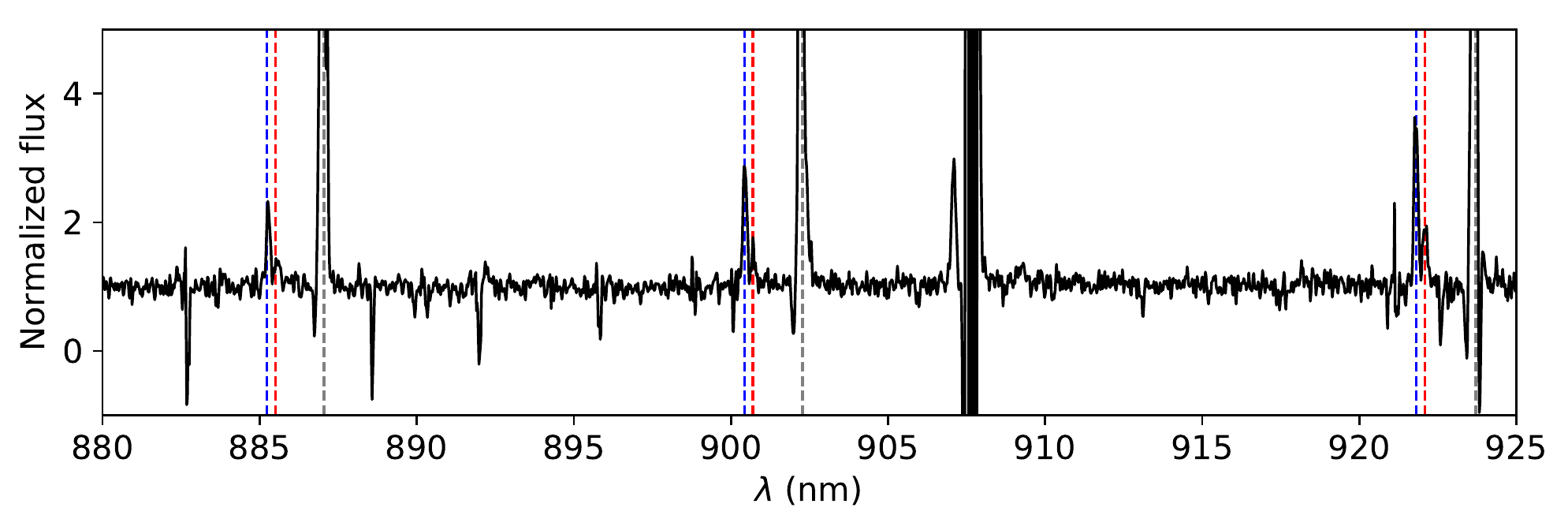}
\caption[Spectrum of S7-B centered on the Pa jump and Pa outflow lines]{The spectrum of S7-B centered around the three Pa lines. With the blue and red vertical dashed lines we denote the locations of the the blueshifted and redshifted peaks of emission at $-355$~\kms\text{} and $-265$~\kms\text{}, respectively. The nebular counterpart of the Pa lines is indicated with the gray vertical lines at $250$~\kms\text{}. The region around $\sim$908~nm is a saturated {[S\,{\sc iii}]} nebular line. Other narrow features are either telluric features or residuals from the nebular subtraction.}
\label{fig:S7-B_blueshift}
\end{figure*}

S5-E is the actual \ac{MYSO} candidate selected to be observed. {It has been identified as a \ac{MYSO} candidate by \citet{Gruendl2009}, and later the young nature of S5-E was confirmed \citep{Seale2009,Jones2017}.} According to \citet{Walborn2013} S5-E becomes apparent from the $K_\mathrm{s}$-band onwards. They estimate a lower limit of $>$19 for the $J$-band magnitude. On the X-shooter slit S5-E shows contamination from S5-A in the \ac{UVB} and \ac{VIS} arms. The continuum of S5-E appears around 1500~nm and is brighter than S5-A from about 2000~nm onwards. Despite the contamination by S5-A, we can recognize weak \ac{CaII}, {H$_2$~2121.8~{nm},} and CO first-overtone bandhead emission. This allows us to classify S5-E as a \ac{MYSO}. 

\subsection{S6}
S6 could represent a case of monolithic massive star formation due to its isolated position \citep{Walborn2002}. \ac{SED} fitting of the \ac{NIR} and \ac{MIR} photometric points results in $L = 3.7 \times 10^{4}~L_{\odot}$, $T_{\text{eff}} = 34000$~K, $M = 18.4~M_{\odot}$, and $A_\mathrm{V}=3.0$~mag \citep{Walborn2013}.
S6 shows an \ac{NIR} excess similar to S4. We detect continuum from about 800~nm onwards, which gets substantially stronger beyond 1500~nm. We detect weak Pa emission features, no other spectral features are visible.  
\citet{Walborn2013} classified S6 as a Class I \ac{MYSO}, where according to the classification scheme of \citet{Megeath2004} it should be a Class II object. Spectroscopically we cannot confirm S6 as a \ac{MYSO}. 

\subsection{S7-A and S7-B}
The complex of S7-A and S7-B is embedded in a large dust pillar oriented towards \ac{R136} \citep{Walborn1999,Walborn2002}. In the $Y$-band S7-A is brighter than S7-B; from the $J$-band onwards S7-B dominates over S7-A \citep{Walborn2013}. Both targets show strong \ac{NIR} excess with S7-B being the third brightest $K_\mathrm{s}$-band target in our sample (after S4 and S6, see Fig.~\ref{fig:photfigure}). Both targets fit on one X-shooter slit. Unfortunately the observations were taken under bad seeing conditions (average 2.2"). 

By \ac{SED} fitting of the \ac{NIR} and \ac{MIR} photometric points of S7-A, \citet{Walborn2013} determined $L = 3.0 \times 10^{4}~L_{\odot}$, $T_{\text{eff}} = 30000$~K, $M = 15.2~M_{\odot}$, and $A_\mathrm{V}=3.2$~mag. \citet{Nayak2016} classify S7-A (J84.695932-69.083807 in their paper) as a {Type~II} \ac{YSO} with $L = 5.62 \times 10^{4}~L_{\odot}$, and $M = 21.8~M_{\odot}$. We detect continuum in the \ac{UVB} arm which gets weaker towards longer wavelengths, eventually almost disappearing around $\sim$1000~nm. It reappears from about $\sim$1500~nm onwards. We detect photospheric Ba and He\,{\sc i} absorption features; H$\alpha$ shows broad emission. From the photospheric absorption features we determine a \ac{RV} of 264.7$\pm$2.3~\kms\text{}, and classify S7-A as a B1~V star. The Pa series shows weak emission features with nebular subtraction residuals superimposed. Additionally we detect broad Br$\gamma$, {Fe\,{\sc ii}~1687.8~{nm}, H$_2$2121.8~{nm}, and weak CO first-overtone bandhead emission. The detected spectral features confirm a \ac{MYSO} nature.}

S7-B is the brightest of the two objects {in the \ac{NIR} and \ac{MIR}}. \citet{Nayak2016} classify S7-B (J84.695173-69.084857 in their paper) as a {Type~II} \ac{YSO} with $L = 5.62 \times 10^{4}~L_{\odot}$, and $M = 19.0~M_{\odot}$. We detect continuum over the entire X-shooter wavelength range. Unfortunately we see no photospheric absorption features like in S7-A. We do detect broad H$\alpha$, Pa series, {Fe\,{\sc ii}~1687.8~{nm}, {Br$\gamma$ (with a red shoulder)}, and H$_2$~2121.8~{nm}} emission. More remarkable are the strong (inversed) Ba and Pa jumps, and the Pa series showing a blueshifted emission component, see Fig.~\ref{fig:S7-B_blueshift}. The latter seems to be double-peaked at velocities of about $-355$~\kms\text{} and $-265$~\kms\text{}. Note that this implies a \ac{RV} of about $-615$~\kms\text{} and $-525$~\kms\text{} in the local reference frame assuming a \ac{RV} of $260$~\kms\text{} for S7-B. This emission may originate from a very high velocity outflow, or may be an effect of binary interaction. Besides the Pa series, this high blueshifed emission seems only weakly visible in the {[Fe\,{\sc ii}]~1643.5~{nm} line}. We classify S7-B as a \ac{MYSO}.   

\subsection{S8}
S8 is a bright \ac{NIR} source surrounded by a cluster of fainter objects \citep{Walborn2002}. In \ac{VMC} observations S8 is resolved into two objects of about equal magnitude whereas in {\it Spitzer} observations S8 is unresolved \citep{Walborn2013}. On the X-shooter slit we are unable to resolve the 2 components despite the relatively good seeing conditions (average 0.8") under which our observations were taken. \citet{Nayak2016} determined $L = 5.62 \times 10^{4}~L_{\odot}$, and $M = 19.0~M_{\odot}$ for S8 (J84.699755-69.069803 in their work){, and classify it as a Type~II \ac{YSO}}.  We detect continuum from about 600~nm onwards, and observe broad H$\alpha$ emission with a red shoulder. {The Pa series shows emission contaminated by nebular subtraction residuals. } Furthermore we see weak \ac{CaII} and weak Br$\gamma$ emission. {We classify S8 as a \ac{MYSO}.}

\subsection{S9}
This red object is the brighter of two extended sources located in the vicinity of the optical multiple system within Knot 2 of \citet{Walborn1999}. This system includes \ac{VFTS}~621, a young massive star of \ac{SpT} O2 V((f*))z \citep{Evans2011,Walborn2014}. S9 is very bright in the $K_\mathrm{s}$ and {\it Spitzer}/\ac{IRAC} 4.5~$\mu$m bands, but is not dominant in the {\it Spitzer}/\ac{IRAC} 8~$\mu$m band. S9 is a \ac{MYSO} candidate according to \citet{Seale2009}, and a water maser associated with S9 has been identified to the north \citep{Ellingsen2010}. {\citet{Nayak2016} classified S9 (J84.703995-69.079110 in their work) as a Type~I \ac{YSO} and} determined $L = 6.81 \times 10^{4}~L_{\odot}$, and $M = 23.9~M_{\odot}$ of S9 . More recently, \citet{Reiter2019} computed $L = 5.01 \times 10^{5}~L_{\odot}$, $T_{\text{eff}} = 21120$~K, and $A_\mathrm{V}=2.46$~mag, and see the CO bandheads in absorption (similar to S3-K in this work). 

In our X-shooter observation the continuum of S9 appears from about 1500~nm onwards and increases only moderately in strength towards longer wavelengths. Some broad weak emission is present around H$\alpha$, {H$_2$2121.8~{nm},} and Br$\gamma$; stronger is the Fe\,{\sc ii}~1687.8~{nm} feature. {We do not detect any} CO bandhead emission/absorption. {S9 is a \ac{MYSO}}.

\subsection{S10-A and S10-BC}
In the {\it Spitzer}/\ac{IRAC} wavelength bands S10 appears as one of the brightest sources in \ac{30Dor}, but in the higher resolution \ac{VMC} images it actually splits up into three considerably fainter sources \citep{Walborn2013}. The system is located within a cavity created by \ac{VFTS}~682, one of the most massive isolated \ac{WR} stars \citep[\ac{SpT} WN5h, $M\sim150~M_\odot$;][]{Bestenlehner2011} which might be a runaway star from \ac{R136} \citep{Renzo2019}. The X-shooter slit was positioned such that all 3 objects would fit within a single exposure. However, we detect only 2 objects on the slit. Whereas S10-A is resolved, S10-B and S10-C are not. The latter two will be discussed under the name S10-BC. 

We detect the continuum of S10-A from about 1600~nm onwards which only increases moderately in strength towards longer wavelengths. We see emission features in the Pa series contaminated by some nebular subtraction residuals. No other spectral features are visible.

The continuum of S10-BC also becomes weakly visible from about 1600~nm onwards and, similar to S10-A, increases moderately in strength towards longer wavelengths. We detect no features in the spectrum of S10-BC. We can neither confirm S10-A nor S10-BC as a \ac{MYSO}. 

\subsection{S10-K}
S10-K is located to the southeast of the S10 region. \citet{Walborn2013} determined $L = 0.7 \times 10^{4}~L_{\odot}$, $T_{\text{eff}} = 25000$~K, $M = 9.1~M_{\odot}$, and $A_\mathrm{V}=4.4$~mag. S10-K steeply raises in brightness from the $J$-band towards the $K_\mathrm{s}$-band, but does not notably increase further in flux in the {\it Spitzer}/\ac{IRAC} bands. In our observation of S10-K we start detecting continuum from about 1500~nm onwards. {Only weak \ac{CaII} emission features and broad weak H$\alpha$ emission is detected. We cannot confirm a \ac{MYSO} nature.}

\subsection{S10-SW-A and S10-SW-B}
This complex was labeled S11 in \citet{Walborn2013} and is unresolved in their {\it Spitzer} images. The unresolved system was classified as a \ac{YSO} candidate by \citet{Seale2009}, and is located on the opposite side of the cavity created by \ac{VFTS}~682 with respect to the S10 and S10-K region. A water maser has been identified at the location of S10-SW-A \citep{Ellingsen2010}. \citet{Nayak2016} determine $L = 3.16 \times 10^{4}~L_{\odot}$, and $M = 14.8~M_{\odot}$ for S10-SW-A (J84.720292-69.077084 in their paper){, and classified it as a Type~I \ac{YSO}}.

\begin{figure*}[t]
\includegraphics[width=0.33\linewidth]{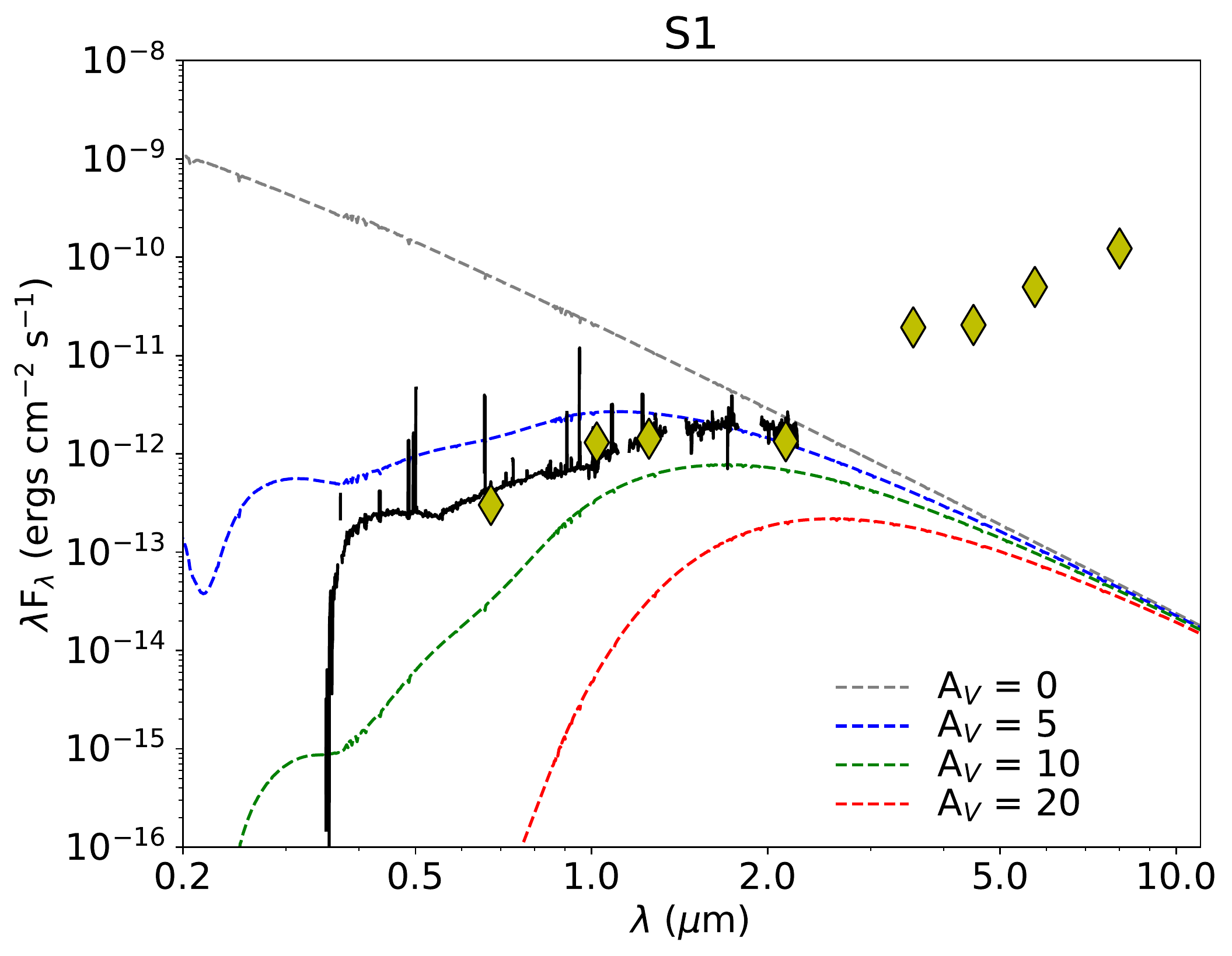}
\includegraphics[width=0.33\linewidth]{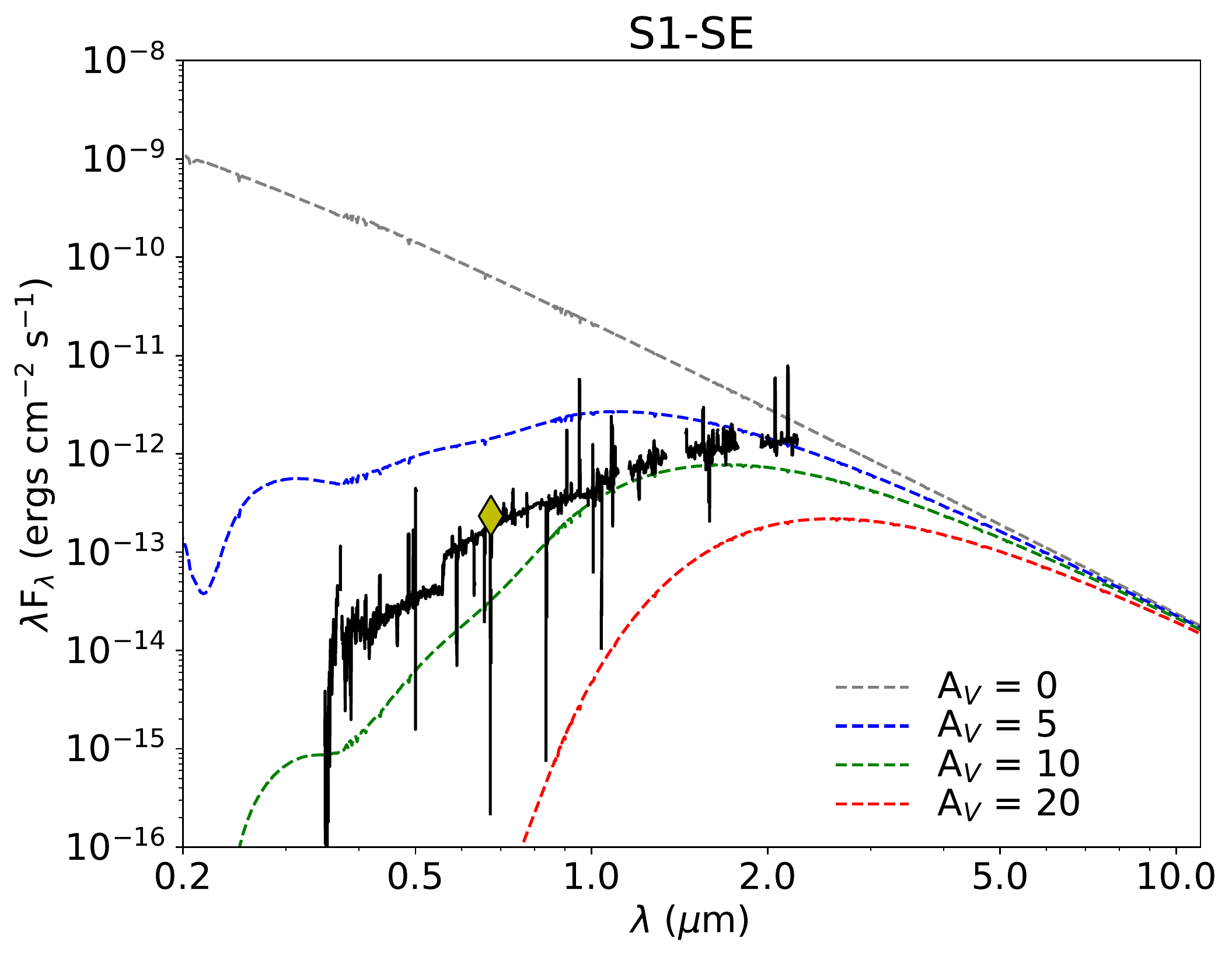}
\includegraphics[width=0.33\linewidth]{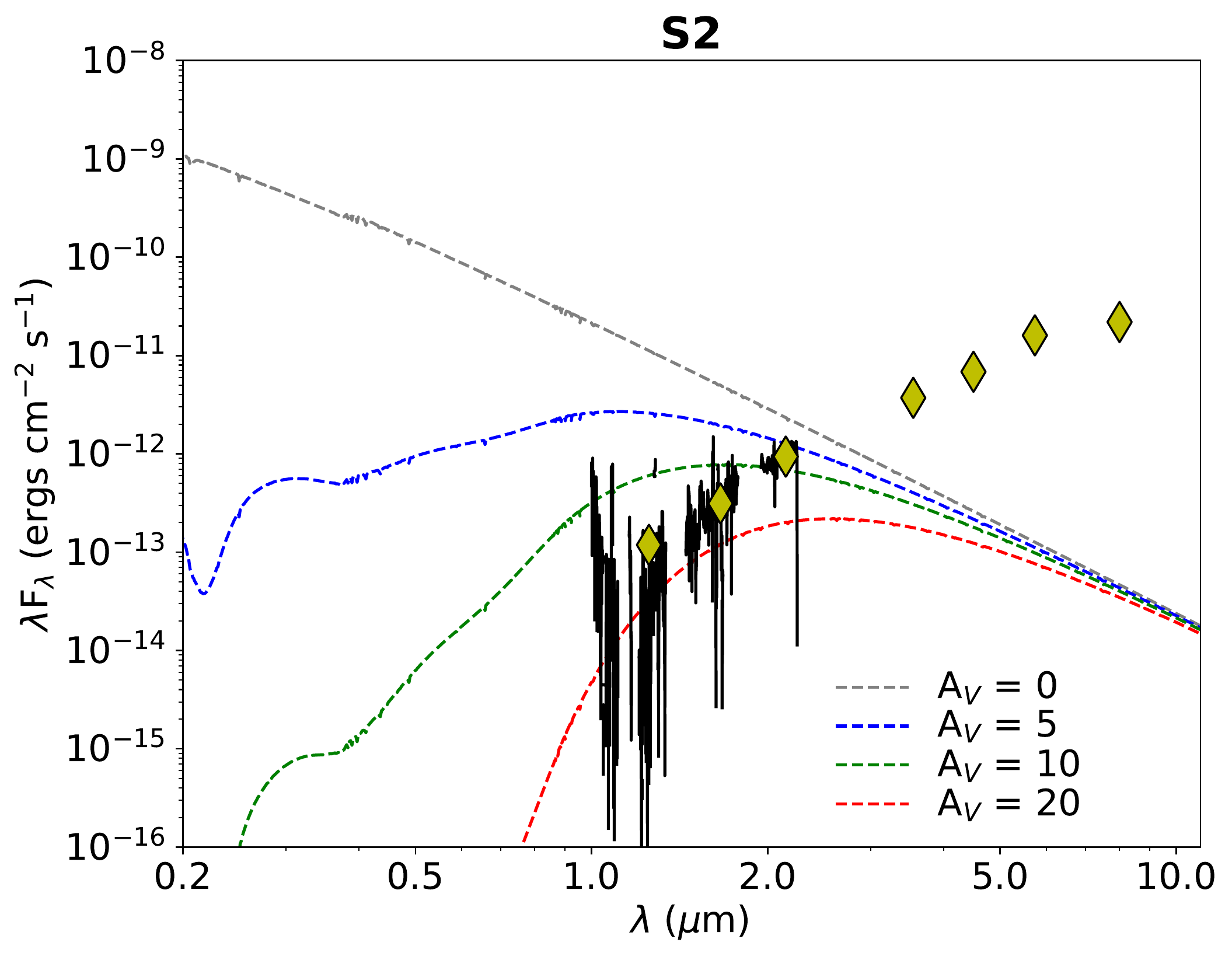}

\includegraphics[width=0.33\linewidth]{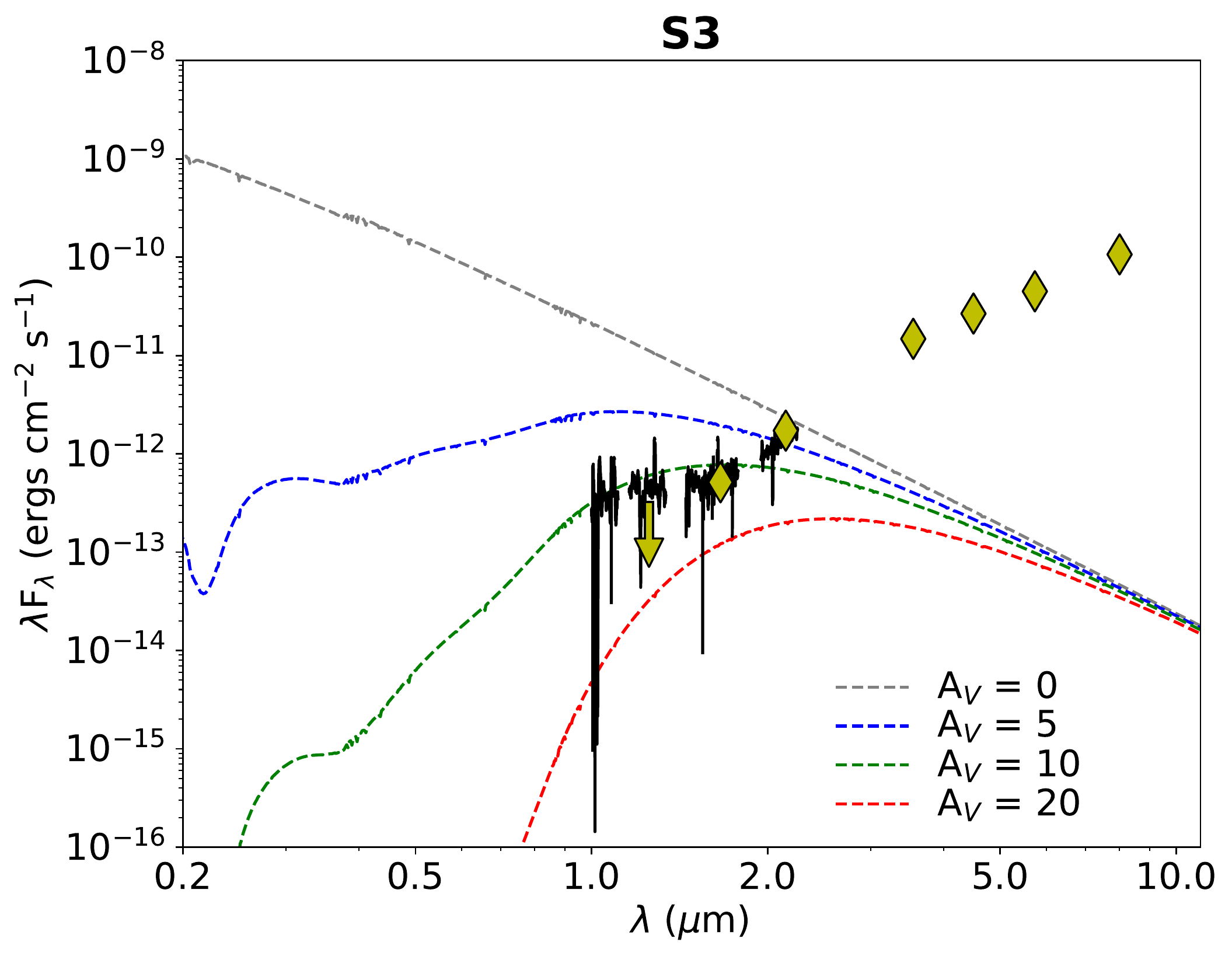}
\includegraphics[width=0.33\linewidth]{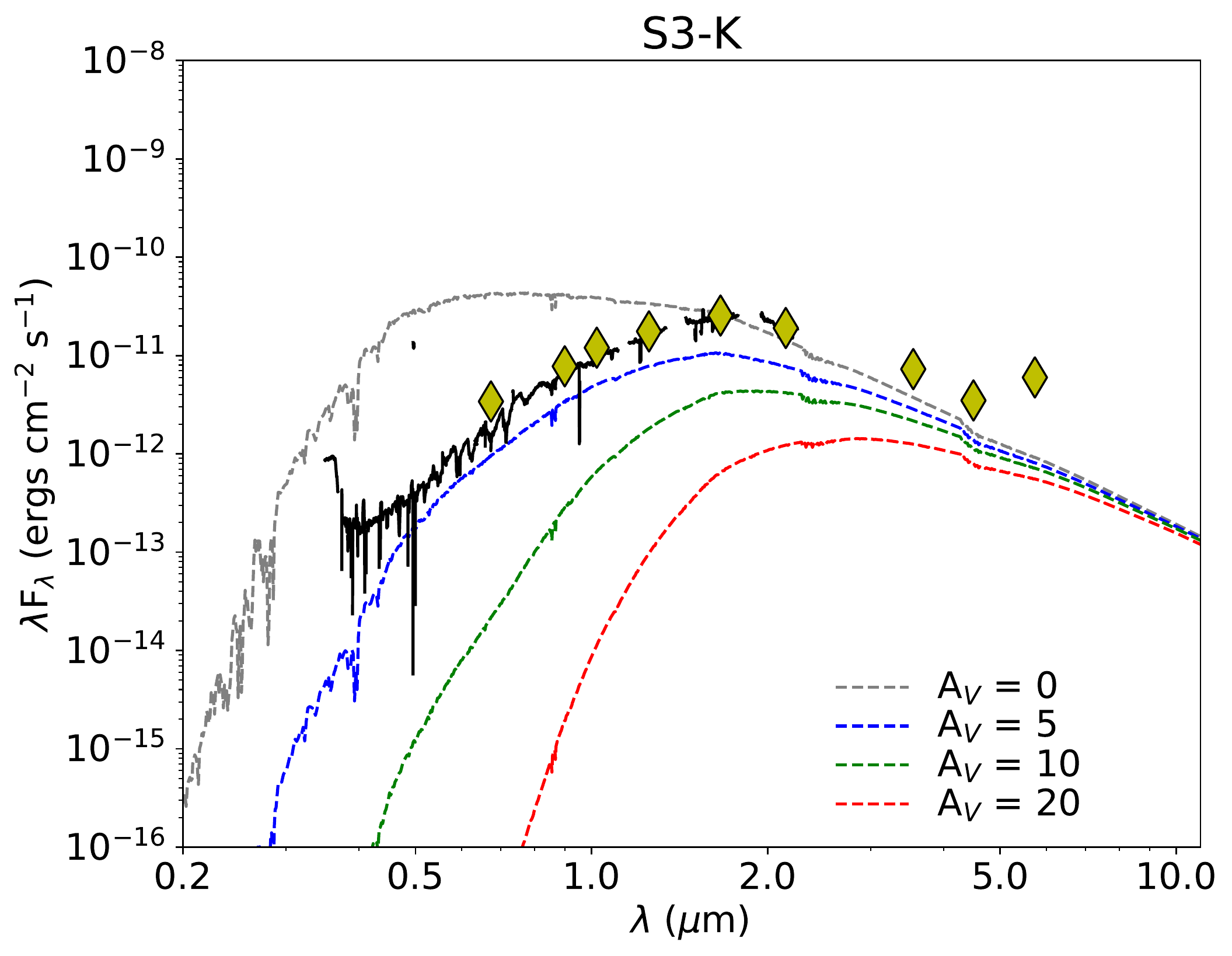}
\includegraphics[width=0.33\linewidth]{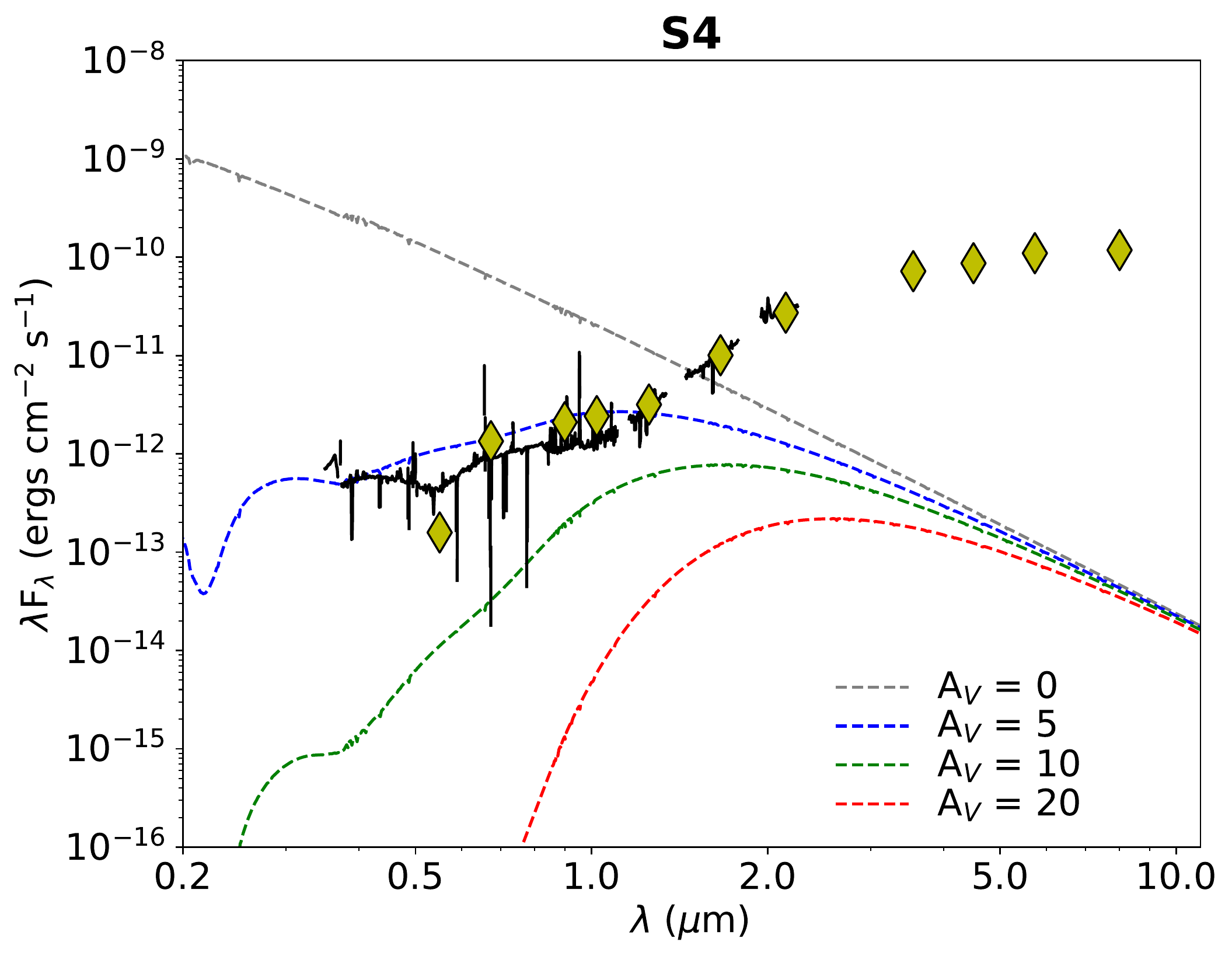}

\includegraphics[width=0.33\linewidth]{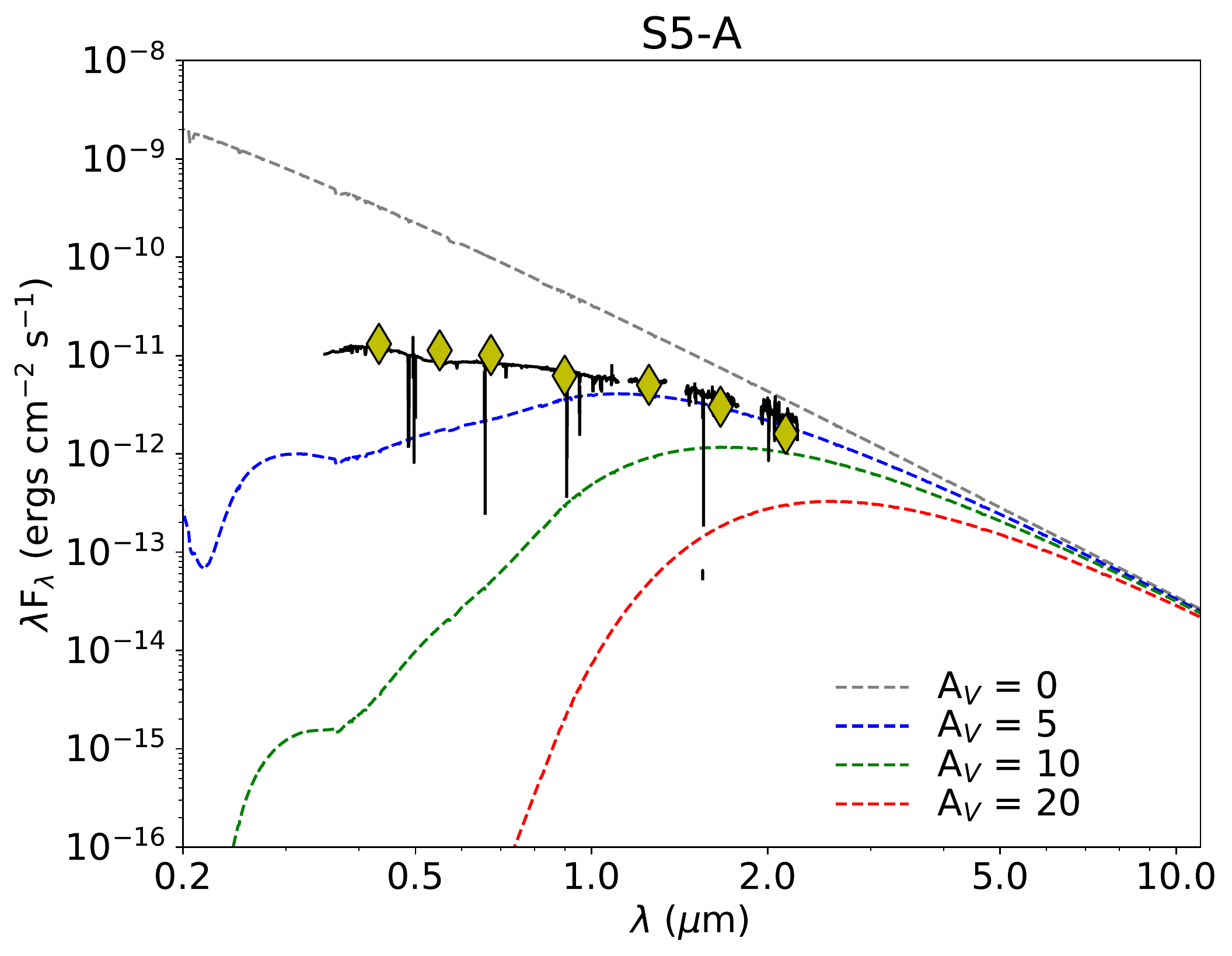}
\includegraphics[width=0.33\linewidth]{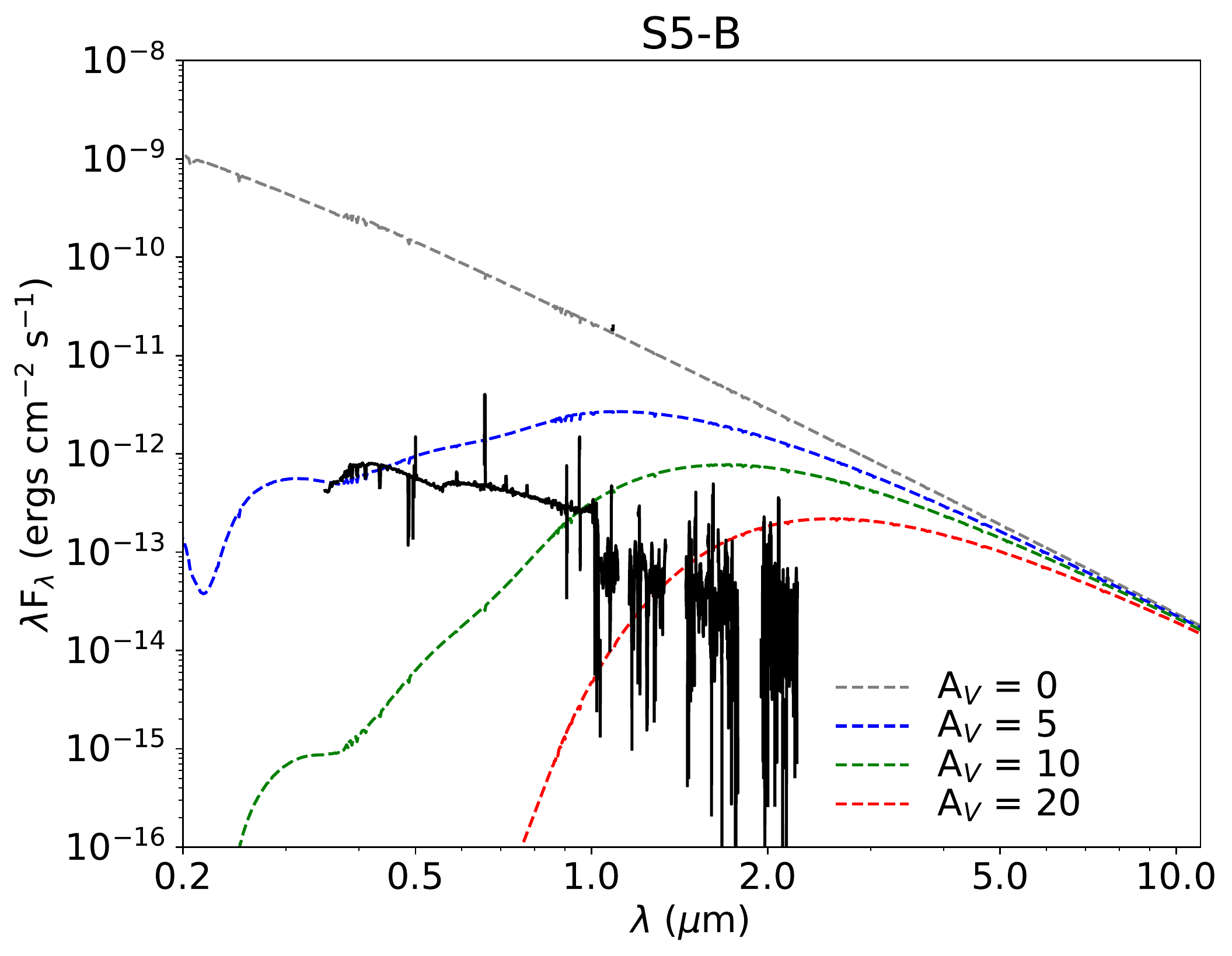}
\includegraphics[width=0.33\linewidth]{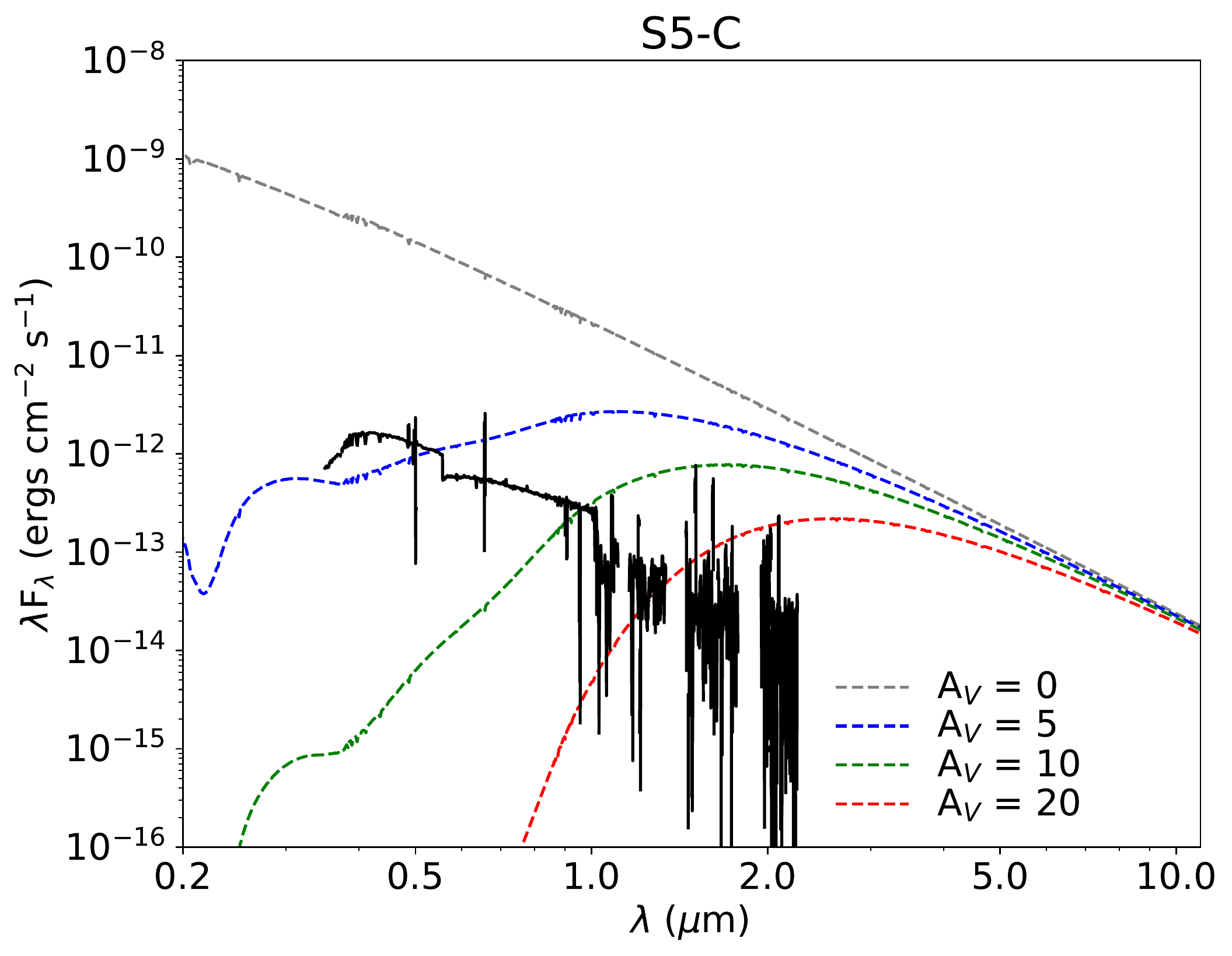}

\includegraphics[width=0.33\linewidth]{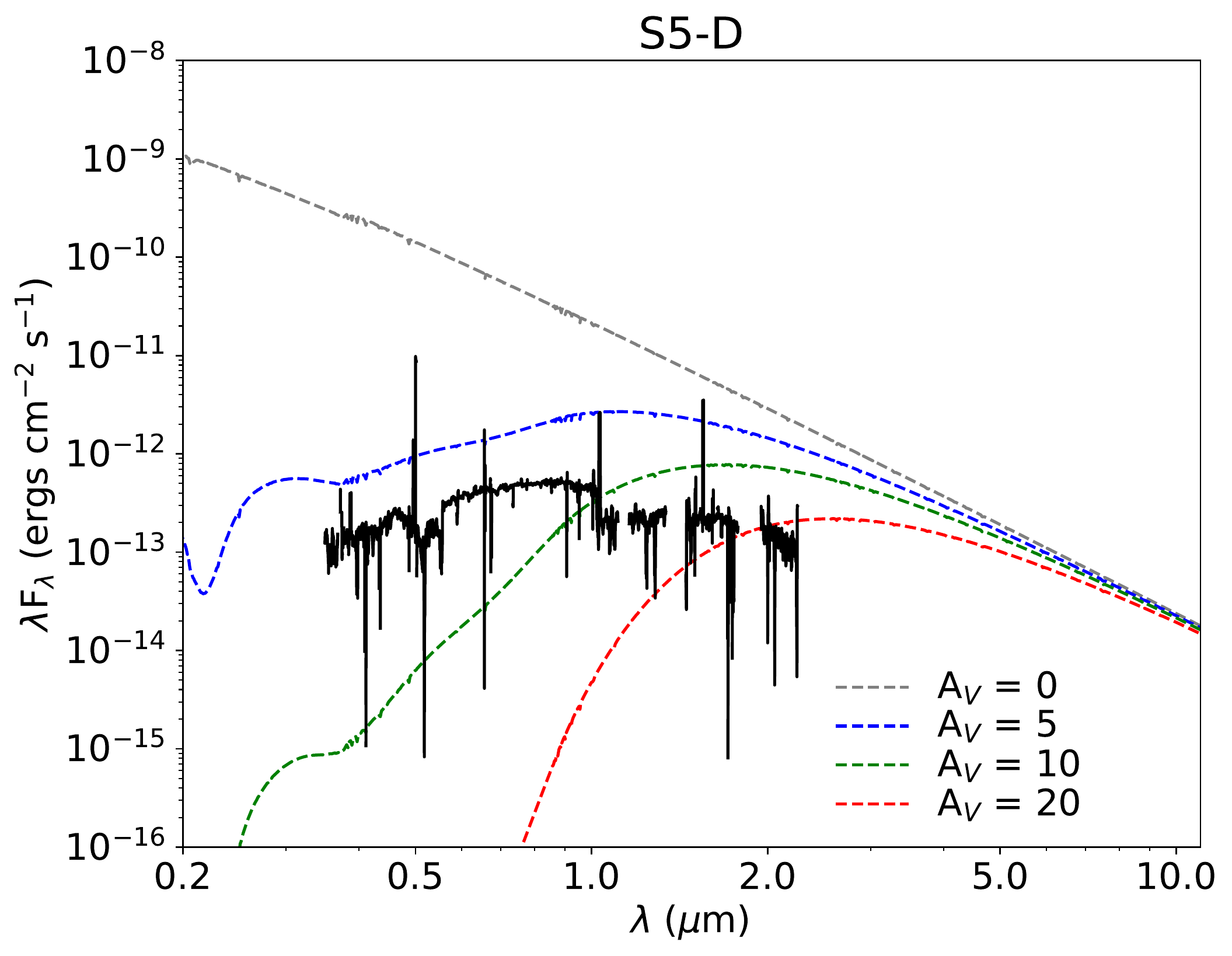}
\includegraphics[width=0.33\linewidth]{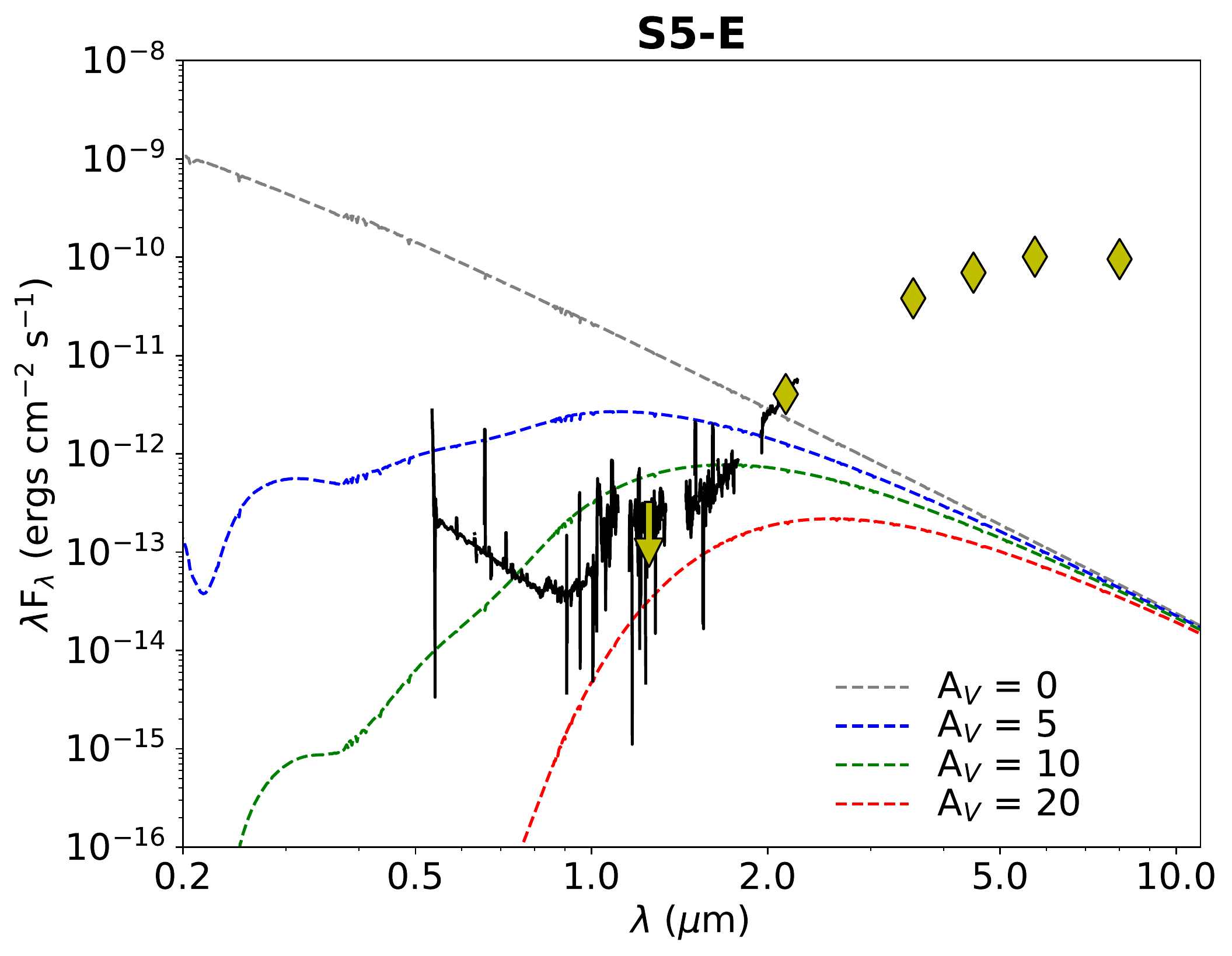}
\includegraphics[width=0.33\linewidth]{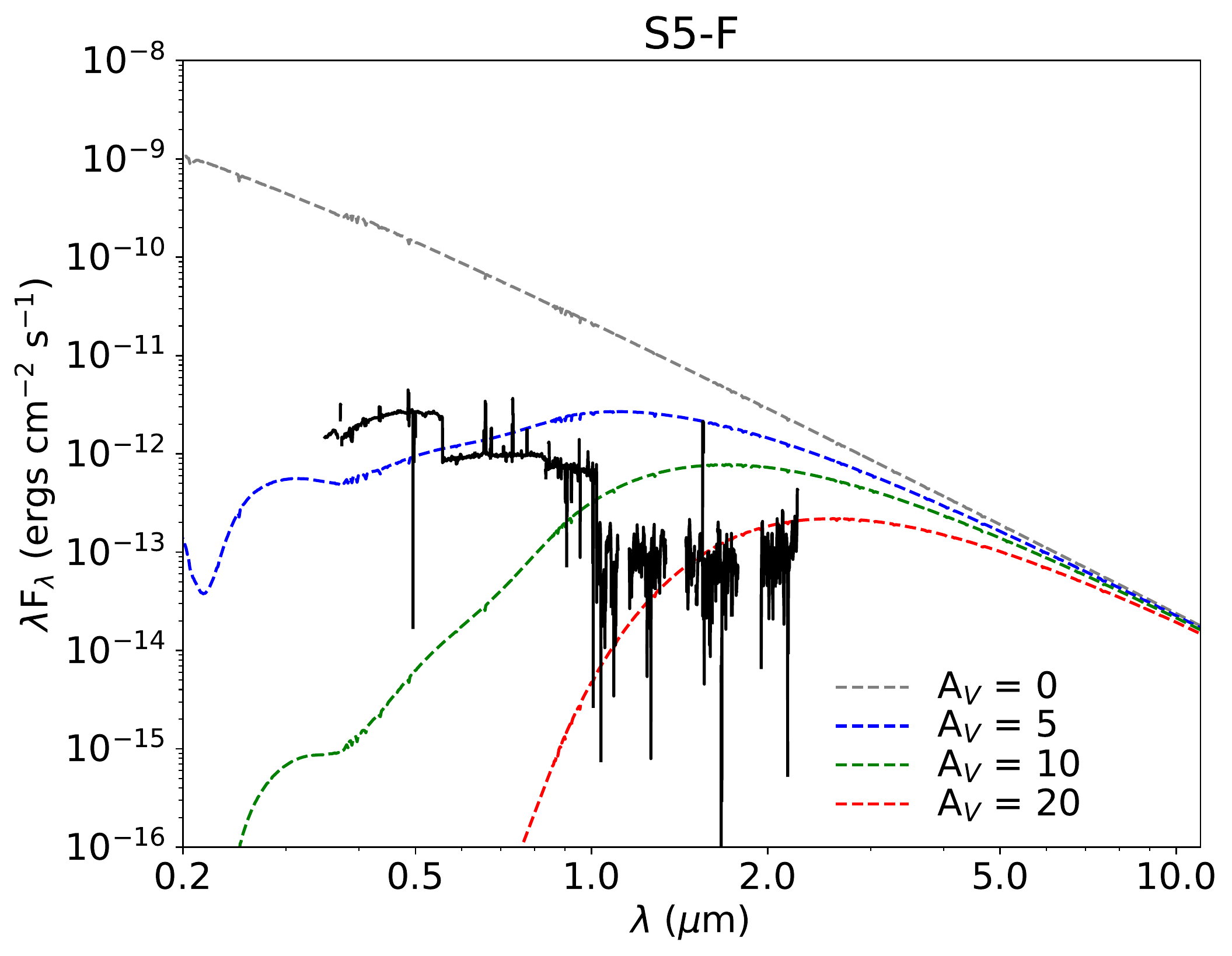}
\caption{{The \ac{SED}s of S1--S5-F. The spectrum of each object is shown in black, and has been smoothed by a factor of 100. The confirmed \ac{MYSO}s are indicated in bold. The objects S5-B, S5-C, S5-D, and S5-F are not corrected for slit loss since we lack photometry of these objects. The telluric absorption bands around 1.1~$\mu$m, 1.5~$\mu$m, and 2.0~$\mu$m are clipped. Additionally, we clipped the spectrum above 2.25~$\mu$m due to sky variations, and at short wavelengths for S2, S3, and S5-E due to the low flux of these objects. Literature photometric points are shown as the yellow diamonds \protect\citep{Parker1992,Cutri2003,Kato2007,Walborn2013,GAIA2016}. Upper limits are indicated with an arrow. With the dashed lines we plot a Castelli $\&$ Kurucz model for various $A_\mathrm{V}$.} }
\label{fig:SEDs_1}
\end{figure*}
\begin{figure*}[t]
\includegraphics[width=0.33\linewidth]{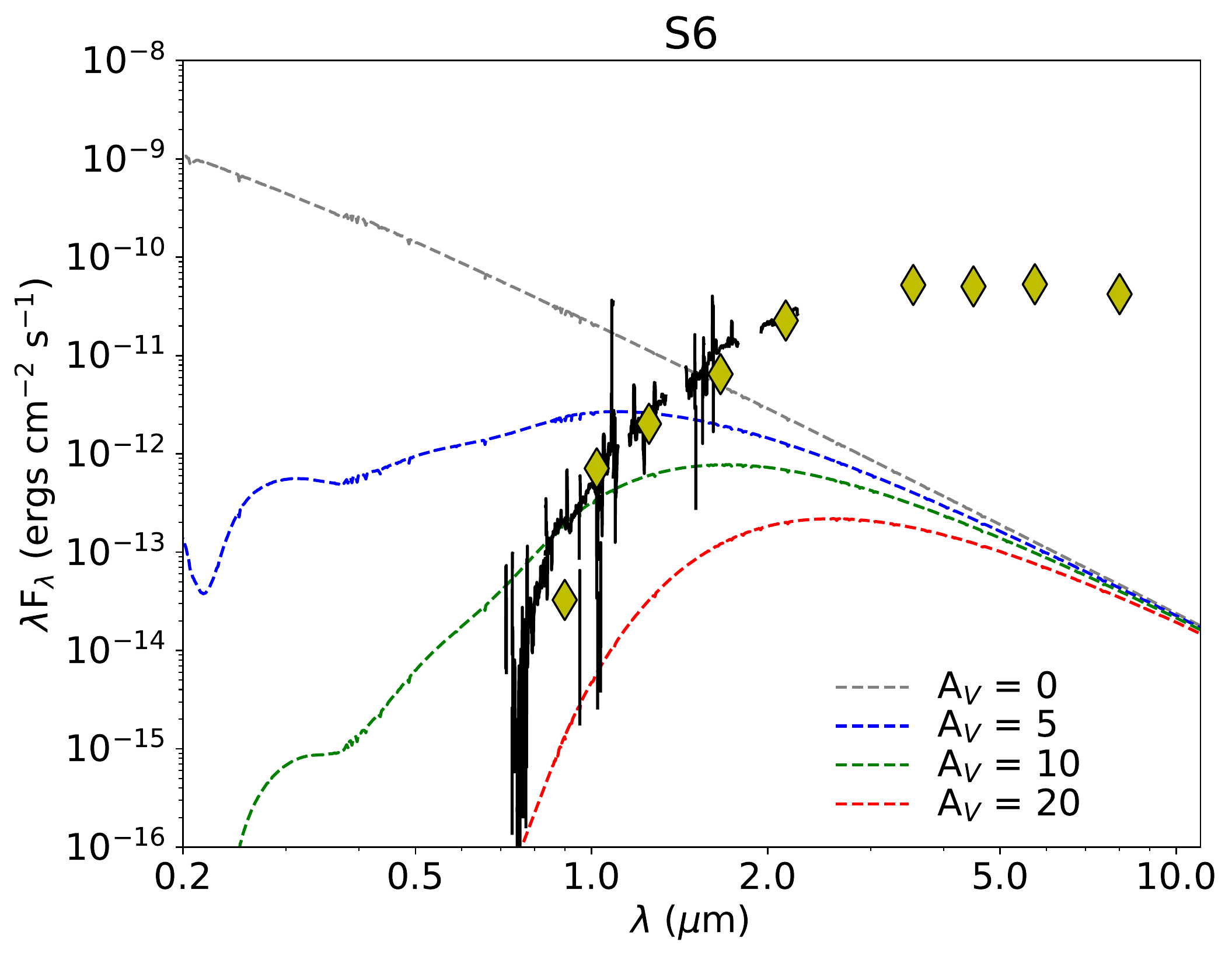}
\includegraphics[width=0.33\linewidth]{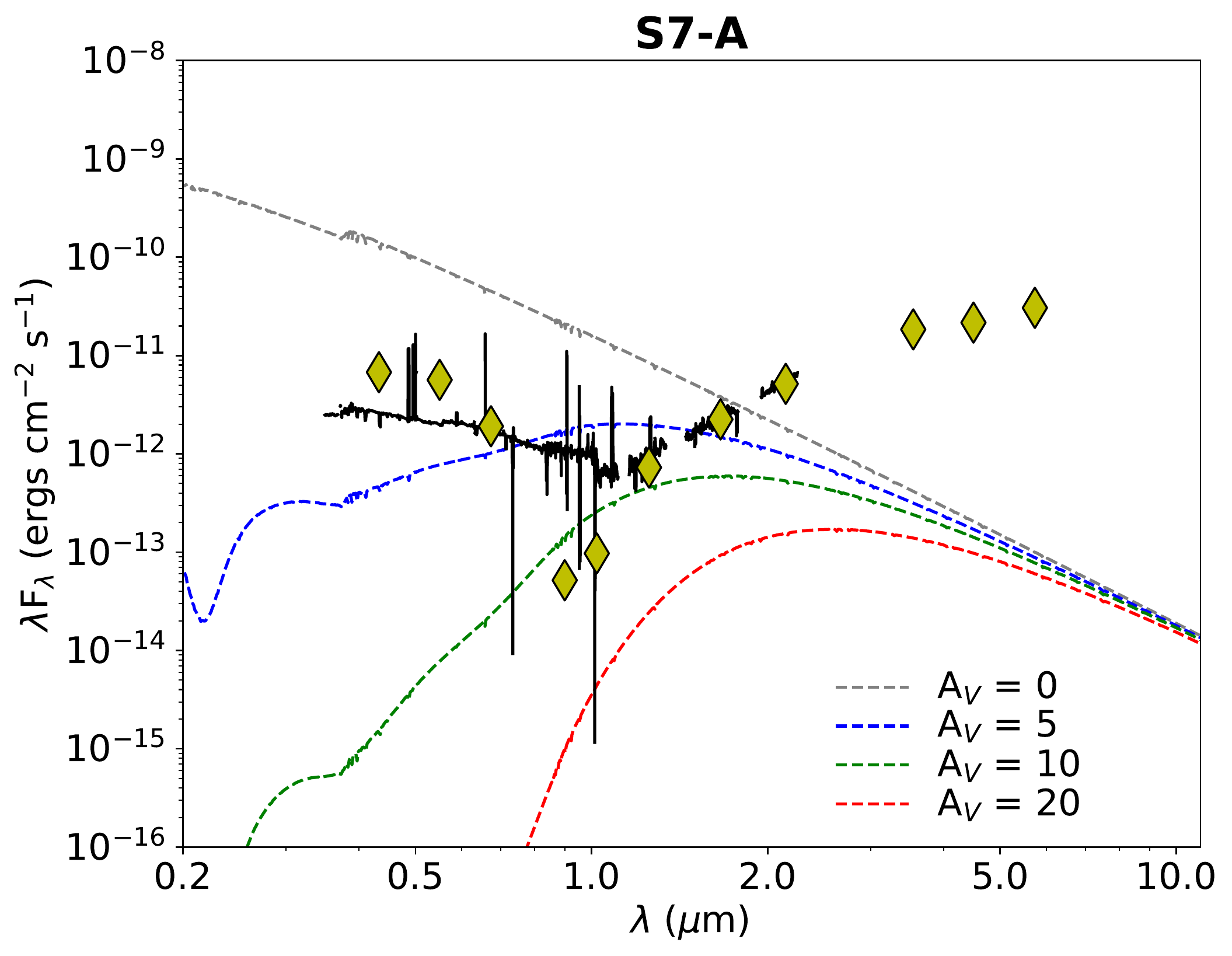}
\includegraphics[width=0.33\linewidth]{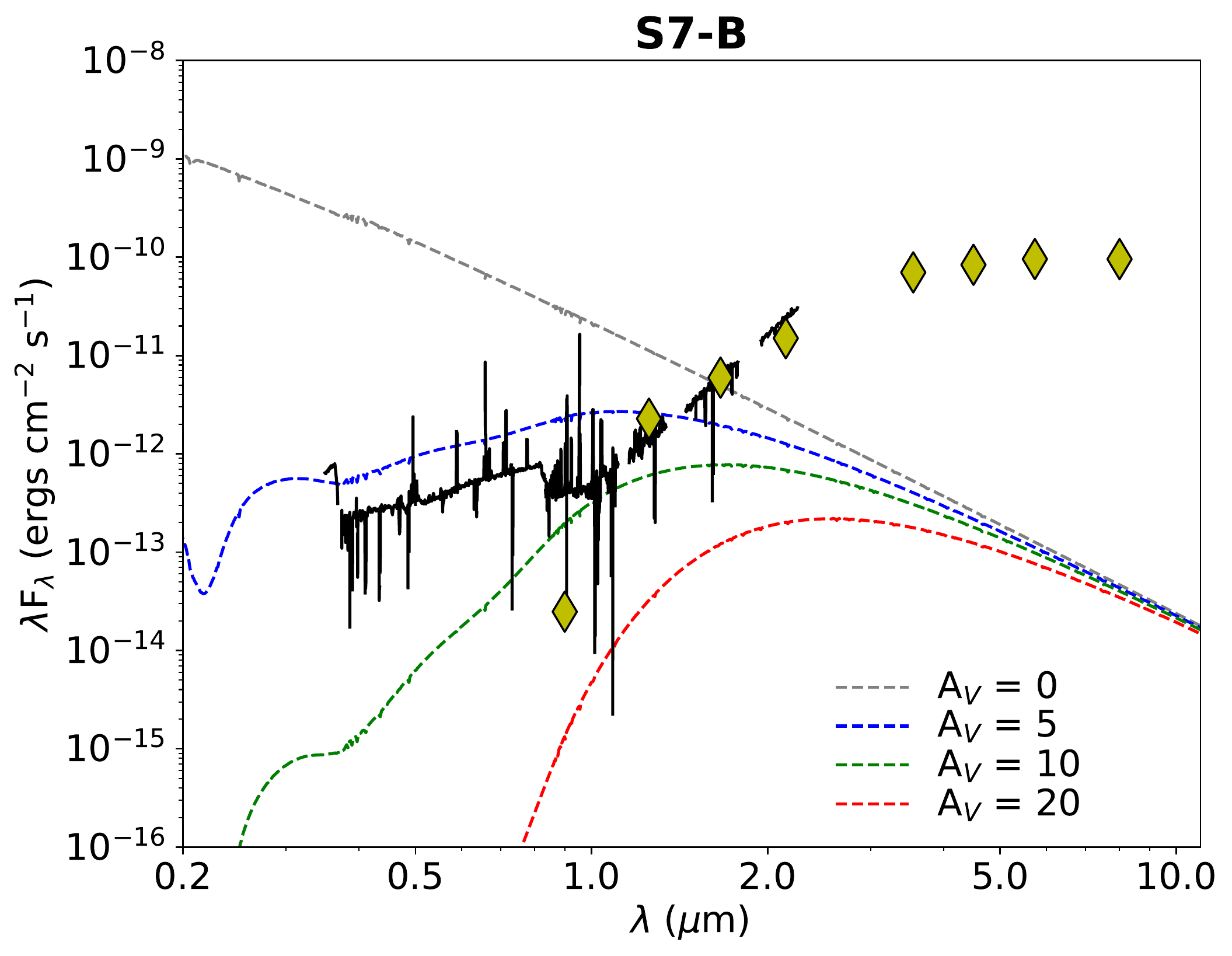}

\includegraphics[width=0.33\linewidth]{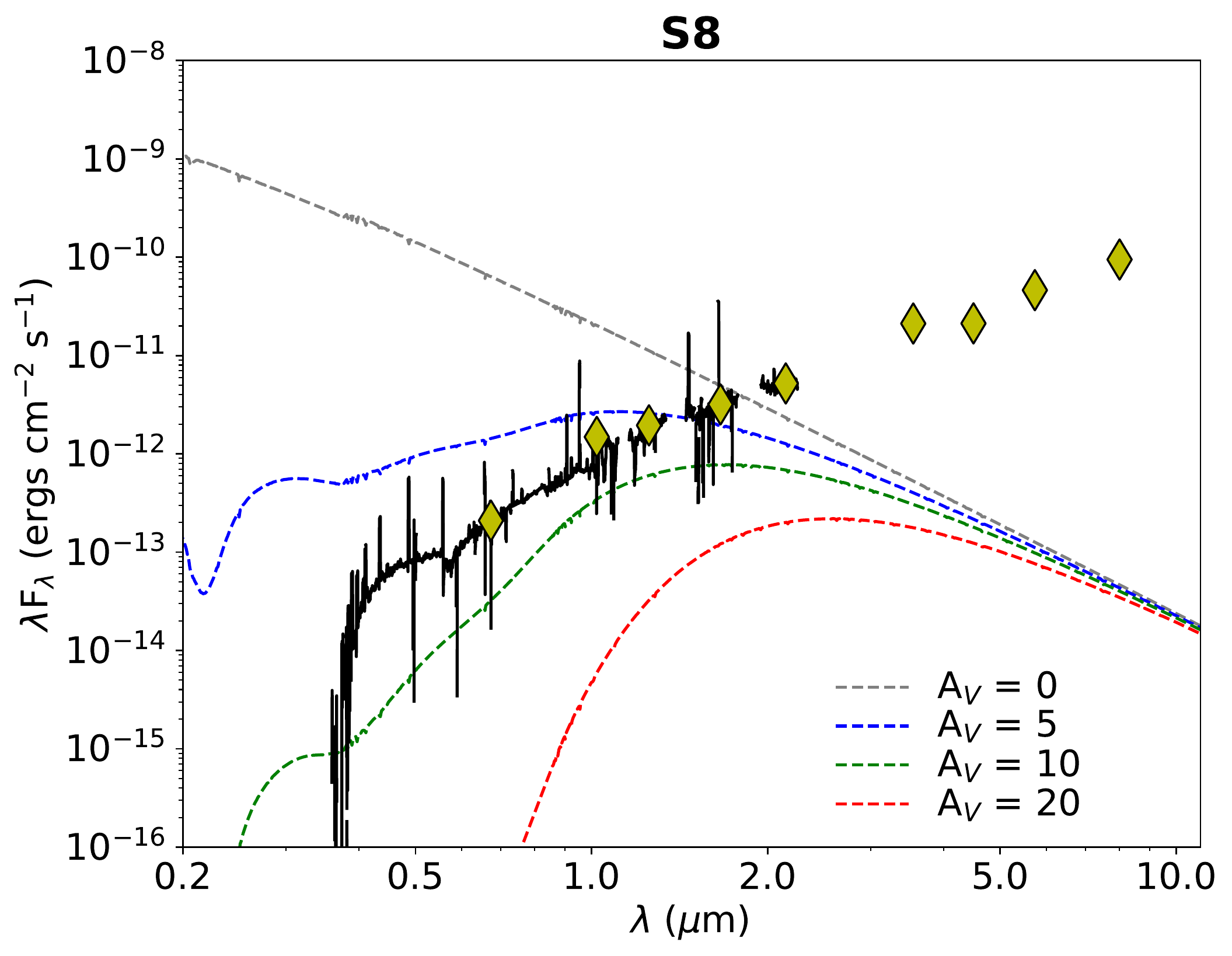}
\includegraphics[width=0.33\linewidth]{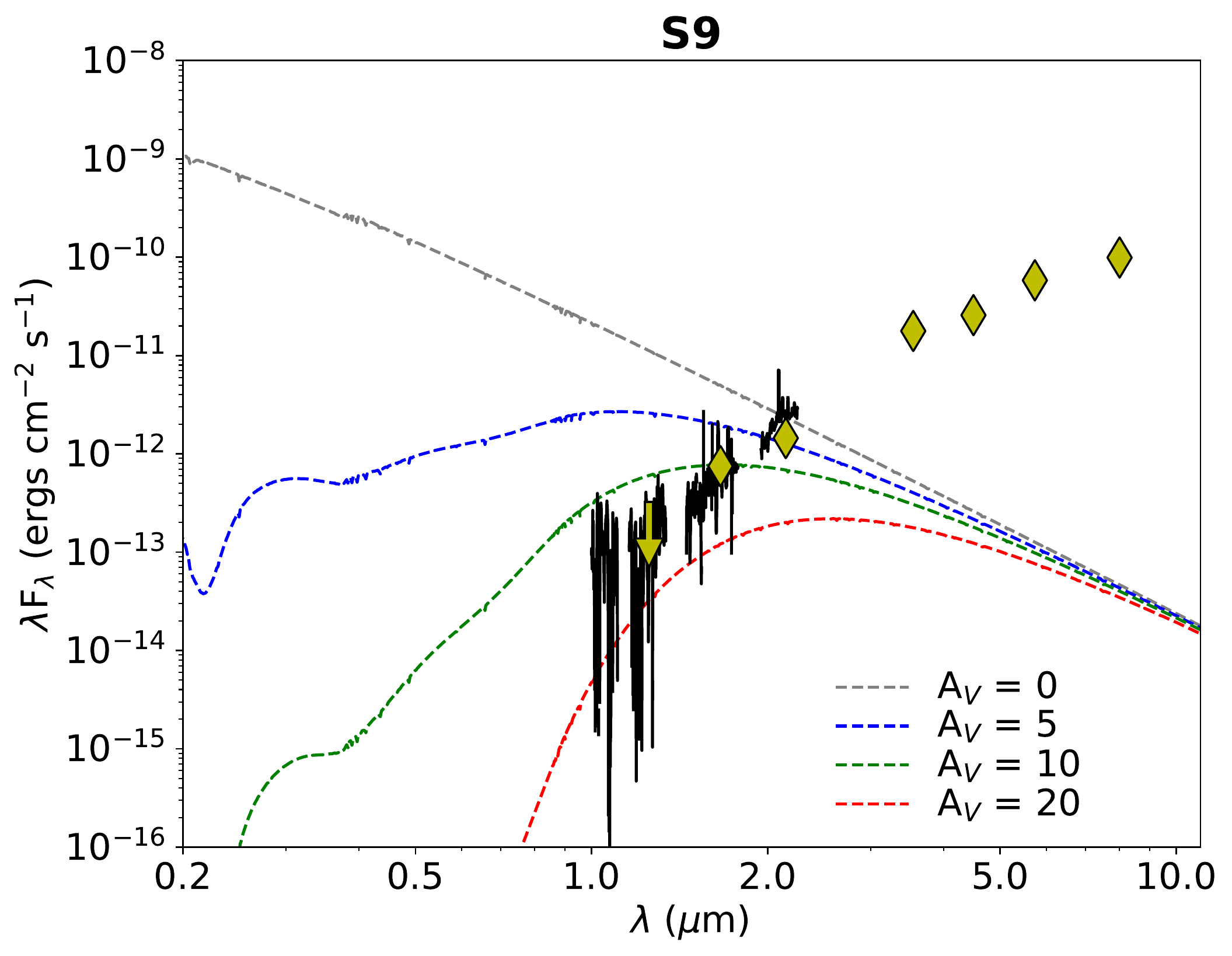}
\includegraphics[width=0.33\linewidth]{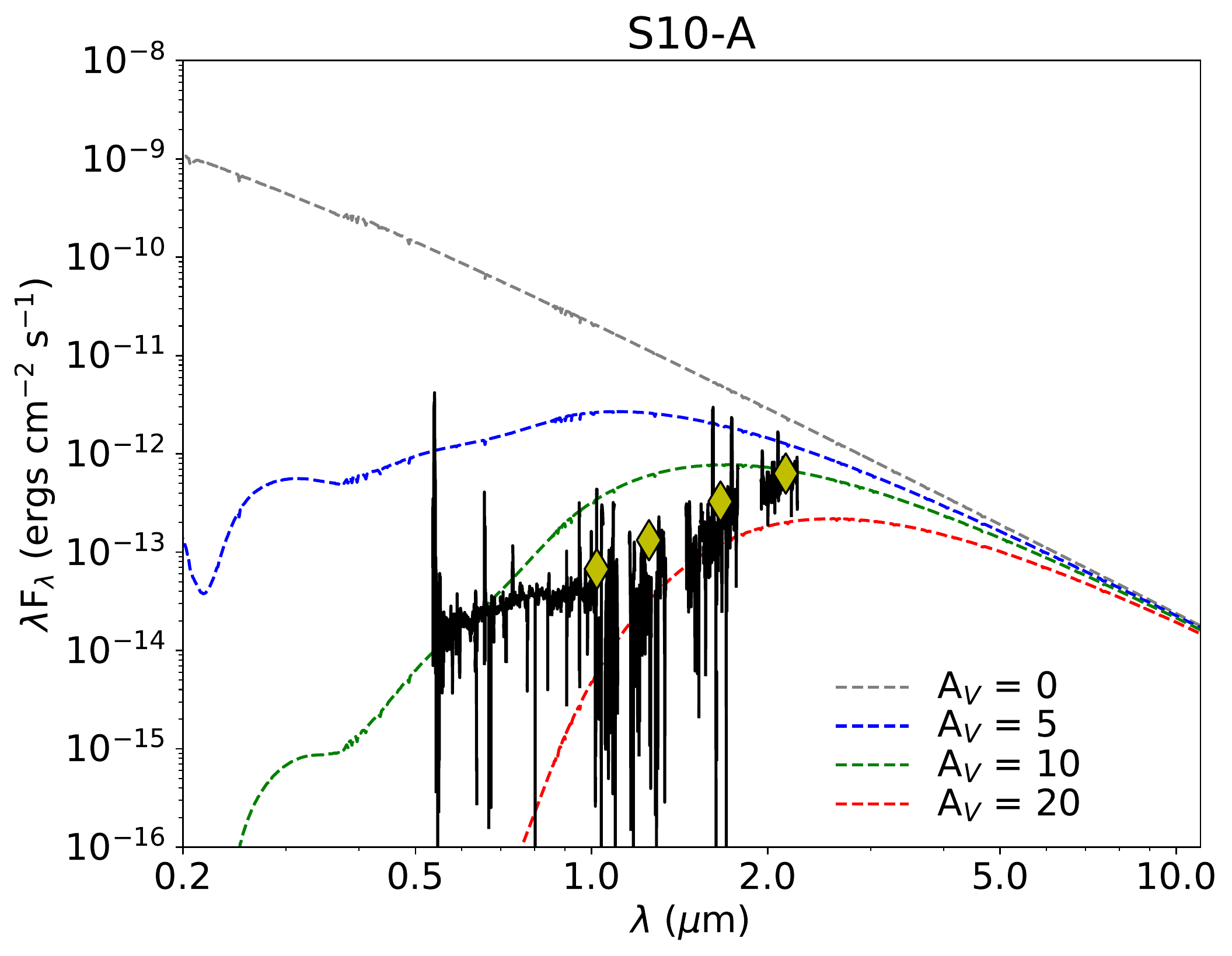}

\includegraphics[width=0.33\linewidth]{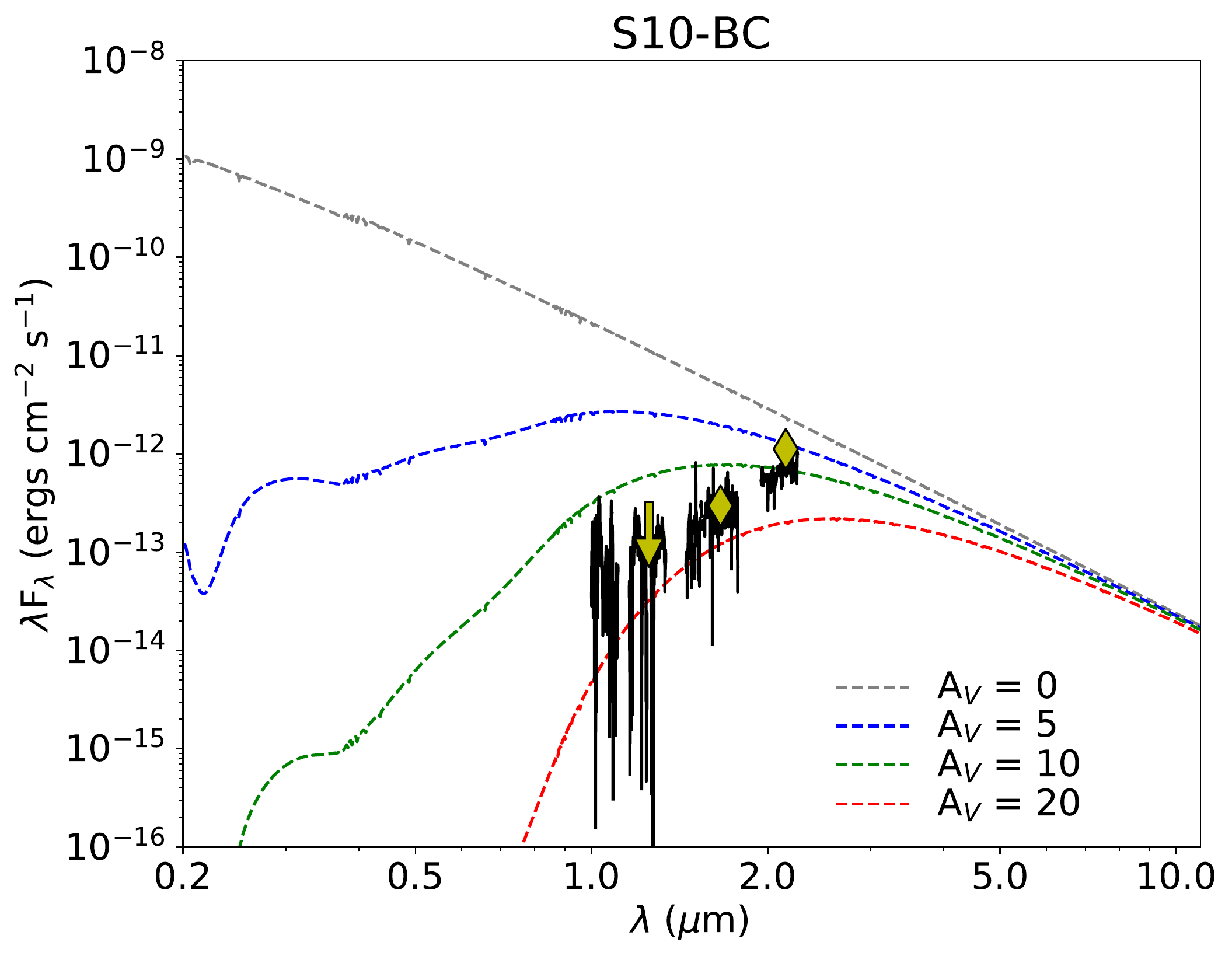}
\includegraphics[width=0.33\linewidth]{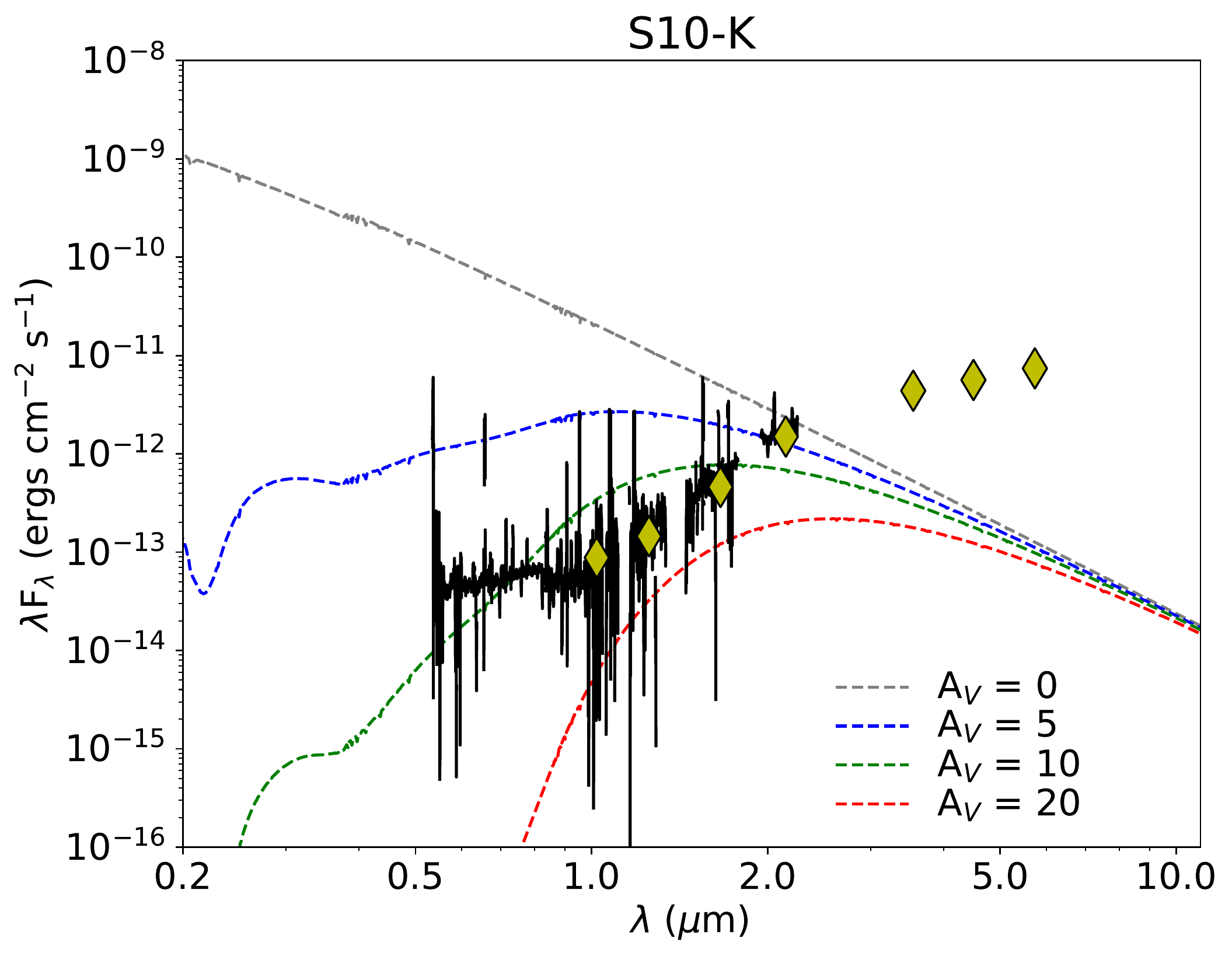}
\includegraphics[width=0.33\linewidth]{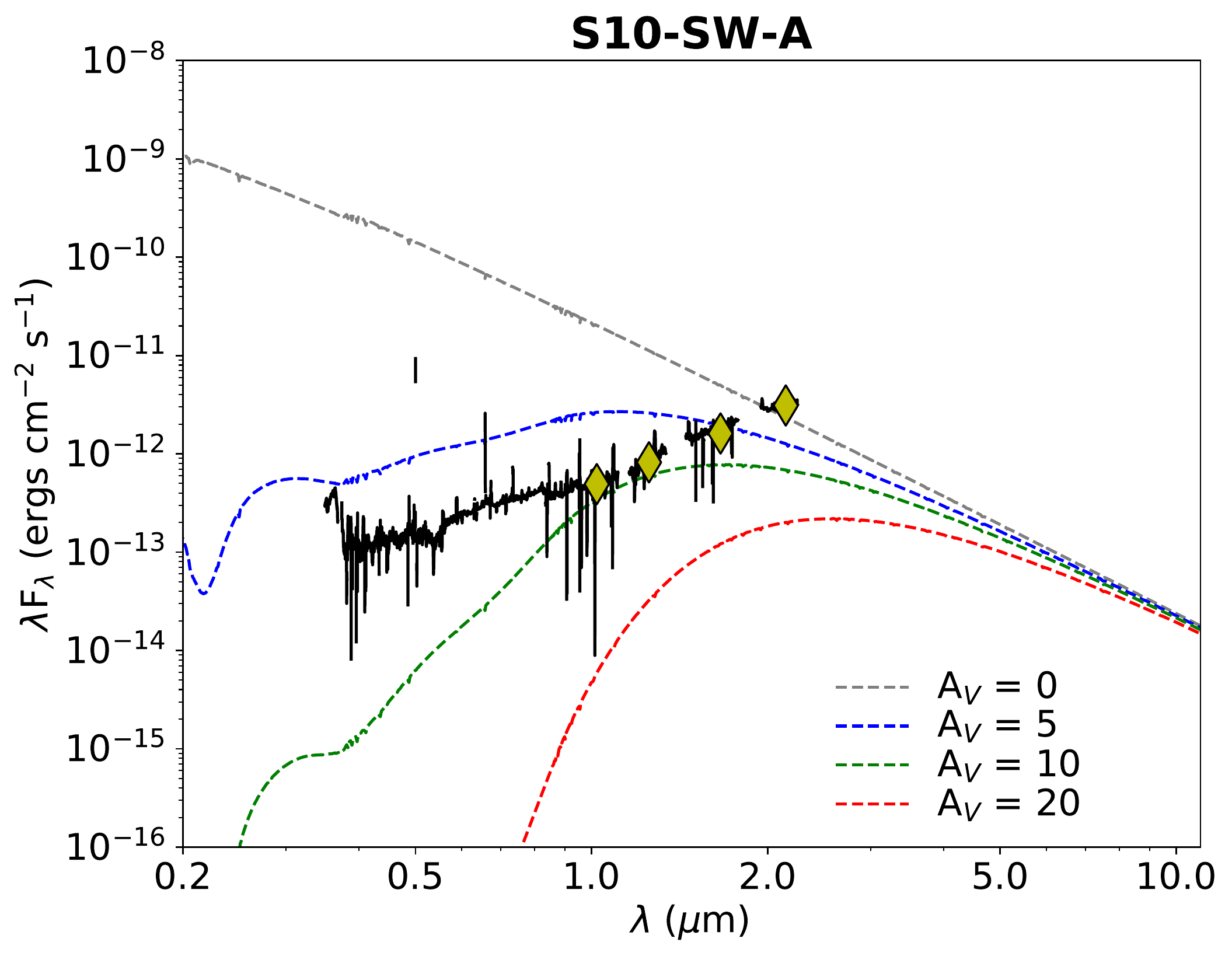}

\includegraphics[width=0.33\linewidth]{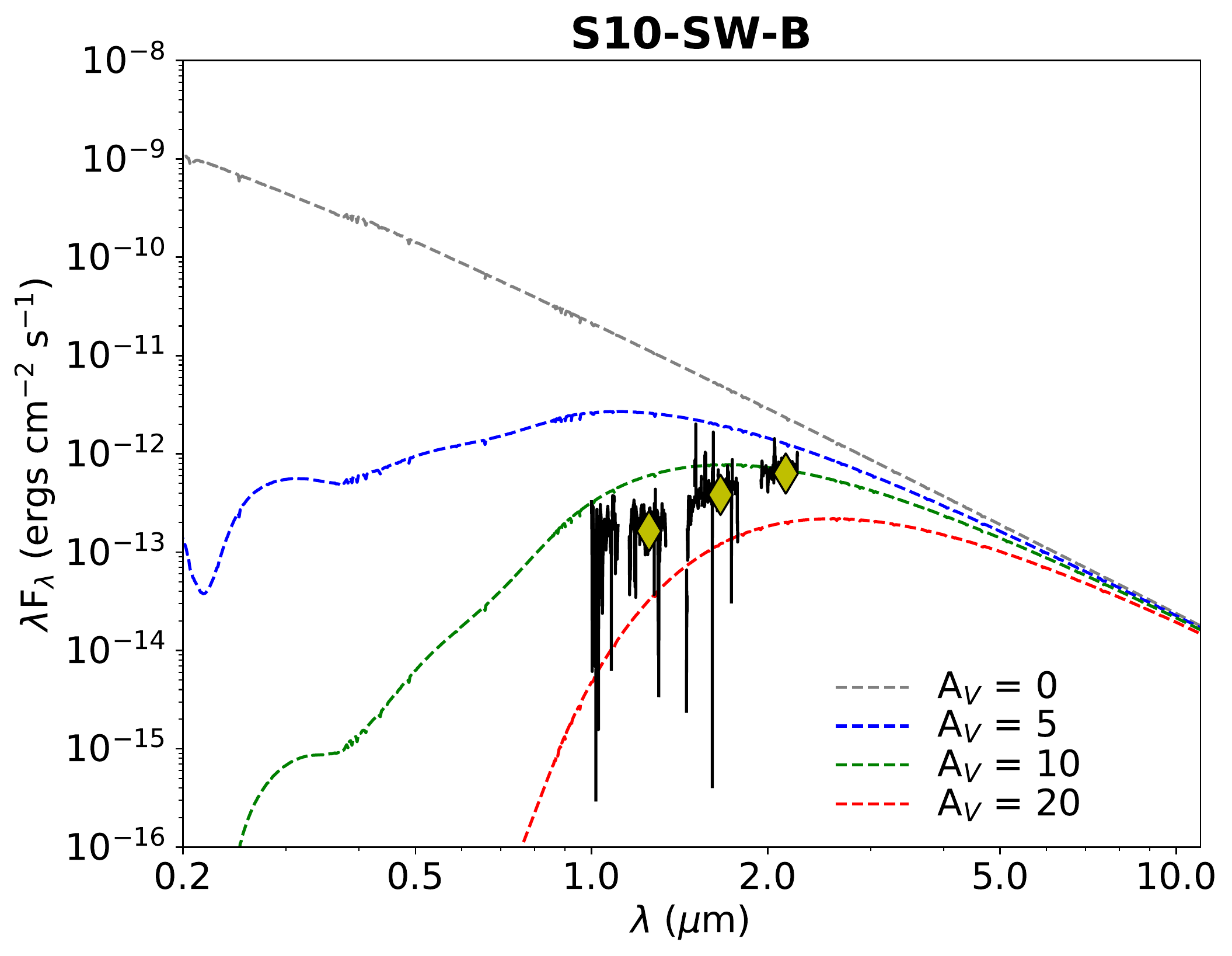}
\includegraphics[width=0.33\linewidth]{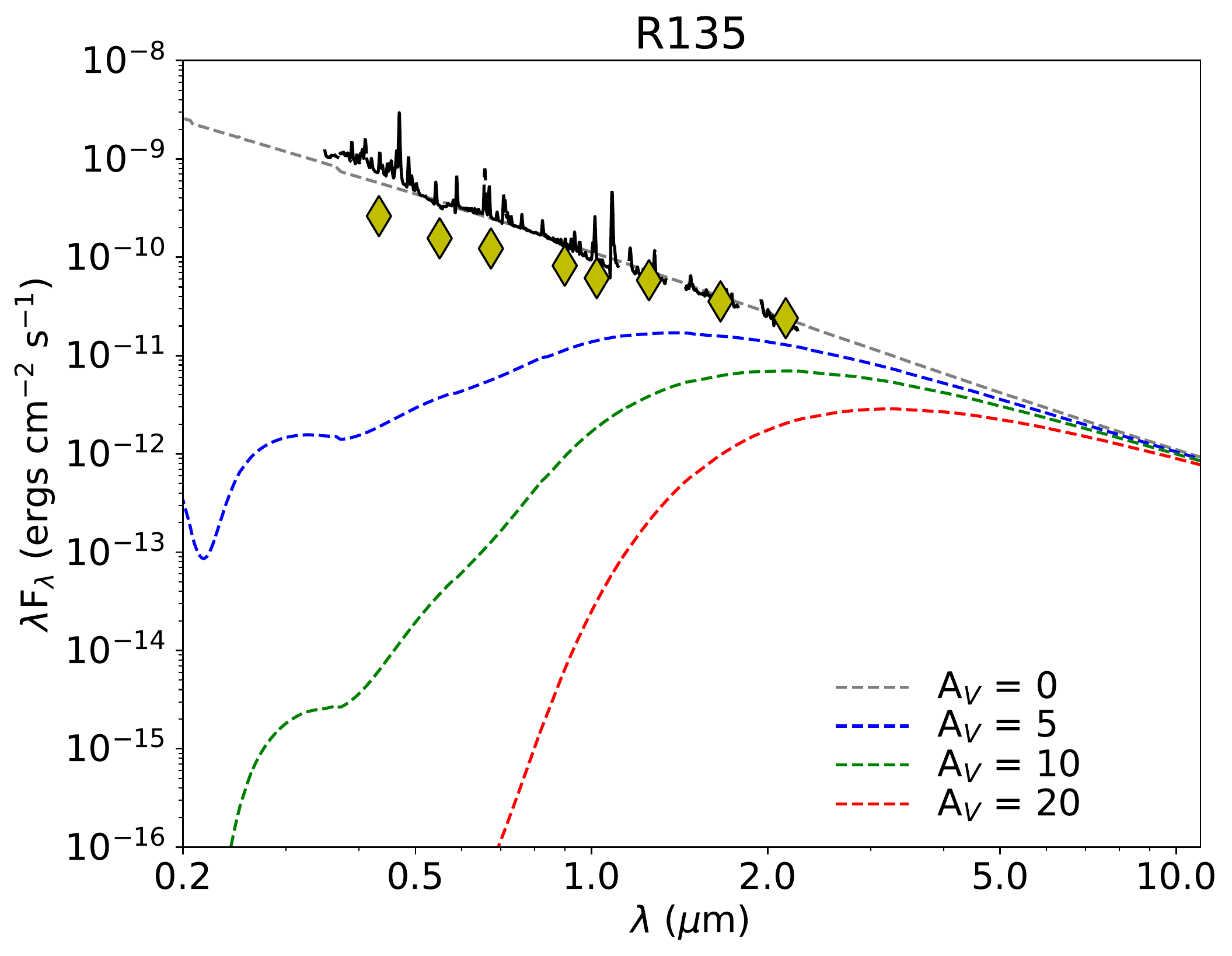}
\caption{Same as Fig.~\ref{fig:SEDs_1} but now for the objects S6 -- R135, where we clipped S6, S9, S10-A, S10-BC, S10-K, and S10-SW-B at short wavelengths.}
\label{fig:SEDs_2}
\end{figure*}

The brightest component across all wavelengths is S10-SW-A, for which we start detecting continuum from about 450~nm onwards increasing moderately in strength towards longer wavelengths. We see broad H$\alpha$ and H$\beta$ emission with a red shoulder. The Pa and Br series similarly show emission but we are unable to detect a red shoulder. Due to the Pa emission we are not able to detect possible \ac{CaII} emission as these lines are superimposed on the Pa series. {Additionally, we detect Fe\,{\sc ii}~1687.8~{nm} and H$_2$~2121.8~{nm} emission.} We classify S10-SW-A as a \ac{MYSO}. 

S10-SW-B becomes visible from about 1700~nm onwards. H$\alpha$ shows weak emission and displays a weak red shoulder. The [O\,{\sc i}]~630.0~{nm} line,{ \ac{CaII}, and H$_2$~2121.8~{nm}} show weak indications of emission. {S10-SW-B is a \ac{MYSO}.}

\subsection{Spectral energy distributions}
In Figs.~\ref{fig:SEDs_1} and \ref{fig:SEDs_2} we show the \acp{SED} of all our targets. We overplotted four Castelli $\&$ Kurucz models corresponding to the \ac{SpT}s derived in the sections above \citep[][]{Kurucz1993, Castelli2004}. For the targets with unknown \ac{SpT} we plotted the Castelli $\&$ Kurucz models of a B0~V star. All model fluxes are scaled to the \ac{30Dor} distance by correcting with a factor $(R_\star/d)^2$ for $R_\star=15$~$R_{\odot}$ and $d=50$~kpc. {For S3-K we use $R_\star=100$~$R_{\odot}$, and for R135 we use the WN model \ac{SED} of \citet{Bestenlehner2014}.} In most of our objects a \ac{NIR} excess is clearly visible in Figs.~\ref{fig:SEDs_1} and \ref{fig:SEDs_2}. Additionally we can deduce from Figs.~\ref{fig:SEDs_1} and \ref{fig:SEDs_2} that the extinction $A_\mathrm{V}$ is between 5 and 10~mag for most targets.

\section{Discussion}
\label{sec:discussion}
\subsection{Near-infrared excess}
All our \ac{MYSO} candidates show a strong \ac{NIR} excess, which suggests that our targets are surrounded by a disk {and/or envelope}. The excess is more than 5~mag for the brightest $K_\mathrm{s}$-band targets. 
The \ac{NIR} photometric points were adopted from \citet{Walborn2013}, who used observations of the \ac{VMC} and fitted \acsingle{PSF}s to the objects. 
{Not all the excess flux of our sources may however be associated with the \ac{MYSO} candidates. Some surrounding nebular (dust) emission may have been included in their photometric computations resulting in an overestimate of the brightness of the \ac{MYSO} candidates. }
Our best angular resolution in terms of seeing was about 0.7", which corresponds to $\sim$35\,000~AU (or $\sim$0.2~pc) in the plane of the sky at the distance of \ac{30Dor}. This means that many of our targets may be blended with surrounding stars. Moreover, since 70\% of the Galactic massive stars reside in close binary or higher order multiples \citep{Sana2012}, many of our targets ought to be unresolved multiple systems. 

\begin{figure*}[t]
\centering
\includegraphics[width=1.0\linewidth]{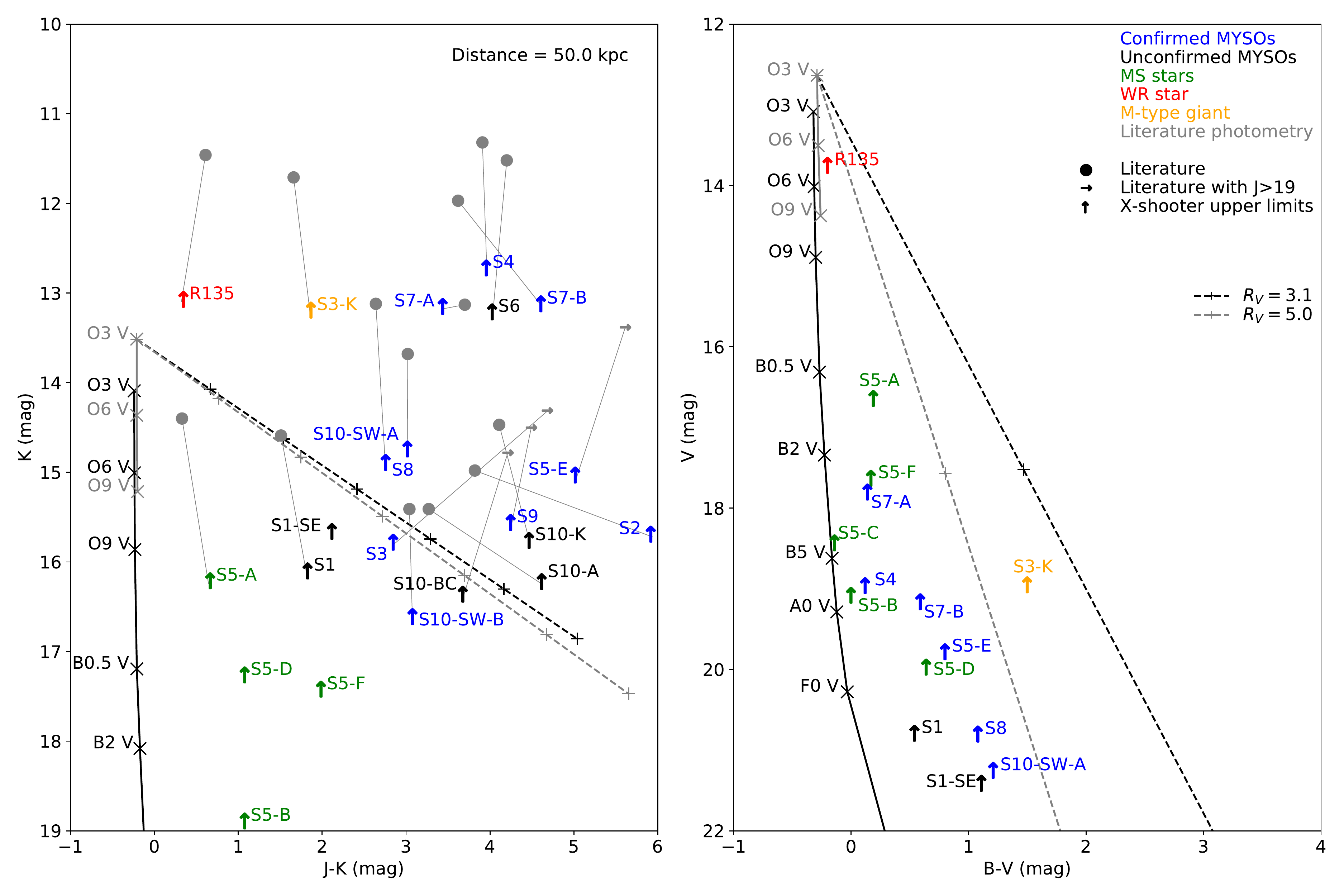}
\caption{{\it Left:} \ac{NIR} color-magnitude diagram of the upper limits derived from our X-shooter spectra (see Appendix~\ref{app:xshoomags}). Error bars are omitted for clarity but may be found in Appendix~\ref{app:xshoomags} (the typical error is $\sim$0.6~mag). {The gray lines link our upper limit with the literature values (gray dots/arrows).} All other lines are the same as in Fig.~\ref{fig:photfigure}. Note tha the color is composed of a subtraction of two upper limits and therefore uncertain. {\it Right:} optical color-magnitude diagram of the upper limit photometric points. In the \ac{VIS} range the photometric upper limits are well below the reddening line. Note that again the color is composed of a subtraction of two upper limits and therefore uncertain.}
\label{fig:xshootphots_fig}
\end{figure*}

We investigated the \ac{NIR} excess with our X-shooter spectra. Our spectra were corrected for slit loss by multiplying our flux calibrated spectra (as acquired from the X-shooter pipeline) with a photometric correction factor in order to match them with the photometric observations of \citet{Walborn2013}. However, with the flux calibrated spectra from the X-shooter pipeline we can determine an upper limit to the photometric points (i.e. lower limit to the brightness) by not correcting for slit loss. To get an unbiased upper limit we also do not correct for the malfunctioning \ac{ADC}s in the \ac{UVB} and \ac{VIS} arms, i.e. by not imposing that the edges of the X-shooter arms overlap. 

We can determine the apparent magnitude $m_i$ in photometric band $i$ by numerically integrating the flux in the band,
\begin{align}\label{eq:appmag}
m_i = -2.5\log_{10} \left( \frac{\int_i \mathcal{F}_{i,\lambda} \lambda S_{i,\lambda} \text{ d}\lambda}{\mathcal{F}_{i,0}\int_i \lambda S_{i,\lambda} \text{ d}\lambda}\right),
\end{align}
where $\mathcal{F}_{i,\lambda}$ is the flux at wavelength $\lambda$, $S_{i,\lambda}$ the filter curve, and $\mathcal{F}_{i,0}$ the flux zero point of the band. For the $B$, $V$, $R$, and $I$-bands, we used the Bessell photometric system \citep{Bessell1990}, for the $G$-band we used the Gaia photometric system \citep{GAIA2016}, and for the $Y$, $J$, $H$ and $K_\mathrm{s}$-bands we used the \acs{VISTA} photometric system\footnote{\url{http://casu.ast.cam.ac.uk/surveys-projects/vista/technical/filter-set}}. In the computation of the magnitudes we clip the regions surrounding the subtracted nebular lines so that any subtraction residuals will not contribute to the calculation. We also clip the edges of the X-shooter arms because of the significant noise and low detector response in these regions, and the part of the spectrum from 2250~nm onwards due to a gradient along the slit of continuum produced in the Earth's atmosphere having a significant impact on the measured fluxes. 

We present the computed upper limits to the magnitudes for all our objects in Appendix~\ref{app:xshoomags}. In the left part of Fig.~\ref{fig:xshootphots_fig} we plot a \ac{NIR} color-magnitude diagram with our upper limits. The color-axis in Fig.~\ref{fig:xshootphots_fig} results from the subtraction of two upper limits and is therefore ambiguous. Nevertheless we still observe a strong \ac{NIR} excess for almost all our \ac{MYSO} targets, though the strength of the excess is on average $\sim$2~mag less compared to Fig.~\ref{fig:photfigure}. In the right part of Fig.~\ref{fig:xshootphots_fig} we plot a \ac{VIS} color-magnitude diagram of the upper limits. Using the \ac{VIS} color-magnitude we can check the validity of Eq.~\eqref{eq:appmag}. If the objects with optical flux would show an excess in the \ac{VIS} color-magnitude diagram, Eq.~\eqref{eq:appmag} would fail to reproduce physical results for sure. All our objects (except for the \ac{WR} star R135) are well below the O3~V reddening lines. 

\subsection{Confirmation of MYSOs}
The confirmation of {ten} \ac{MYSO}s with X-shooter data marks the first spectroscopic confirmation of most of these \ac{MYSO}s. All our targets were selected based on the top 10 {\it Spitzer} \ac{MYSO} candidates as identified by \citet{Walborn2013}. They determined the mass of their \ac{MYSO} candidates using (mostly) \ac{NIR} and \ac{MIR} photometric points and the \ac{YSO} models of \citet{Robitaille2006}. At these wavelengths the radiation emerges mostly from the accretion disk and (inner) envelope; the stellar mass thus had to be estimated from disk and envelope properties. Since the relation between disk, envelope, and star is not well established, the mass estimate will be rather uncertain. They report S4 as the most massive \ac{MYSO} ($M=27~M_\odot$). In our observations S4 is the most prominent \ac{MYSO}. Due to the lack of photospheric lines we cannot provide a mass estimate of this source. 

{According to the classification scheme of \citet{Megeath2004}, S6 should be a Class II object. Spectroscopically we do not confirm S6, and the lack of any optical emission suggests that if S6 is a \ac{MYSO}, it is still deeply embedded. \citet{Nayak2016} classify S7-A as a Type II \ac{MYSO}. We confirm S7-A as a \ac{MYSO} and detect photospheric lines, which suggests that it might be a Class II object. S7-B is also a Type II \ac{MYSO} in their work; however, the lack of photospheric lines in this work hints at a Class 0/I nature rather than a Class II nature. All other confirmed \ac{MYSO}s, S2, {S3}, S4, S8, {S9}, S10-SW-A, and {S10-SW-B} are Type~I \ac{YSO}s \citep{Nayak2016}. Furthermore, two \ac{MYSO} candidates identified with {\it Spitzer}/\ac{IRS} (i.e. S5-E and S10-SW-A) are confirmed here as \ac{MYSO}s \citep{Seale2009,Jones2017}. 

S9 was also a {\it Spitzer}/\ac{IRS} \ac{MYSO} candidate, and was the most massive \ac{MYSO} in the sample of \citet[][although they may have mixed S9 with the massive young star VFTS~621 since they refer to S9 as the object with \ac{SpT} O2]{Nayak2016}. 
{More remarkably is the absence of the CO bandhead absorption observed by \citet{Reiter2019}.
Our observations were taken about 3 years earlier; however, variability on timescales of $\sim1$~yr have been observed for e.g. FU~Orionis type stars \citep{ContrerasPena2017}.} 

\begin{table}[t]
\begin{center}
\caption{Lower limit luminosities and masses computed from X-shooter upper limits to the magnitudes.}
\footnotesize{
\begin{tabular}{@{\extracolsep{-1pt}}lllll@{}}
\hline
\hline
Object & $J$ (mag) & $\log_{10}(L_\star/L_\odot)$ & $\log_{10}(L_\star/L_\odot)$ & $M_\star/M_\odot$ \\
 & This work & & This work & This work\\
\hline
S1       & 17.9$\pm$0.2 & 4.8$\pm$0.1 & 4.1$\pm$0.1 & 15 \\
S1-SE    & 17.8$\pm$0.2 & --  & 4.2$\pm$0.1 & 16\\
\textbf{S2}       & 21.6$\pm$3.4 & 3.8$\pm$0.2 & 2.6$\pm$1.3 & 6 \\
\textbf{S3}       & 18.6$\pm$0.4 & 3.7\tablefootmark{1}  & 3.8$\pm$0.1 & 13 \\
S3-K     & 15.0$\pm$0.2 & 3.3$\pm$0.1 & 2.7$\pm$0.1 & 6 \\
\textbf{S4}       & 16.7$\pm$0.2 & 5.2$\pm$0.1 & 4.6$\pm$0.1 & 21 \\
S5-A     & 16.9$\pm$0.2 & 5.7$\pm$0.1\tablefootmark{2} & 4.9$\pm$0.1 & 25 \\
S5-B     & 19.9$\pm$0.6  & --  & 3.3$\pm$0.2 & 9  \\
S5-C     & 20.1$\pm$0.6 & --  & 3.2$\pm$0.2 & 8 \\
S5-D     & 18.3$\pm$0.3 & --  & 4.0$\pm$0.1 & 14 \\
\textbf{S5-E}     & 20.0$\pm$0.5 & 3.7\tablefootmark{1} & 3.3$\pm$0.2 & 9 \\
S5-F     & 19.4$\pm$0.5 & --  & 3.5$\pm$0.2 & 10 \\
S6       & 17.2$\pm$0.2 & 5.0$\pm$0.1 & 4.4$\pm$0.1 & 18 \\
\textbf{S7-A}     & 16.6$\pm$0.2 & 4.3$\pm$0.1 & 4.4$\pm$0.1 & 18 \\
\textbf{S7-B}     & 17.7$\pm$0.2 & 5.0$\pm$0.1 & 4.2$\pm$0.1 & 16 \\
\textbf{S8}       & 17.6$\pm$0.3 & 5.0$\pm$0.1\tablefootmark{3} & 4.2$\pm$0.1 & 16 \\
\textbf{S9}       & 19.8$\pm$0.8 & 3.7\tablefootmark{1}  & 3.4$\pm$0.3 & 9  \\
S10-A    & 20.8$\pm$1.8 & 3.8$\pm$0.1 & 3.0$\pm$0.7 & 7 \\
S10-BC   & 20.0$\pm$0.7 & 3.7\tablefootmark{1} & 3.3$\pm$0.3 & 9 \\
S10-K   & 20.2$\pm$0.8 & 3.8$\pm$0.1 & 3.2$\pm$0.3 & 8 \\
\textbf{S10-SW-A} & 17.7$\pm$0.2 & 4.6$\pm$0.1 & 4.2$\pm$0.1 & 16 \\
\textbf{S10-SW-B} & 19.7$\pm$0.4 & 3.9$\pm$0.1 & 3.4$\pm$0.2 & 9 \\
R135 & 13.4$\pm$0.2 & 6.9$\pm$0.1\tablefootmark{2,4} & 6.4$\pm$0.1\tablefootmark{4} & -- \\
\hline
\end{tabular}
}
\tablefoot{The objects shown in bold are the \ac{MYSO}s confirmed in this work. The luminosity in the third column computed from the $J$-band magnitude of \citet{Walborn2013}. Since we use magnitude upper limits in this work to compute the luminosity and mass, the values in this table are lower limits. \\
\tablefoottext{1}{Calculated for a $J>19$ lower limit, and therefore an upper limit to the luminosity.} \tablefoottext{2}{Using the magnitude from the 2MASS database \citep{Cutri2003}.} \tablefoottext{3}{Computed for the combined magnitude of both sources.} \tablefoottext{4}{For $A_\mathrm{V}=0$}.}
\label{tab:luminosities}
\end{center}
\end{table}
}

We can determine whether our confirmed \ac{MYSO}s are indeed massive by deriving their luminosity and using a mass-luminosity relation to estimate the mass. For this we use the $J$-band since there the disk does not (yet) completely dominate over the central star, and because the extinction in this band is rather low. {We compute the luminosity both for the magnitudes reported by \citet{Walborn2013} and for the magnitude upper limits presented in this work (see Appendix~\ref{app:xshoomags}). Note that in the latter case the resulting luminosity and mass is a lower limit. The results of the calculation are shown in Table~\ref{tab:luminosities}, where we used $A_\mathrm{V}=5$, and the bolometric correction (BC$_\mathrm{J}$) following \citet{Martins2006} for the corresponding \ac{SpT} (B0~V was used for the targets with unknown \ac{SpT}).} Typically, our luminosity lower limits are about 0.5~dex lower than the luminosities derived from the photometric points of \citet{Walborn2013}. The mass is estimated using a typical \ac{ZAMS} L-M relation ($L\propto M^{3.5}$). We do not compute errors on the mass lower limits since our estimates are based on a proportionality. All confirmed \ac{MYSO}s except S2 show luminosities and masses consistent with a massive star nature. Note that S2 may still be a massive star because the mass estimate is a lower limit and based on the assumption of the source already being on the \ac{ZAMS}.

\subsection{Comparison to {other samples}}
Strong emission lines such as the \ac{CaII} and Br$\gamma$ are indicative of inflow of circumstellar material. We detect \ac{CaII} emission towards 50\% of the confirmed \ac{MYSO}s, which agrees with the Galactic star forming region M17 \citep[66\%;][]{Macla2017}. Our detection rate of Br$\gamma$ is 80\%, which is consistent with the high detection rate in other \ac{LMC} samples {\citep{Ward2016,Ward2017PhD,Reiter2019}}, \ac{SMC} samples \citep[][]{Ward2017,Reiter2019}, and larger Galactic samples \citep{Cooper2013,Pomohaci2017}. 
{In Fig.~\ref{fig:Lbr-AbsK} we show the Br$\gamma$ luminosity of our confirmed \ac{MYSO}s against the absolute $K$-band magnitude (using the $K_\mathrm{s}$-band magnitudes of \citet{Walborn2013} and $A_\mathrm{V}=5$; for the \ac{LMC} MYSOs of \citet{Reiter2019} we find lower luminosities with their Br$\gamma$ fluxes). The main difference between Galactic and Magellanic \ac{MYSO}s is the difference in metallicity. \citet{Ward2017} suggested that the Br$\gamma$ luminosity (which is a probe of the accretion luminosity) increases with decreasing metallicity. 
However, the spread of the data points of the Magellanic Clouds in Fig.~\ref{fig:Lbr-AbsK} is too large to see any significant correlation. }

\begin{figure*}
    \centering
    \includegraphics[width=1.0\linewidth]{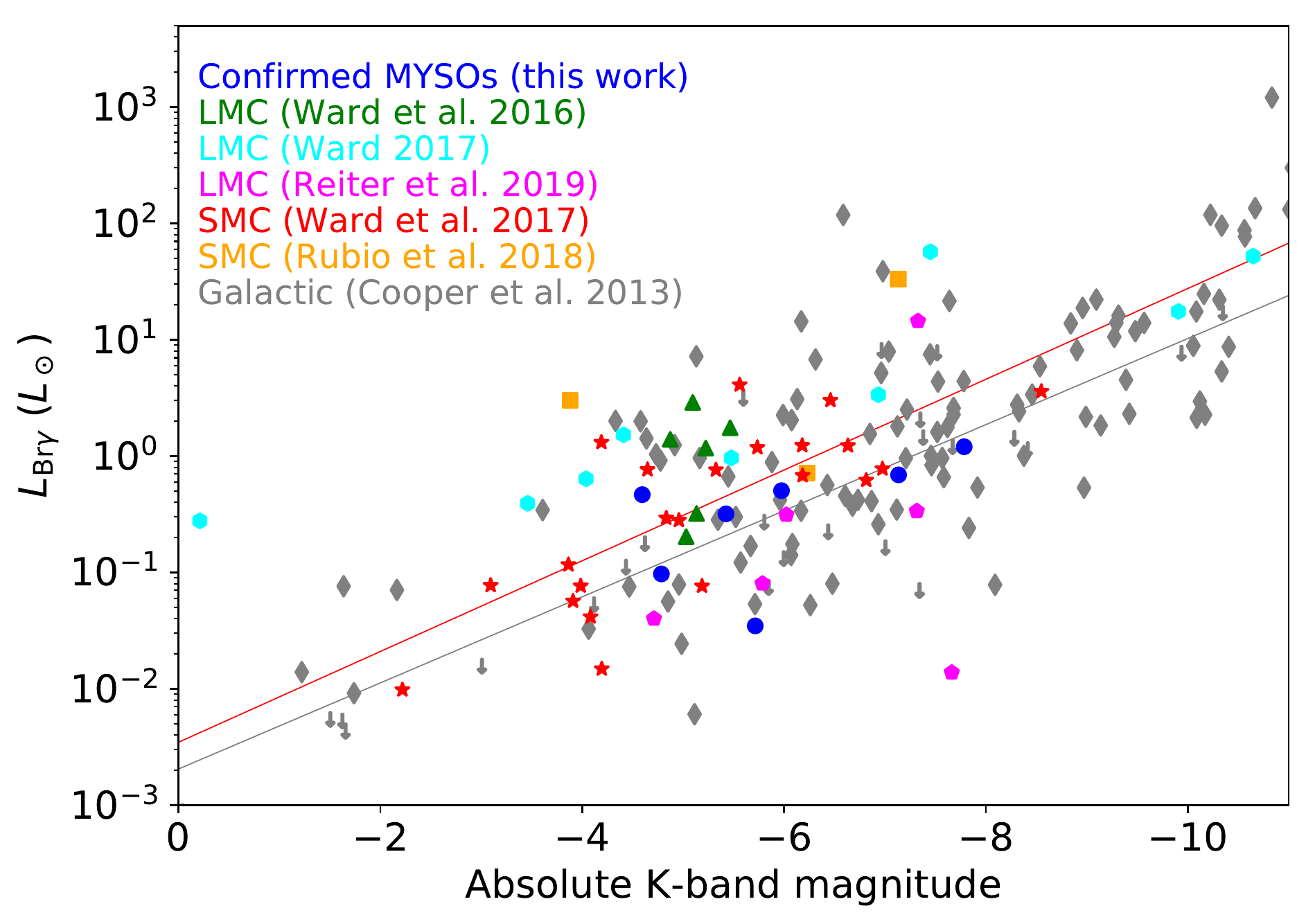}
    \caption{{The luminosity of Br$\gamma$ plotted against the absolute $K$-band magnitude. Overplotted are various \ac{MYSO} samples in the \ac{LMC}, \ac{SMC}, and our Galaxy. With the gray and red lines we indicate the empirically derived relations of \citet{Cooper2013} and \citet{Ward2017} for our Galaxy and the \ac{SMC}, respectively.}}
    \label{fig:Lbr-AbsK}
\end{figure*}

Another metallicity dependent observable might be the detection rate of fluorescent Fe\,{\sc ii} and CO bandheads. {Both these are tracers of an accretion disk. In total 70\% of our confirmed \ac{MYSO}s show either fluorescent Fe\,{\sc ii} and CO bandheads, which is higher than the Galactic rate of $\sim$40\% \citep{Cooper2013,Ellerbroek2013_1,Macla2017,Pomohaci2017}.} However, we have a small sample and are biased towards brighter targets. Adding the \ac{LMC} samples of \citet[][only CO bandheads]{Ward2016}, {\citet{Ward2017PhD}}, and \citet{Reiter2019} gives a combined detection rate of {47\% for either}  Fe\,{\sc ii} or CO bandheads (or both). This is consistent with the Galactic rate, yet still the combined \ac{LMC} sample is significantly smaller.

\subsection{Outflows}
Outflows are thought to be common in \ac{MYSO}s \citep[e.g.][]{Zhang2001, Zhang2005}. They can be characterized by e.g. {[O\,{\sc i}]~630.0~nm}, H$_2$ or [Fe\,{\sc ii}] showing emission, {occasionally} with an offset \ac{RV} of up to a few hundred~\kms\text{} \citep{Ellerbroek2013_1}. {Of our confirmed \ac{MYSO}s, 80\% shows outflow signatures. {[O\,{\sc i}]~630.0~{nm} emission is detected for 40\% of the \ac{MYSO}s; however, the identification was often hampered by residuals of the nebular subtraction.} S2 and S8 are the only confirmed \ac{MYSO}s which do not show any H$_2$~2121.8~nm emission.} The \ac{RV} of H$_2$~2121.8~{nm} in all sources shows no {significant} offset from the assumed systemic velocity (250--260~\kms). [Fe\,{\sc ii}]~1643.5~{nm} is only detected towards S4 and S7-B.  Whereas the \ac{RV} of [Fe\,{\sc ii}]~1643.5~{nm} in S4 is around the systemic velocity, S7-B shows double-peaked [Fe\,{\sc ii}]~1643.5~{nm} and Pa series emission with a \ac{RV} of $-615$~\kms\text{} and $-525$~\kms\text{} in the local frame of reference. The two velocities might indicate so-called bullets in the outflow, where it shows enhancements in density and temperature at certain positions compared to the rest of the outflow.

Another measure of bipolar outflow is the presence of H$_2$O maser emission. Water masers coinciding with the locations of S9 and S10-SW-A have been reported by \citet{Ellingsen2010}. This is consistent with the detection of H$_2$~2121.8~{nm} for these sources in this work. 

\section{Summary}
\label{sec:conclusion}
{We present the results on a spectroscopic analysis of the top {10} {\it Spitzer} \ac{MYSO} candidates in \ac{30Dor} of \citet{Walborn2013}. These targets are resolved into in $\sim$ 20 sources.} We took advantage of the unparalleled spectral resolution (R$\sim$4000-17\,000) and wavelength coverage (300-2500~nm) of \ac{VLT}/X-shooter to detect spectral features characteristic for \ac{MYSO}s. 

All VLT/X-shooter spectra of the \ac{MYSO} candidates were contaminated by nebular emission. We used a scaling method developed in this work to subtract this nebular contamination from our spectra, revealing the spectral features intrinsic to the \ac{MYSO}s. 

Photometric observations from the literature suggest that our objects possess a strong \ac{NIR} excess indicating the presence of an accretion disk. We computed photometric upper limits using our X-shooter spectra. These limits still argue for the presence of a strong \ac{NIR} excess. This indicates that our targets are surrounded by a large amount of circumstellar dust.

We spectroscopically confirm S2, S3, S4, S5-E, S7-A, S7-B, S8, S9, S10-SW-A, and S10-SW-B as \ac{MYSO}s by the detection of features such as the \ac{CaII}, Br$\gamma$, {fluorescent Fe\,{\sc ii}}, {H$_2$}, and CO first-overtone bandhead emission. We computed luminosity and mass lower limits for our targets which support a massive star nature for all confirmed \ac{MYSO}s except S2. S7-A shows photospheric lines which hint at a Class II \ac{MYSO} origin whereas all other confirmed \ac{MYSO}s seem to have a Class 0/I origin. 

We computed the Br$\gamma$ luminosity for all confirmed \ac{MYSO}s, and find that these are consistent with other samples in the \ac{LMC} and \ac{SMC} {\citep{Ward2016,Ward2017,Ward2017PhD,Rubio2018,Reiter2019}}. {Due to the large scatter in datapoints, no clear correlation was seen between the Br$\gamma$ luminosity (i.e. accretion luminosity) and metallicity.}
Combining our {detection rate of disk tracers such as fluorescent Fe\,{\sc ii} and CO bandhead} with those of other \ac{LMC} samples of \citet{Ward2016} and \citet{Reiter2019} is consistent with the Galactic rate \citep[$\sim$40\%;][]{Cooper2013}

{We detect signatures of an outflow in 80\% of \ac{MYSO}s through the detection of [O\,{\sc i}]~630~nm, H$_2$~2121.8~nm, and [Fe\,{\sc ii}]~1643.5~nm.} {S7-B might show so-called bullets (i.e. enhancements in density and temperature) in the outflow.} Future analysis of this outflow is required to confirm its nature and composition.

With this still rather small sample of \ac{MYSO}s we {studied} massive star formation {in the most extreme massive star forming region in the Local Group}. Modelling the emission and \ac{NIR} excess to disk models will provide insight in how these massive stars form and evolve. In particular, the effect of metallicity on {e.g. the accretion luminosty} can be studied by comparing these models to their Galactic counterparts. This sample contains several \ac{MYSO}s which are still deeply embedded in their birth clouds. 
With current sub-mm telescopes such as {the} \ac{ALMA} we can study the gaseous molecular content of these \ac{MYSO}s \citep[e.g. CO;][]{Nayak2016}. The angular resolution of \ac{ALMA} is, however, still insufficient to spatially resolve these \ac{MYSO}s (0.02" resolution corresponds to $\sim1000$~AU in \ac{30Dor}). Future optical and \ac{NIR}/\ac{MIR} facilities such as the \acl{JWST} and \acl{ELT}  will be vital in further characterizing the still poorly understood process of formation of massive stars, both in our Galaxy and in the Magellanic Clouds.

\begin{acknowledgements}
{We would like to thank the anonymous referee for his/her useful comments on the manuscript.}

Based on observations collected at the European Southern Observatory under ESO program 088.D-0850(A) and 090.C-0346(A). We thank the ESO support staff (including former DG Tim de Zeeuw) for carrying out the observations. 

LK and JJ acknowledge support from NOVA and an NWO-FAPESP grant for advanced instrumentation in astronomy. 

The work of M.S. is based upon work supported by NASA under award number 80GSFC17M0002.

This work made use of observations made with the NASA/ESA Hubble Space Telescope, and were obtained from the Hubble Legacy Archive, which is a collaboration between the Space Telescope Science Institute (STScI/NASA), the Space Telescope European Coordinating Facility (ST-ECF/ESA) and the Canadian Astronomy Data Centre (CADC/NRC/CSA).

\end{acknowledgements}


\bibliographystyle{aa}
\bibliography{refs}

\clearpage

\begin{appendix}
{ 
\section{Nebular line subtraction}
\subsection{Nebular line models}
\label{app:nebcor}
The main model used is the \acf{GD}
\begin{align}
    y_{\text{GD}}(\lambda)  & = N \exp\left(-\frac{(\lambda-\lambda_0)^2}{2\sigma^2}\right), 
\end{align}
where $N$ is the peak flux, $\lambda_0$ is the central wavelength, $\sigma$ is the width. Additionally, if needed (due to e.g. lower detector response or bad seeing conditions during the observation), we used a \acl{FGD} \citep[\acs{FGD};][]{Blazquez2008}, or \acl{MD} \citep[\acs{MD};][]{Moffat1969}, 
\begin{align}
    y_{\text{FGD}}(\lambda) & = N \exp\left(-\frac{(\lambda-\lambda_0)^2}{2\sigma_1^2} - \frac{(\lambda-\lambda_0)^4}{2\sigma_2^4}\right), \\
    y_{\text{MD}}(\lambda)  & = N \left(1+\frac{(\lambda-\lambda_0)^2}{\alpha^2}\right)^{-\beta},
\end{align}
where $N$ is the peak flux, $\lambda_0$ is the central wavelength of the corresponding distribution, $\sigma_{1,2}$ represent the width of the \ac{FGD} and $\alpha$ and $\beta$ are seeing dependent variables. 

Since the \ac{30Dor} nebula consist of multiple velocity components \citep{Torres-Flores2013,Mendes2017}, estimating the nebular contribution required up to 3 of the distributions above depending on the line of sight. 

\subsection{Nebular line plots}
\label{app:nebplot}
In Fig.~\ref{fig:nebsubplots} we present a few plots showing a nebular line fitted to a triple Gaussian model. The top and middle panels show lines from the lower ionized species N\,{\sc ii} and S\,{\sc ii}, and the bottom panel a line of a higher ionized species: Ar\,{\sc iii}. As mentioned in Section~\ref{subsubsec:nebmodeling}, the lower ionized species (e.g. N\,{\sc ii} and S\,{\sc ii}) varied differently along the X-shooter slit than higher ionized species (e.g. Ar\,{\sc iii}). This also becomes evident from Fig.~\ref{fig:nebsubplots}, where the [N\,{\sc ii}]~654.8~{nm} and [S\,{\sc ii}]~671.6~{nm} lines show similar nebular line shapes (with three distinct Gaussians) where the [Ar\,{\sc iii}]~713.6~{nm} line has one main Gaussian and two much weaker side Gaussians. 

\begin{figure}[p]
\centering
\includegraphics[width=1\linewidth]{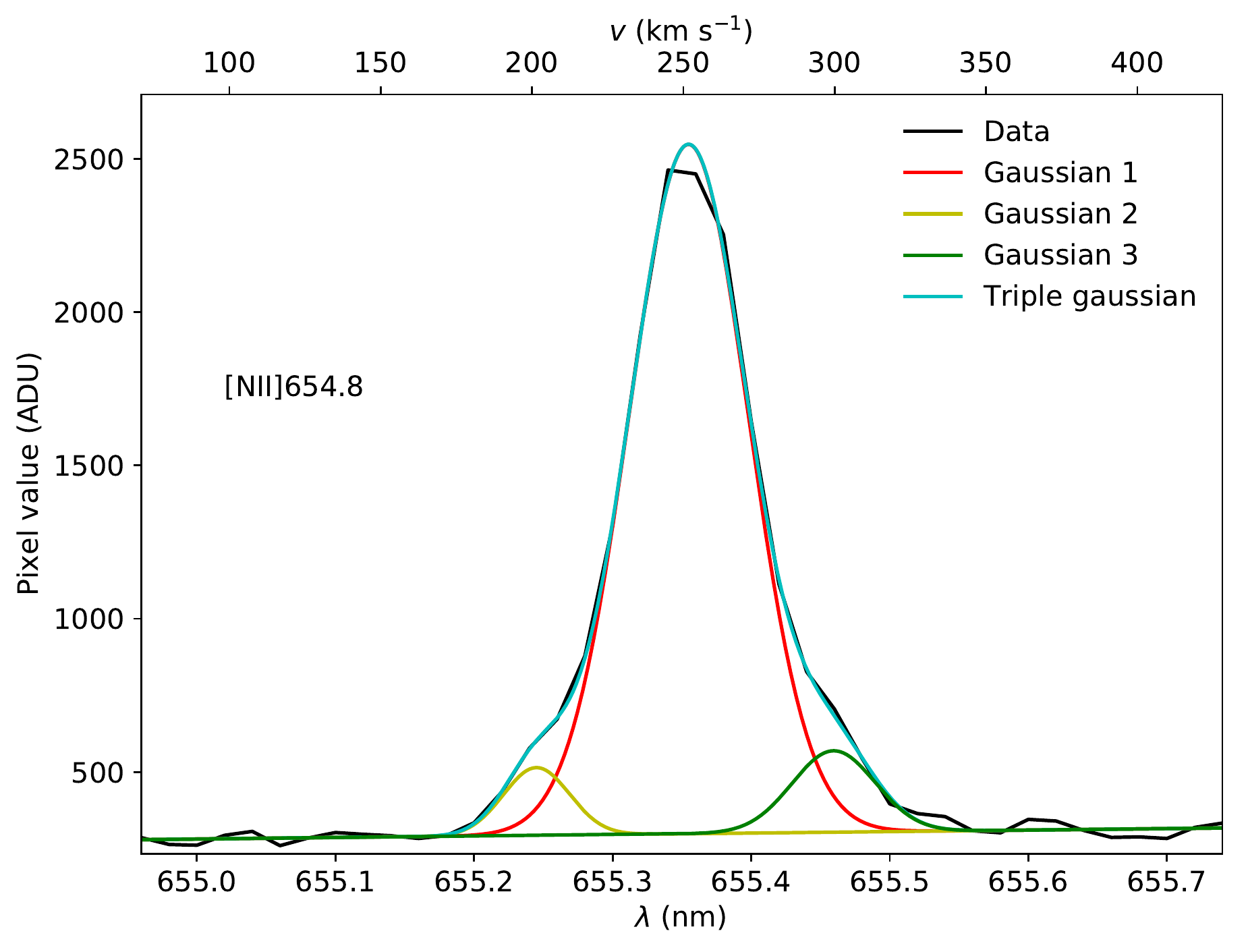}
\includegraphics[width=1\linewidth]{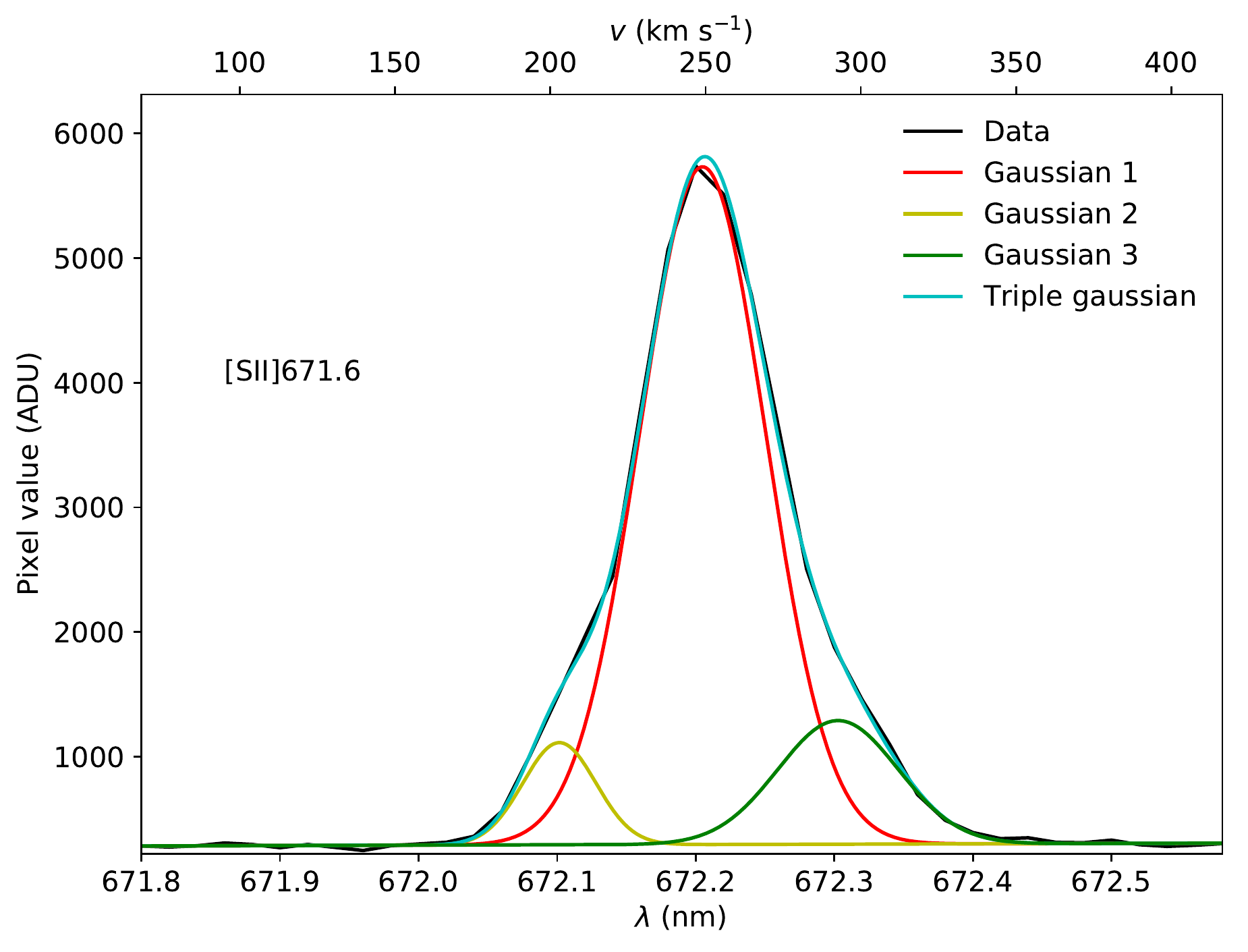}
\includegraphics[width=1\linewidth]{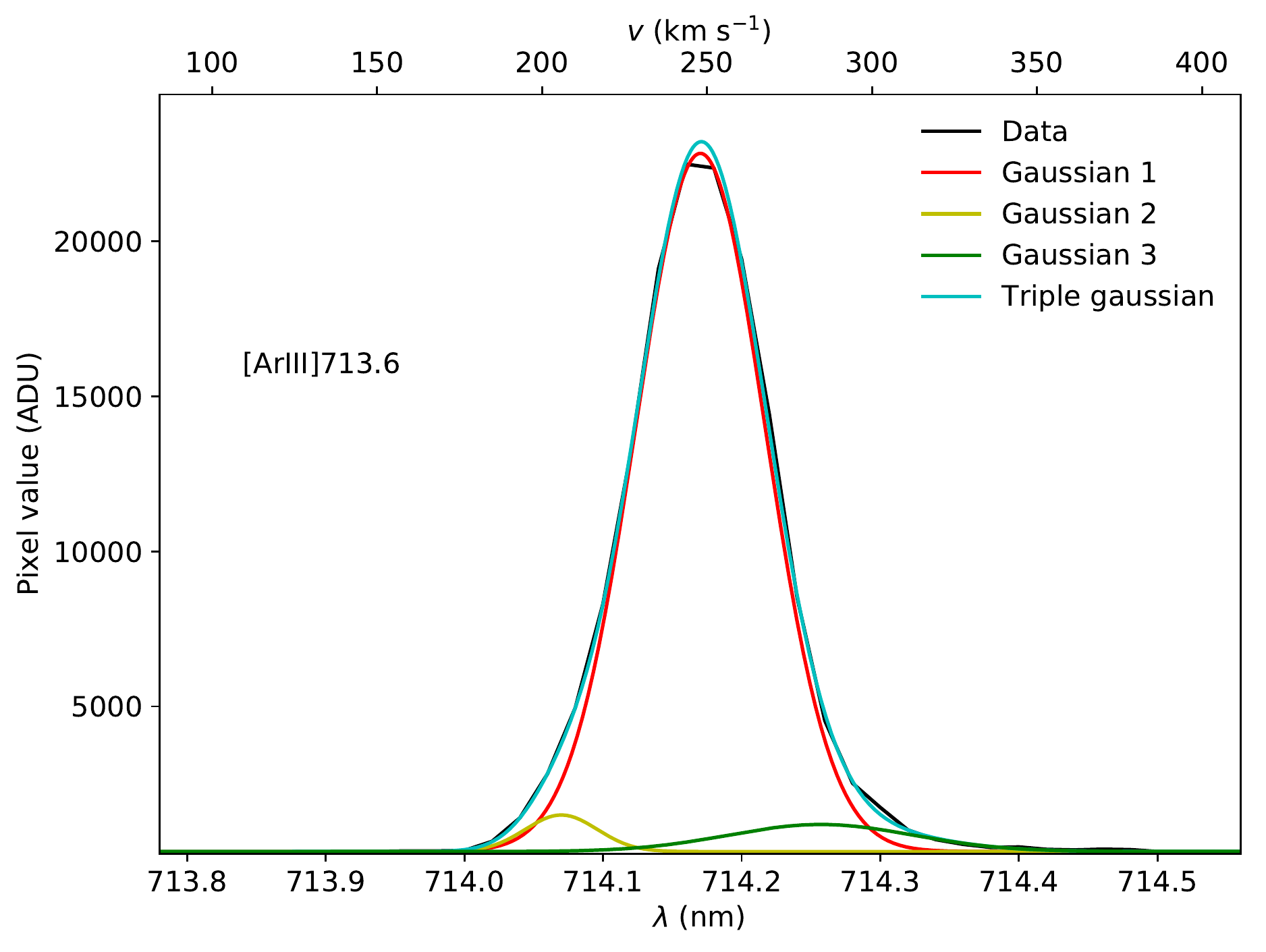}
\caption{{The [N\,{\sc ii}]~654.8 (top), [S\,{\sc ii}]~671.6 (middel), and [Ar\,{\sc iii}]~713.6 (bottom) nebular lines fitted with a triple Gaussian model. The model was extracted from the -5.1" position in Fig.~\ref{fig:S671var}.}}
\label{fig:nebsubplots}
\end{figure}
}
\clearpage

\section{Line plots}
\label{app:lineplots}

\noindent\begin{minipage}{\textwidth}
    \begin{center}
    \includegraphics[width=0.9\hsize]{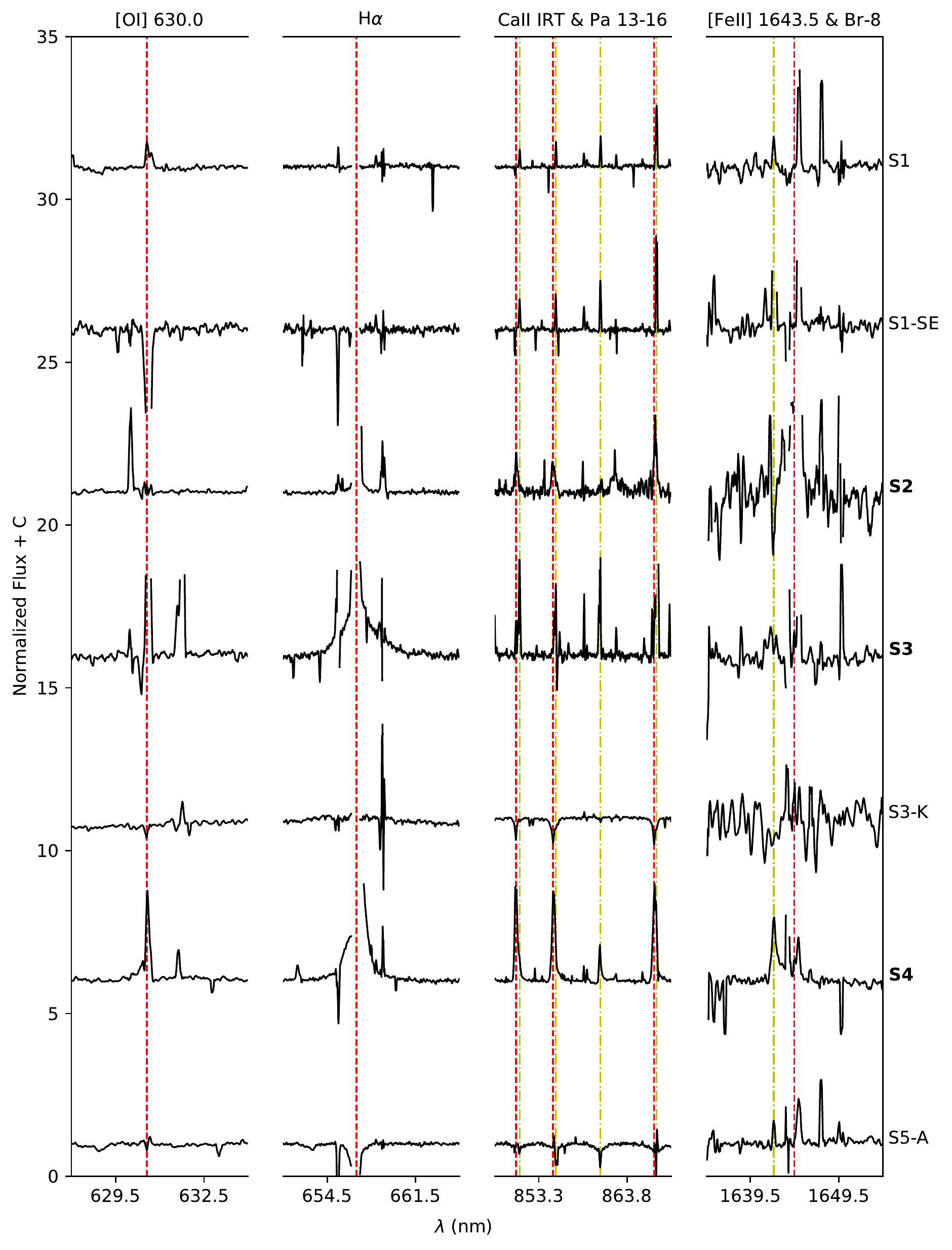}
    \captionof{figure}{The [O\,{\sc i}]~630.0~{nm}, H$\alpha$, \ac{CaII} and Pa-13--16, and {[Fe\,{\sc ii}]~1643.5~nm} regions shown for S1--S5-A. {For clarity, the [Fe\,{\sc ii}]~1643.5~nm region is enhanced}. The confirmed \ac{MYSO}s are boldfaced. We indicate the positions of the transitions with the red dashed lines (shifted with 260~\kms\text{} with respect to the heliocentric frame). {The Paschen series and Br-8 lines are marked by yellow dash-dotted lines for clarification.} The center of H$\alpha$ was saturated due to nebular emission and has been clipped. All narrow features are either telluric lines or residuals from the nebular or sky subtractions. \label{fig:spectra_targets_1}}
    \end{center}
\end{minipage}

\begin{figure*}[p]
\centering
\includegraphics[width=1\linewidth]{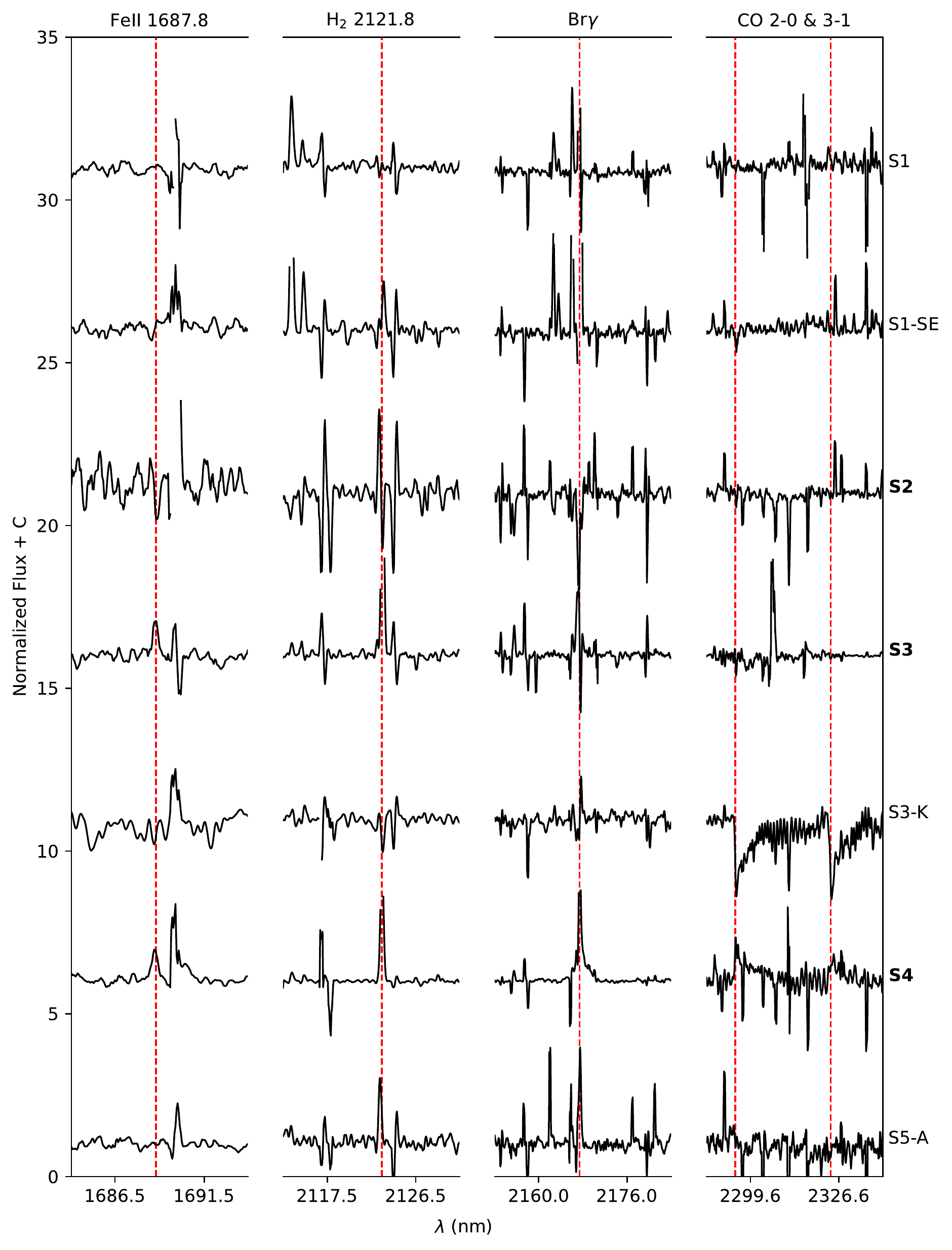}
\caption{The {Fe\,{\sc ii}~1687.7~{nm}}, {H$_2$~2121.8~{nm}}, Br$\gamma$, and the 2--0 and 3--1 CO bandhead regions shown for S1--S5-A. {All spectral regions} are enhanced for clarity. All other lines and features are the same as in Fig.~\ref{fig:spectra_targets_1}.}
\label{fig:spectra_targets_2}
\end{figure*}

\begin{figure*}[p]
\centering
\includegraphics[width=1\linewidth]{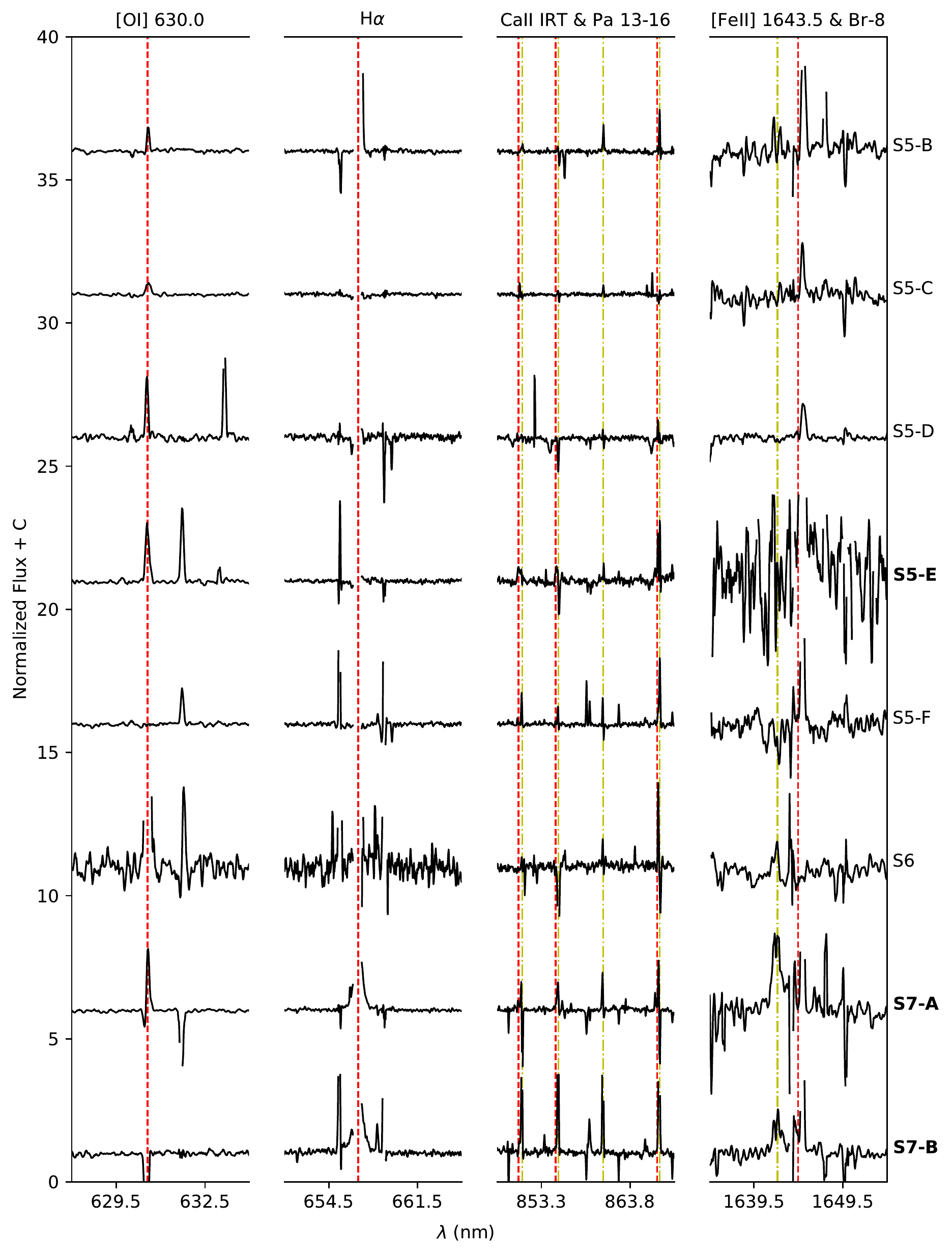}
\caption{Same as in Fig.~\ref{fig:spectra_targets_1} but now for S5-B--S7-B.}
\label{fig:spectra_targets_3}
\end{figure*}

\begin{figure*}[p]
\centering
\includegraphics[width=1\linewidth]{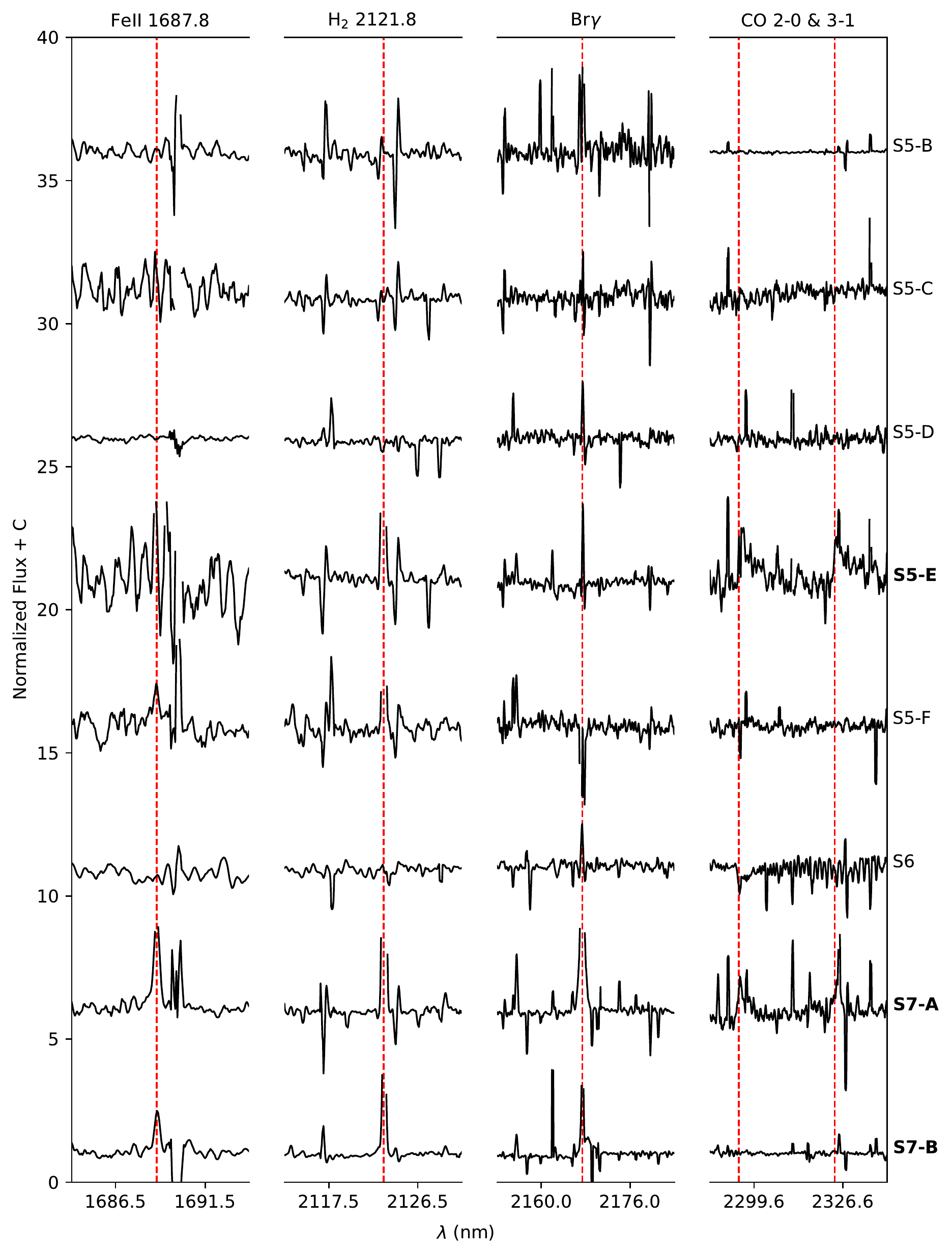}
\caption{Same as in Fig.~\ref{fig:spectra_targets_2} but now for S5-B--S7-B.}
\label{fig:spectra_targets_4}
\end{figure*}

\begin{figure*}[p]
\centering
\includegraphics[width=1\linewidth]{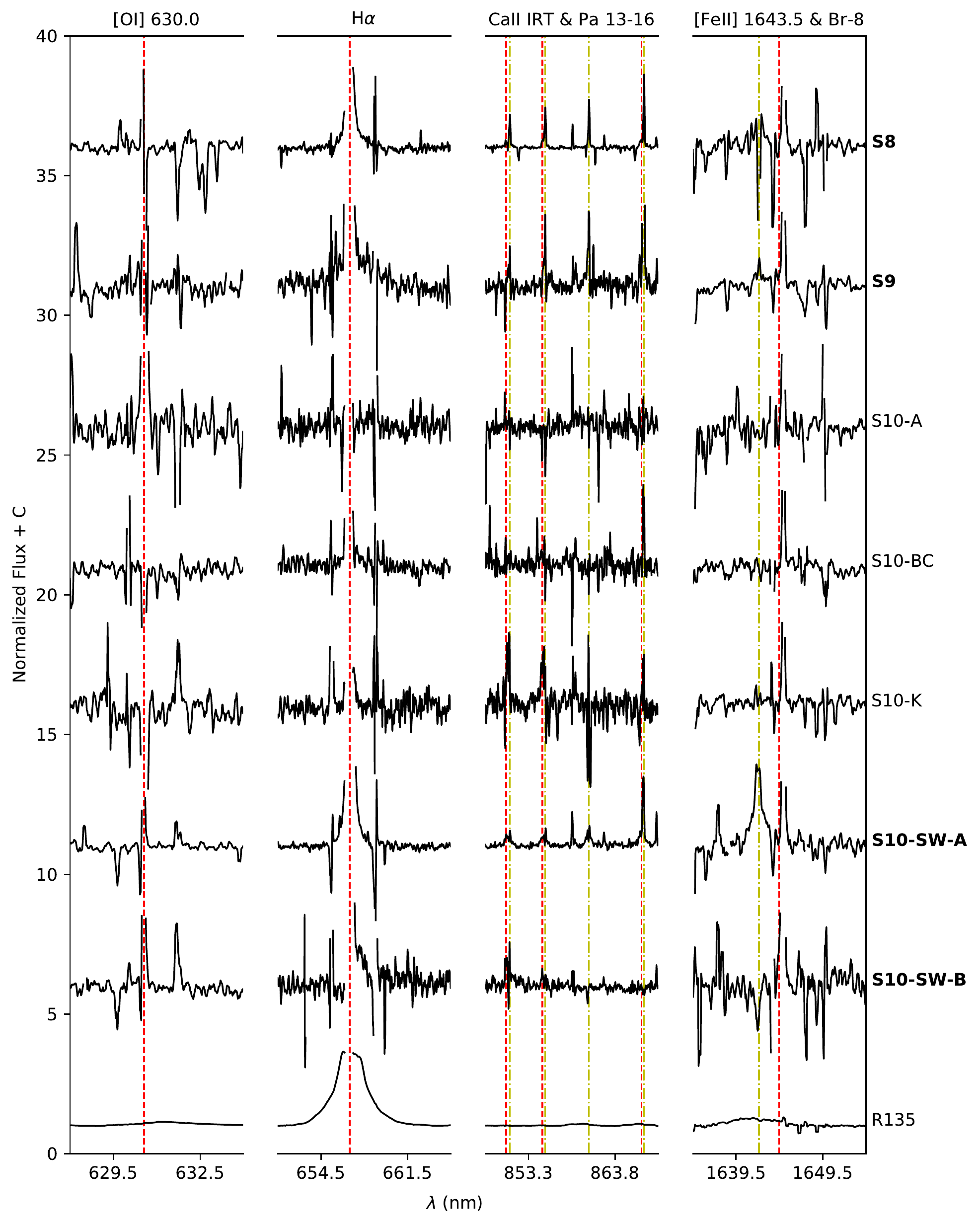}
\caption{Same as in Fig.~\ref{fig:spectra_targets_1} but now for S8--S10-SW-B, and R135.}
\label{fig:spectra_targets_5}
\end{figure*}

\begin{figure*}[p]
\centering
\includegraphics[width=1\linewidth]{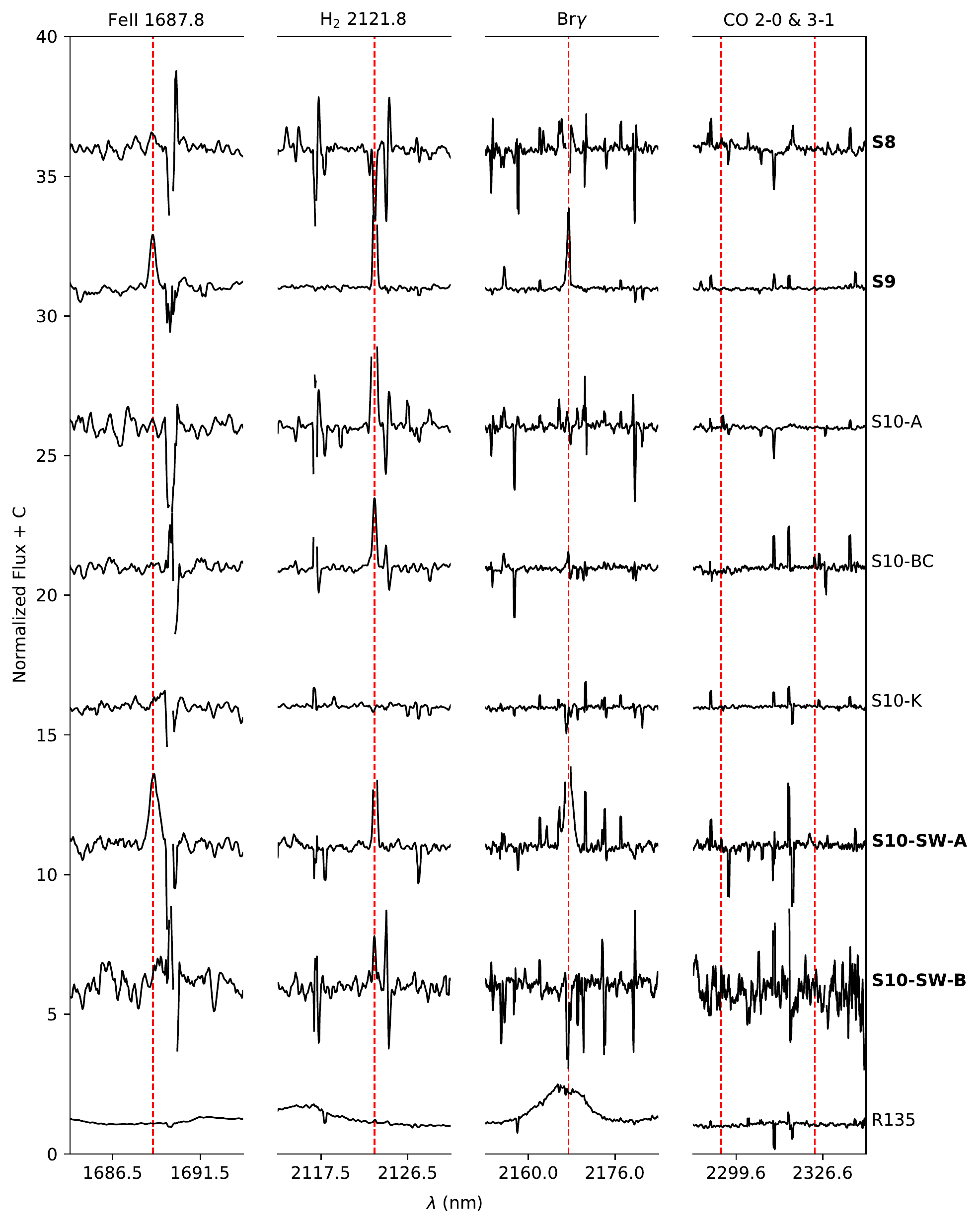}
\caption{Same as in Fig.~\ref{fig:spectra_targets_2} but now for S8--S10-SW-B, and R135.}
\label{fig:spectra_targets_6}
\end{figure*}

\clearpage

\section{X-shooter magnitudes}
\label{app:xshoomags}

\noindent\begin{minipage}{\textwidth}
\begin{center}
\captionof{table}{Photometric upper limits in multiple bands computed from our X-shooter spectra. }
\footnotesize{
\begin{tabular*}{\linewidth}{L{0.1\linewidth}C{0.07\linewidth}C{0.07\linewidth}C{0.07\linewidth}C{0.07\linewidth}C{0.07\linewidth}C{0.07\linewidth}C{0.07\linewidth}C{0.07\linewidth}C{0.07\linewidth}}
\hline
\hline
Object & $B$ & $V$ & $G$ & $R$ & $I$ & $Y$ & $J$ & $H$ & $K_\mathrm{s}$ \\
\hline
S1 & 21.3$\pm$0.4 & 20.8$\pm$0.4 & 19.9$\pm$0.3 & 19.7$\pm$0.4 & 18.7$\pm$0.5 & 18.6$\pm$0.3 & 17.9$\pm$0.2 & 16.8$\pm$0.4 & 16.1$\pm$0.3 \\
S1-SE & 22.5$\pm$0.7 & 21.4$\pm$0.4 & 20.3$\pm$0.3 & 20.1$\pm$0.4 & 18.9$\pm$0.5 & 18.6$\pm$0.2 & 17.8$\pm$0.2 & 16.6$\pm$0.5 & 15.7$\pm$0.3 \\
\textbf{S2} & -- & -- & -- & -- & -- & -- & 21.6$\pm$3.4 & 17.9$\pm$3.3 & 15.7$\pm$0.4 \\
\textbf{S3} & -- & -- & 19.7$\pm$0.5 & 19.0$\pm$0.5 & 18.7$\pm$0.6 & 18.9$\pm$0.4 & 18.6$\pm$0.4 & 17.4$\pm$0.5 & 15.8$\pm$0.5 \\
S3-K & 20.4$\pm$0.4 & 19.0$\pm$0.3 & 17.4$\pm$0.3 & 17.3$\pm$0.3 & 15.7$\pm$0.4 & 15.3$\pm$0.3 & 15.0$\pm$0.2 & 13.7$\pm$0.3 & 13.2$\pm$0.3 \\
\textbf{S4} & 19.1$\pm$0.4 & 19.0$\pm$0.4 & 18.2$\pm$0.3 & 18.0$\pm$0.4 & 17.1$\pm$0.5 & 17.3$\pm$0.3 & 16.7$\pm$0.2 & 14.5$\pm$0.3 & 12.7$\pm$0.3 \\
S5-A & 16.8$\pm$0.4 & 16.6$\pm$0.4 & 16.2$\pm$0.4 & 15.9$\pm$0.4 & 15.5$\pm$0.4 & 16.1$\pm$0.3 & 16.9$\pm$0.2 & 16.4$\pm$0.4 & 16.2$\pm$0.3 \\
S5-B & 19.1$\pm$0.5 & 19.1$\pm$0.4 & 18.7$\pm$0.4 & 18.5$\pm$0.5 & 18.2$\pm$0.4 & 19.0$\pm$0.3 & 19.9$\pm$0.6 & 19.6$\pm$2.4 & 18.9$\pm$3.1 \\
S5-C & 18.3$\pm$0.4 & 18.4$\pm$0.4 & 18.2$\pm$0.6 & 18.4$\pm$0.5 & 18.1$\pm$0.4 & 19.0$\pm$0.3 & 20.1$\pm$0.6 & 19.7$\pm$3.1 & 20.5$\pm$5.1 \\
S5-D & 20.6$\pm$0.6 & 20.0$\pm$0.4 & 19.1$\pm$0.3 & 18.7$\pm$0.4 & 17.9$\pm$0.4 & 18.3$\pm$0.3 & 18.3$\pm$0.3 & 17.2$\pm$3.9 & 17.3$\pm$0.5 \\
\textbf{S5-E} & 20.6$\pm$0.6 & 19.8$\pm$0.5 & 19.5$\pm$0.6 & 19.0$\pm$0.6 & 19.3$\pm$0.4 & 19.4$\pm$0.3 & 20.0$\pm$0.5 & 18.0$\pm$0.4 & 15.0$\pm$0.3 \\
S5-F & 17.8$\pm$0.4 & 17.6$\pm$0.5 & 17.5$\pm$0.6 & 17.7$\pm$0.4 & 17.2$\pm$0.6 & 18.1$\pm$0.4 & 19.4$\pm$0.5 & 18.9$\pm$2.5 & 17.4$\pm$0.8 \\
S6 & -- & 17.9$\pm$0.7 & 18.1$\pm$1.1 & -- & 20.2$\pm$1.0 & 18.4$\pm$0.4 & 17.2$\pm$0.2 & 14.8$\pm$0.5 & 13.2$\pm$0.3 \\
\textbf{S7-A} & 17.9$\pm$0.4 & 17.8$\pm$0.4 & 17.5$\pm$0.4 & 17.4$\pm$0.5 & 17.2$\pm$0.5 & 17.5$\pm$0.3 & 16.6$\pm$0.2 & 14.8$\pm$0.3 & 13.2$\pm$0.3 \\
\textbf{S7-B} & 19.7$\pm$0.4 & 19.2$\pm$0.4 & 18.7$\pm$0.3 & 18.4$\pm$0.4 & 17.8$\pm$0.6 & 18.6$\pm$0.6 & 17.7$\pm$0.2 & 15.3$\pm$0.3 & 13.1$\pm$0.3 \\
\textbf{S8} & 21.9$\pm$0.5 & 20.8$\pm$0.4 & 19.5$\pm$0.3 & 19.4$\pm$0.4 & 18.0$\pm$0.5 & 17.8$\pm$0.3 & 17.6$\pm$0.3 & 16.2$\pm$0.5 & 14.9$\pm$0.4 \\
\textbf{S9} & -- & 21.1$\pm$0.5 & 20.9$\pm$0.5 & 20.8$\pm$0.5 & 20.0$\pm$0.6 & 20.1$\pm$0.6 & 19.8$\pm$0.8 & 17.4$\pm$2.3 & 15.6$\pm$3.7 \\
S10-A & -- & 21.1$\pm$0.5 & 20.9$\pm$0.5 & 21.0$\pm$0.6 & 20.0$\pm$0.7 & 20.9$\pm$1.3 & 20.8$\pm$1.8 & -- & 16.2$\pm$0.5 \\
S10-BC & -- & 21.0$\pm$0.5 & 20.8$\pm$0.5 & 20.6$\pm$0.6 & 20.0$\pm$0.6 & 20.4$\pm$0.7 & 20.0$\pm$0.7 & 18.0$\pm$0.6 & 16.4$\pm$0.4 \\
S10-K & -- & 21.1$\pm$0.6 & 21.1$\pm$0.7 & 21.2$\pm$1.2 & 20.5$\pm$0.7 & 21.5$\pm$2.9 & 20.2$\pm$0.8 & 17.5$\pm$2.8 & 15.8$\pm$0.4 \\
\textbf{S10-SW-A} & 22.4$\pm$0.7 & 21.2$\pm$0.4 & 20.1$\pm$0.4 & 19.7$\pm$0.4 & 18.8$\pm$0.6 & 18.6$\pm$0.3 & 17.7$\pm$0.2 & 16.1$\pm$0.5 & 14.7$\pm$0.3 \\
\textbf{S10-SW-B} & -- & 22.3$\pm$0.8 & 22.4$\pm$1.2 & 22.8$\pm$2.4 & 23.3$\pm$6.7 & 21.0$\pm$1.3 & 19.7$\pm$0.4 & 17.9$\pm$0.4 & 16.6$\pm$0.4 \\
R135 & 13.5$\pm$0.5 & 13.7$\pm$0.4 & 13.2$\pm$0.5 & 12.9$\pm$0.5 & 12.9$\pm$0.4 & 13.4$\pm$0.2 & 13.4$\pm$0.2 & 13.2$\pm$0.3 & 13.1$\pm$0.3 \\
\hline
\end{tabular*}
}
\tablefoot{All magnitudes are upper limits computed using Eq.~\eqref{eq:appmag}. For the $B$, $V$, $R$, and $I$-bands we used the Bessell photometric system \citep{Bessell1990}, for the $G$-band we used the Gaia photometric system \citep{GAIA2016}, and for the $Y$, $J$, $H$ and $K_\mathrm{s}$-bands we used the \acs{VISTA} photometric system. The empty band values resulted from a negative integrated flux in the corresponding band. The objects shown in bold are the \ac{MYSO}s confirmed in this work.} 
\label{tab:xshootphots_tab}

\end{center}
\end{minipage}

\clearpage

\end{appendix}

\end{document}